\documentclass[12pt]{extarticle}
\usepackage[utf8]{inputenc}
\usepackage[a4paper]{geometry}
\usepackage{color}
\usepackage{amsmath}
\usepackage{amsthm}
\usepackage{amssymb}
\usepackage{graphicx}
\usepackage[authoryear]{natbib}
\usepackage[unicode=true,
 bookmarks=false,
 breaklinks=false,pdfborder={0 0 1},backref=section,colorlinks=true]
 {hyperref}
\hypersetup{
 citecolor=blue,linkcolor=red}

\makeatletter

\providecommand{\tabularnewline}{\\}

\usepackage{lineno}
\usepackage{amsfonts}
\usepackage{enumerate}
\usepackage{psfrag}
\usepackage{color}
\usepackage{bbm}
\usepackage{dcolumn}
\usepackage{algorithm}
\usepackage{algcompatible}
\usepackage{psfrag}
\usepackage{color}
\usepackage{fancyhdr}
\usepackage{amsthm}
\usepackage{mathrsfs}
\usepackage{indentfirst}
\usepackage{epsfig}
\usepackage{graphics}
\usepackage{lscape}

\newtheorem{theorem}{THEOREM}\newtheorem{corollary}{COROLLARY}\newtheorem{remark}{REMARK}

\@ifundefined{showcaptionsetup}{}{%
 \PassOptionsToPackage{caption=false}{subfig}}
\usepackage{subfig}
\makeatother

\begin{document}
\title{Bi-integrative analysis of two-dimensional heterogeneous panel data
model}
\author{Wei Wang\thanks{Email Address: wangwei\_0115@outlook.com.}
\and Xiaodong Yan\thanks{Co-first author. Zhongtai Securities Institute for Financial Studies, Shandong University, China, Email Address: yanxiaodong@sdu.edu.cn.}  \and Yanyan Ren\thanks{School of Economics, Shandong University, China. Email Address: ryy1996@163.com. }
\and Zhijie Xiao \thanks{Corresponding author. Department of Economics, Boston College, USA, Email Address:zhijie.xiao@bc.edu.}}
\maketitle
\begin{abstract}
\baselineskip=18.0pt Heterogeneous panel data models that allow the
coefficients to vary across individuals and/or change over time have
received increasingly more attention in statistics and econometrics.
This paper proposes a two-dimensional heterogeneous panel regression
model that incorporate a group structure of individual heterogeneous
effects with cohort formation for their time-variations, which allows
common coefficients between nonadjacent time points. A bi-integrative
procedure that detects the information regarding group and cohort
patterns simultaneously via a doubly penalized least square with concave
fused penalties is introduced. We use an alternating direction method
of multipliers (ADMM) algorithm that automatically bi-integrates the
two-dimensional heterogeneous panel data model pertaining to a common
one. Consistency and asymptotic normality for the proposed estimators
are developed. We show that the resulting estimators exhibit oracle
properties, i.e., the proposed estimator is asymptotically equivalent
to the oracle estimator obtained using the known group and cohort
structures. Furthermore, the simulation studies provide supportive
evidence that the proposed method has good finite sample performence.
A real data empirical application has been provided to highlight the
proposed method.

\textbf{Keywords:} Panel Data, Bi-integration, Two-dimensional heterogeneity,
Group Structure, Cohort Structure, Fused penalty. 
\end{abstract}
\baselineskip=18.0pt

\section{Introduction}

Panel (or longitudinal) data models have been widely-used in economics,
finance, and many other fields. Panel models exhibit various advantages
in combining useful cross-sectional and time series information in
the data. Traditional homogeneous panel data model assumes that the
slope coefficients are constant across individuals and periods. However,
homogeneous assumption maybe too restrictive in many applications.
In practice, both cross-sectional and time domain variations are observed.
Many panel datasets cover lots of individuals coming from different
backgrounds, such as different experimental methods, distinct crowds
or geographic locations (census, tract, county, state, etc.), external
classification, observable explanatory categories, nested (hierarchical)
or non-nested datasets, which lead to heterogeneity across individuals.
In addition, the heterogeneity usually represents some individual
characteristics that are unobserved, such as the ability of individuals
\citep{BH2002} in the labor market, the loan willingness for banks
\citep{cornett2011}. On the other side, time-specified coefficients
captures unobserved time-varying behavior such as the historical events
in the process of democratization\citep{BM2015}, technological progress,
institutional transformation, or economic transition. For these and
other reasons, it is important to take into account for both cross-sectional
and temporal heterogeneity in many applications.

Over the last few years, there is a fast growing literature on panel
data model with heterogeneous slope coefficients, along two directions.
One direction of research assumes that the regression coefficients
are time-varying. In this case, the regression coefficients are assumed
to be functions of time trend under the nonparametric framework \citep{Li2011,Pei2018}.
In order to deal with the incidental parameter problem, some researchers
assume that there exist unknown common breaks \citep{Bai2010,Kim2011}
or multiple structural breaks \citep{QianSu2016,Li2017} of regression
coefficients in prior, so that the difference of regression coefficients
are successively pairwise sparse in time-dimension. Most of these
works use shrinkage or fused penalty method to detect and estimate
multiple change points simultaneously. The other direction assumes
that the slope coefficients are heterogeneous across individuals,
due to some individual-specific characteristics \citep{Bester2016,FGPZ2017}.
This literature includes complete heterogeneity and group-based heterogeneity.
Complete heterogeneity assumes that slope coefficients are different
across each individuals, and are usually modelled by random coefficient
panel data models \citep{Wooldridge2005,MW2008}, nonparametric model
\citep{Boneva2015,Vogt2017} or other complete heterogeneous settings,
see \citet{Pesaran2006}, \citet{CPT2011}, \citet{KHJW2017}, among
others. Group heterogeneous models assume that individuals can be
classified into different groups, where the regression coefficients
are the same within each group but heterogeneous across groups. Another
approach uses finite mixture models in discrete choice panel data
models with parametric\citep{KS2009} or nonparametric method \citep{BC2010}.
A third method is clustering or integrating the group membership and
estimating parameters simultaneously by solving the penalized objective
function added the penalty term of slope coefficients between different
individuals, such as the C-lasso \citep{Su2016,SuJu2018,Huang2018},
Panel-CARDS\citep{WangSu2018} and others.

An important issue in practice is that individual heterogeneity and
time variation may occur simultaneously. For example, the saving-retention
coefficients or saving-investment relations are heterogeneous across
countries and periods in the famous Feldstein-Horioka puzzle. For
this reason, research attention has been recently shifted to panel
data model under two-dimensional heterogeneity. The main challenge
in this situation is the increasing number of unknown parameters along
with N and T. Thus, dimensional decomposition strategy becomes popular
for reducing the number of coefficients along the two-dimensions in
parametric model. \citet{Baltagi2016} extend the common correlated
estimation(CCE) with common structural break in the individual-specific
slope coefficients. \citet{Smith2018} develops a new Bayesian approach
to estimate non-common structural breaks in panel regression models.
\citet{Neal2018}, \citet{LuSu2019} decomposed the two-dimensional
heterogeneous regression coefficients into three parts additively,
i.e., the average component, individual-specific and time-varying
component respectively. \citet{chernozhukov2018} considered interactive
pattern of two-dimensional heterogeneous regression coefficients.
Another technique identifying the amounts of unknown coefficients
with two-dimensions is to assume block structures, for example, \citet{OW2020,RobinOkuiWang2020}
consider block-based structural slope coefficients with structural
breaks and time-invariant grouped structure simultaneously and utilize
fused lasso to detect the true pattern. In addition, \citet{Su2019}
use nonparametric method to allow slope coefficients as a smooth function
of time, synchronously considering heterogeneity across units.

In this paper, we propose a method of estimation and inference of
panel data model with a more flexible heterogeneous structure. We
argue that the concave pairwise fusion approach, proposed by \citet{Ma2016,MaHuang2017},
can be extended to conduct a Bi-integrative analysis of Group and
Cohort Recovery (BIGCORE) for two-dimensional heterogeneous panel
structure models. In particular, the BIGCORE method deals with panel
structural model where the regression coefficients have block structure,
in which the coefficients of observations within the same block are
identical, but distinct across blocks in the rectangular arrangement
of the two dimensional heterogeneous coefficients. The two-dimensional
heterogeneous panel model is general, not only including the existing
homogeneous panel data model, panel structural model with pure grouped
structure or structural breaks, but also time-varying grouped structure
with multiple structural breaks, diverse multiple structural breaks
with time-varying grouped structure, and others. Therefore, our model
in this paper is more general than the previous research and has great
potential in empirical analysis for panel data with grouped and structural
changes. Compared to the existing methods, our approach do not require
strong assumptions that the coefficients have specific sparse structure,
such as common structural breaks, invariant group membership and others
and similarly, our approach also need not to determine the number
of groups and change points in prior.

We use the ADMM algorithm to solve the optimization problem with double
fused penalties. We establish that the estimators are consistent and
asymptotically normal. We prove that the estimators have the oracle
property in the sense that it is asymptotically equivalent to the
infeasible estimator with known two-dimensional heterogeneous structure.
Monte Carlo simulations are conducted and show nice sampling properties
of our estimators in finite sample. Finally we illustrate the potential
of our methods by an empirical application.

The rest of this paper is organized as follows. In Section 2, the
two-dimensional panel structure model and the proposed estimation
method are presented. In Section 3, an estimation procedure based
on the ADMM algorithm is given to solve the optimization problem.
In Section 4, we derive the asymptotic properties of the estimator.
Section 5 discusses the determination of the penalty parameters and
initial values. Section 6 conducts a Monte Carlo simulation. In section
7, we apply the proposed approach to a real dataset. Finally, we conclude
the paper.

\textbf{Notation.} we introduce following notations that will be used
throughout this paper. $\otimes$ is the Kronecker product, $\circ$
is the Hadamard product, $vec$ is the vectorization operator, $\gg$
denote much greater, the superscript $\top$ denote the transpose
of a matrix, $\|\cdot\|$ stands for the Euclidean norm for vector,
$\|\cdot\|_{F}$ denote the Frobenius norm of matrix. $\langle a,b\rangle=a^{\top}b$
be the inner product of two vectors a and b with the same dimension.
$A^{+}$ denotes a vector obtained from row sums of matrix A. For
a given vector $b=(b_{1},\ldots,b_{t})\in\mathbb{R}^{t}$ and a symmetric
matrix $A_{t\times t}$, define $\|b\|_{\infty}=\max_{1\leq s\leq t}|b_{s}|$,
$\|A\|_{\infty}=\max_{1\leq i\leq t}\sum_{j=1}^{t}|A_{ij}|$, $\|A\|=\|A\|_{2}=\max_{b\in\mathbb{R}^{t},\|b\|=1}\|Ab\|$
and $\|A\|_{2,\infty}=\max_{1\leq i\leq t}\|A_{i,}\|$, where $A_{i,}$
denotes vector of $i$th row of $A$. $\gamma_{\min}(A)$ and $\gamma_{\max}(A)$
be the smallest and largest eigenvalues of $A$ respectively. $\overset{D}{\rightarrow}$
denotes convergence in distribution.

\section{The Model}

Giving a panel dataset $\{(y_{it},z_{it}):i=1,\cdots,N;t=1,\cdots,T\}$,
where $N$ and $T$ correspond to the total number of individuals
and periods respectively, we consider the following heterogeneous
regression model with two-way varying coefficients: 
\begin{equation}
y_{it}=\mu_{it}+\boldsymbol{z}_{it}^{\top}\boldsymbol{\eta}_{it}+\epsilon_{it},\qquad i=1,\cdots,N;t=1,\cdots,T,\label{model1}
\end{equation}
where $y_{it}\in\mathbb{R}^{1}$ is the dependent variable, $\mu_{it}$
is the time-varying individual fixed effect, $\boldsymbol{z}_{it}=(z_{it(1)},\cdots,z_{it(P-1)})^{\top}$
is $P-1$ dimensional regressors with slope coefficients $\boldsymbol{\eta}_{it}={(\boldsymbol{\eta}_{it1},\cdots,\boldsymbol{\eta}_{it(P-1)})^{\top}}\in\mathbb{R}^{(P-1)}$
that are potentially heterogeneous in both individual and temporal
dimensions and $P$ is fixed. $\epsilon_{it}$'s are independent random
errors with mean zero and standard error $\sigma$. In this model,
the fixed effect $\mu_{it}$ and the slope coefficients $\boldsymbol{\eta}_{it}$
may vary in both individual and temporal dimensions.

In order to identify the unknown two-dimensional regression coefficients,
we assume the following block structure, which combines grouped pattern
among individuals with cohort structure across time. The time cohort
structure is flexible, it allows the existence of common coefficients
between nonadjacent time points and also includes structural breaks
as special cases.

Let $\boldsymbol{\beta}_{it}=(\mu_{it},\boldsymbol{\eta}_{it}^{\top})^{\top}$,
the true unknown block structure can be characterized in the following
form: 
\begin{equation}
\boldsymbol{\beta}_{it}=\begin{cases}
\boldsymbol{\alpha}_{1}, & \mbox{if}\;(i,t)\in\mathcal{A}_{1},\\
\boldsymbol{\alpha}_{2}, & \mbox{if}\;(i,t)\in\mathcal{A}_{2},\\
\ \vdots & \qquad\vdots\\
\boldsymbol{\alpha}_{L}, & \mbox{if}\;(i,t)\in\mathcal{A}_{L},
\end{cases}\label{block}
\end{equation}
where $L$ is the unknown number of blocks with unknown partition
of rectangle $\{\mathcal{A}_{l}:1\leq l\leq L\}$.

\begin{figure}[th]
\centering \subfloat[constant group memberships with common structural break]{\includegraphics[width=5cm,height=5cm]{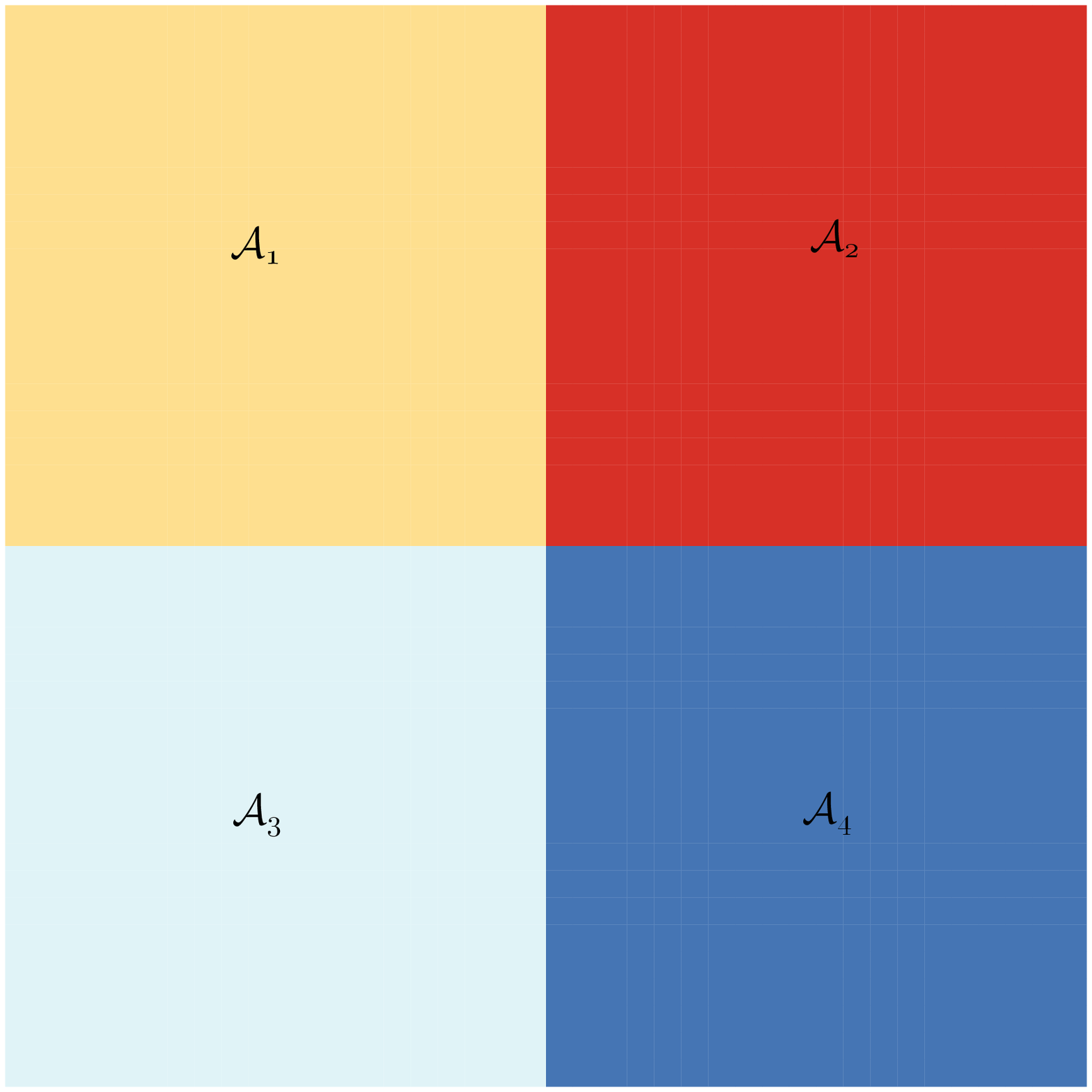}\label{subfig:1}

}\hspace{30pt} \subfloat[time-varying group memberships with common structural break]{\includegraphics[width=5cm,height=5cm]{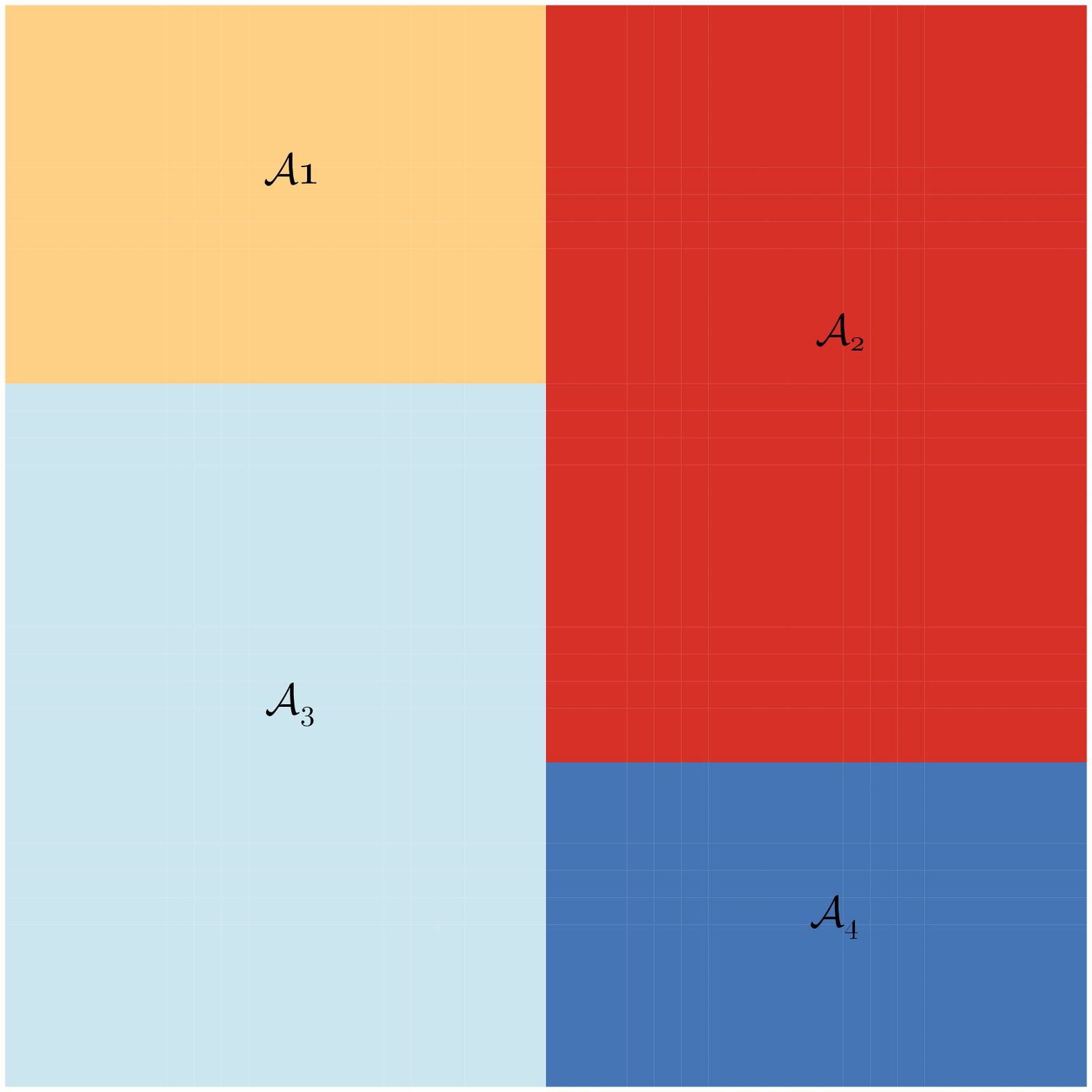}\label{subfig:2}

}\\
 \subfloat[constant group memberships with noncommon structural break]{\includegraphics[width=5cm,height=5cm]{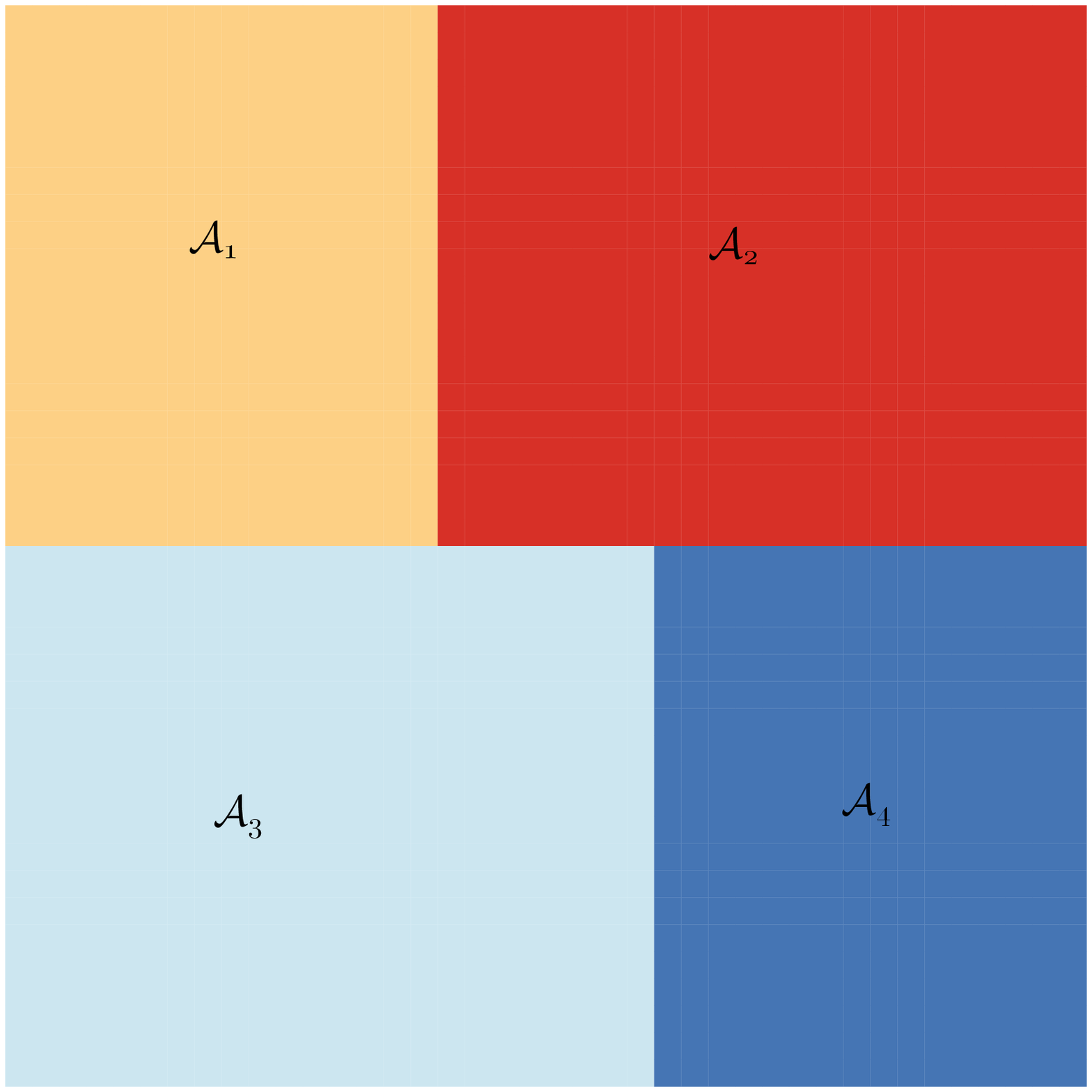}\label{subfig:3}

}\hspace{30pt} \subfloat[time-varying group memberships with noncommon structural breaks]{\includegraphics[width=5cm,height=5cm]{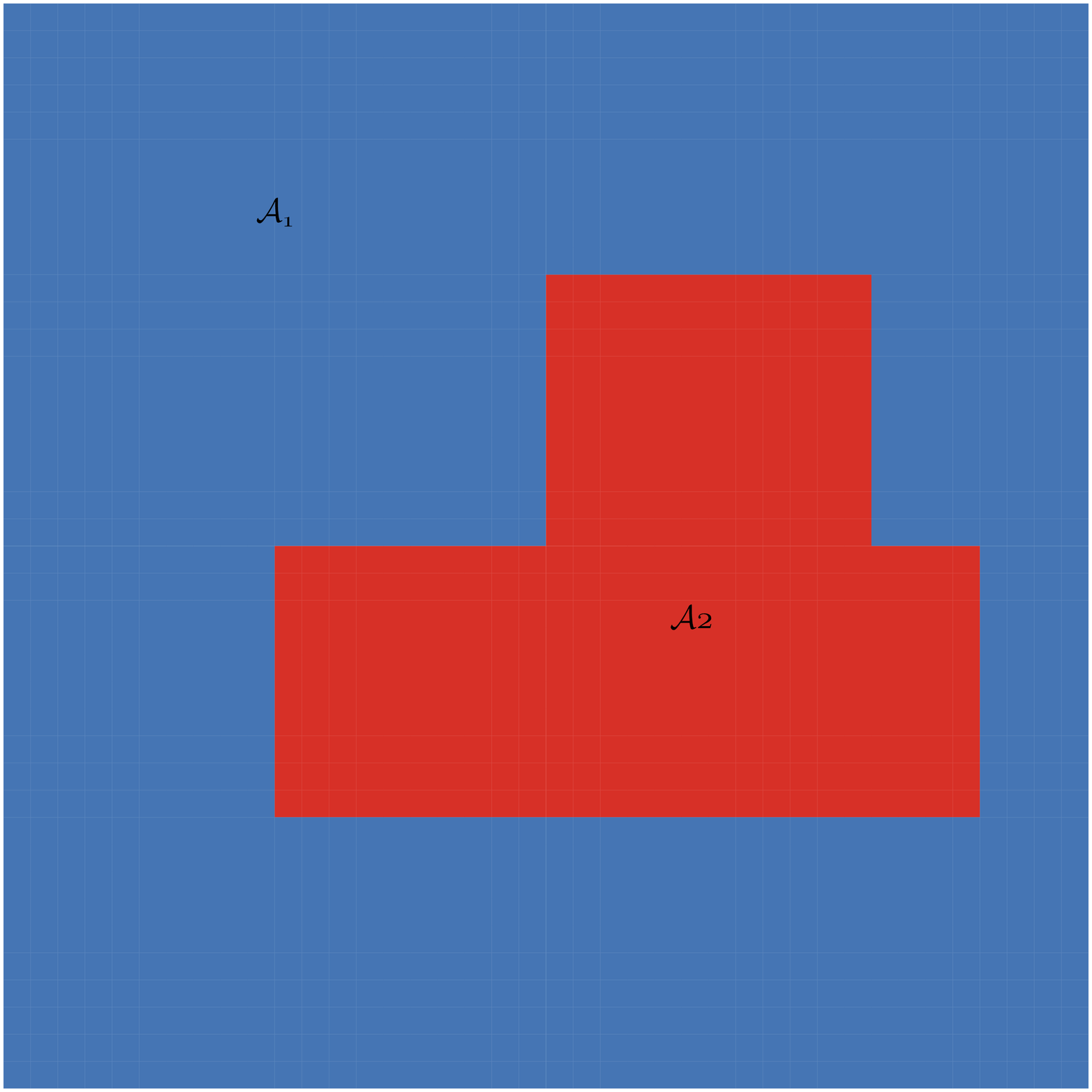}\label{subfig:4}

}\caption{Block structures on regression coefficients: the rows represent individuals
and the columns represent the periods.}
\label{RealPlot} 
\end{figure}

The block structure on regression coefficients given by \eqref{block}
is quite general. Apparently, it includes many classical structures,
such as: homogeneous constant with $\boldsymbol{\beta}_{it}=\boldsymbol{\alpha}$
for all $i=1,\cdots,N$ and $t=1,\cdots,T$; grouped pattern corresponding
to $\boldsymbol{\beta}_{it}=\boldsymbol{\alpha}_{l}$, $(i,t)\in\mathcal{A}_{l}$
for all $t=1,\cdots,T$; and structural breaks with $\boldsymbol{\beta}_{it}=\boldsymbol{\alpha}_{l}$,
$(i,t)\in\mathcal{A}_{l}$ for all $i=1,\cdots,N$ and $t_{l1}\leq t\leq t_{lk_{l}}$,
where $t_{lj}$ for $j=1,\cdots,k_{l}$ is the time to event in the
vertical ordinate of $\mathcal{A}_{l}$. Furthermore, it also includes
some irregular heterogeneous structures depicted Figure (\ref{RealPlot}),
which illustrates some values of coefficients matrix, where the vertical
ordinate represents the individuals and the horizontal ordinate represents
temporal points. The block structure in our paper given by \eqref{block}
under two-dimensional heterogeneity also includes the time-varying
group memberships with common structural break in Figure (\ref{subfig:2}),
constant group memberships with non-common structural in Figure (\ref{subfig:1}),
identical group memberships with non-common structural breaks that
depicted in Figure (\ref{subfig:3}), which is the same as the block
structure on regression coefficients considered by \citet{OW2020}.
The block structure in Figure (\ref{subfig:4}) is more complex: group
memberships are time-varying in individual dimension and the cohorts\footnote{The pattern that homogeneous coefficient exits in adjacent or nonadjacent
temporal dimension is called cohort structure in our paper, which
is similar to the grouped structure. Therefore, structural breaks
can be regarded as a special case of cohort structure. } are different across groups. Individuals are divided into three groups,
where the first quarter and the last quarter belong to identical group
with constant and the second quarter and third quarter consist of
different cohorts, where parts of nonadjacent time periods have common
values. In reality of economics, there may exist other complex block
structure under two-dimensional heterogeneity that also can be included
in the setting of \eqref{block}. To find out the above block structure,
we need to estimate three sets of parameters: the block-specified
coefficients $\boldsymbol{\alpha}$, the membership $\mathcal{A}_{l}$'s
of individuals and time, and the number of blocks $L$.

In the traditional homogeneous panel data models with individual or
time fixed effects, a commonly-used technique to tackle incidental
parameter problem is 'difference'. However, this strategy of eliminating
the heterogeneous fixed effects is invalid whenever the regression
slop coefficients are heterogeneous, no matter one-dimensional (i.e.,
$\boldsymbol{\eta}_{i}$ or $\boldsymbol{\eta}_{t}$) or two-dimensional
heterogeneous (i.e., $\boldsymbol{\eta}_{it}$).

In this paper, to estimate the panel regression model with two-dimensional
heterogeneous structure given by (\ref{block}), we propose a bi-integrative
procedure via doubly penalized least square with concave fused penalties.
Penalized procedures are commonly used for parameter estimation and
sparsity structure recovery.

To estimate the parameters $\boldsymbol{\beta}=\left(\boldsymbol{\beta}_{11}^{\top},\boldsymbol{\beta}_{12}^{\top},\cdots,\boldsymbol{\beta}_{1T}^{\top},\cdots,\boldsymbol{\beta}_{N1}^{\top},\boldsymbol{\beta}_{N2}^{\top},\cdots,\boldsymbol{\beta}_{NT}^{\top}\right)^{\top}$,
and recover block structure under the fused sparse assumption $\Vert\boldsymbol{\beta}_{it}-\boldsymbol{\beta}_{jt^{\prime}}\Vert=0$
for $(i,t)$ and $(j,t^{\prime})$ belonging to common block $\mathcal{A}_{l}$
for $l=1,\cdots,L$, we consider the following double penalized least
squares objective function: 
\begin{equation}
\ell_{p}(\boldsymbol{\beta};\gamma,\lambda)={\displaystyle \frac{1}{2}\sum\limits _{i=1}^{N}\sum\limits _{t=1}^{T}(y_{it}-\boldsymbol{x}_{it}^{\top}\boldsymbol{\beta}_{it})^{2}+\sum_{t=1}^{T}\sum\limits _{i<j}\mathcal{P}_{\lambda}(\Vert\boldsymbol{\beta}_{it}-\boldsymbol{\beta}_{jt}\Vert)+\sum_{i=1}^{N}\sum\limits _{t<t^{\prime}}\mathcal{P}_{\gamma}\left(\left\Vert \boldsymbol{\beta}_{it}-\boldsymbol{\beta}_{it^{\prime}}\right\Vert \right)},
\end{equation}
where $\mathcal{P}_{\lambda}(\cdot)$ and $\mathcal{P}_{\gamma}(\cdot)$are
pairwise concave penalty functions, for example, SCAD penalty\citep{Fan2001}
with tuning parameters $\lambda$ 
\[
\mathcal{P}_{\lambda}(\kappa)=\lambda\int_{0}^{\kappa}\left(1-x/(\lambda\pi)\right)_{+}dx,
\]
and MCP penalty\citep{Zhang2010} with tuning parameters $\gamma$,
\[
\mathcal{P}_{\gamma}(\kappa)=\gamma\int_{0}^{\kappa}\text{min}\{1,(\pi-x/)_{+}/(\pi-1)\}dx,
\]
where the fixed parameter $\pi$ controls the concavity of the penalty
function, $\kappa$ represents the pairwise term between individuals
or periods.

Notice that the penalty functions in our objective function are composed
of two parts: $\mathcal{P}_{\lambda}(\cdot)$ and $\mathcal{P}_{\gamma}(\cdot)$,
where $\mathcal{P}_{\lambda}(\cdot)$ classifies individuals into
the grouped structure and $\mathcal{P}_{\gamma}(\cdot)$ is used to
integrate observations across different (adjacent and nonadjacent)
periods into the cohort structure, apparently including detection
of structural breaks. $\lambda$, $\gamma\geq0$ are tuning parameters
that control the amount of penalty on $\Vert\boldsymbol{\beta}_{it}-\boldsymbol{\beta}_{jt}\Vert$'s
and $\Vert\boldsymbol{\beta}_{it}-\boldsymbol{\beta}_{it^{\prime}}\Vert$'s,
respectively and determine an estimation path of the coefficient matrix
$\boldsymbol{\beta}$, in which it can shrink $\Vert\boldsymbol{\beta}_{it}-\boldsymbol{\beta}_{jt}\Vert$'s
and $\Vert\boldsymbol{\beta}_{it}-\boldsymbol{\beta}_{it^{\prime}}\Vert$'s
towards zero with large enough values of $\lambda$ or $\gamma$.

For given $\lambda$ and $\gamma$, we define 
\begin{equation}
\widehat{\boldsymbol{\beta}}(\lambda,\gamma)=\mathrm{argmin}_{\boldsymbol{\beta}\in\mathbb{R}^{NTP\times1}}\ \ell_{p}(\boldsymbol{\beta};\gamma,\lambda),\label{eq:path}
\end{equation}
and the values of $\lambda$ and $\gamma$ can be selected via a properly
constructed Bayesian Information Criterion in the following sections.
Specifically, for $\gamma\in[\gamma_{\mathrm{min}},\gamma_{\mathrm{max}}]$,
$\lambda\in[\lambda_{\mathrm{min}},\lambda_{\mathrm{max}}]$, let
the values of $\gamma$ and $\lambda$ be from a grid $\gamma_{\mathrm{min}}=\gamma_{0}<\ldots<\gamma_{M}=\gamma_{\mathrm{max}}$
and $\lambda_{\mathrm{min}}=\lambda_{0}<\ldots<\lambda_{W}=\lambda_{\mathrm{max}}$,
respectively. Then for given $\gamma_{m}$, we compute the solution
path $\widehat{\boldsymbol{\beta}}(\gamma_{m},\lambda_{w})$ based
on the initial value $\widehat{\boldsymbol{\beta}}(\gamma_{m},\lambda_{w-1}).$
Using $\lambda_{w}$ and $\gamma_{m}$, we can compute the $\widehat{L}(\gamma_{m},\lambda_{w})$
distinct values of $\widehat{\boldsymbol{\beta}}_{it}(\gamma_{m},\lambda_{w})$,
corresponding to $\{\widehat{\boldsymbol{\alpha}}_{1},\ldots,\widehat{\boldsymbol{\alpha}}_{\widehat{L}(\gamma_{m},\lambda_{w})}\}$.
Then we select optimal $\widehat{\gamma}$ and $\widehat{\lambda}$
minimizing a data-driven criterion BIC defined later in (\ref{tuningparas}),
i.e., $(\widehat{\gamma},\widehat{\lambda})=\arg\min_{\gamma_{m},\lambda_{w}}\mathrm{BIC}(\gamma_{m},\lambda_{w})$.
Given $\widehat{\gamma}$ and $\widehat{\lambda}$, we can calculate
the estimates $\widehat{\boldsymbol{\beta}}=\widehat{\boldsymbol{\beta}}(\widehat{\gamma},\widehat{\lambda})$.
Thus all the observations can be separated into $\widehat{L}=\widehat{L}(\widehat{\gamma},\widehat{\lambda})$
blocks accordingly, for example, $\widehat{\mathcal{A}}_{l}=\{(i,t):\widehat{\boldsymbol{\beta}}_{it}=\widehat{\boldsymbol{\alpha}}_{l},1\leq l\leq L\}$,
and $\{\widehat{\mathcal{A}}_{1},\ldots,\widehat{\mathcal{A}}_{\widehat{L}}\}$
is a mutually exclusive partition of $\{(i,t):i=1,\cdots,N,t=1,\cdots,T\}$.

The construction of solution path with varying double tuning parameters
uses the ``bottom up''\ strategy - an important and necessary tactic
in the literature of fusion penalty method, because the way of block
structure recovery shares similarity as that of dendrogram for agglomerative
hierarchical clustering.

\section{The Estimation Procedure}

Since the objective function does not have a closed-form solution,
we use the Alternating Direction Method of Multipliers (ADMM) to solve
the optimization problem. Let $\boldsymbol{\rho}_{ij,t}=\boldsymbol{\beta}_{it}-\boldsymbol{\beta}_{jt}$
be the difference of two individual-specified coefficients at a given
period, and let $\boldsymbol{\delta}_{i,tt^{\prime}}=\boldsymbol{\beta}_{it}-\boldsymbol{\beta}_{it^{\prime}}$
represent the difference of two period-specified coefficients under
a given individual, then the objective function is equivalent to 
\begin{equation}
\tilde{\ell}_{p}(\boldsymbol{\beta},\boldsymbol{\rho},\boldsymbol{\delta})={\displaystyle \frac{1}{2}\sum\limits _{i=1}^{N}\sum\limits _{t=1}^{T}(y_{it}-\boldsymbol{x}_{it}^{\top}\boldsymbol{\beta}_{it})^{2}+\sum_{t=1}^{T}\sum\limits _{i<j}\mathcal{P}_{\lambda}(\Vert\boldsymbol{\rho}_{ij,t}\Vert){\displaystyle +\sum_{i=1}^{N}\sum\limits _{t<t^{\prime}}\mathcal{P}_{\gamma}(\Vert\boldsymbol{\delta}_{i,tt^{\prime}}\Vert),}}\label{Objec}
\end{equation}
\[
s_{\cdot}t_{\cdot}\quad\boldsymbol{\rho}_{ij,t}=\boldsymbol{\beta}_{it}-\boldsymbol{\beta}_{jt}\quad and\quad\boldsymbol{\delta}_{i,tt^{\prime}}=\boldsymbol{\beta}_{it}-\boldsymbol{\beta}_{it^{\prime}},
\]
where $\boldsymbol{\rho}=\{\boldsymbol{\rho}_{ij,t}^{\top},i<j,t=1,\cdots T\}^{\top}$
and $\boldsymbol{\delta}=\{\boldsymbol{\delta}_{i,tt^{\prime}}^{\top},t<t^{\prime},i=1,\cdots,N\}^{\top}$.
Under the constraints, the augmented Lagrangian objective function
is given by 
\[
\begin{array}{lll}
Q(\boldsymbol{\beta},\boldsymbol{\rho},\boldsymbol{\delta},\boldsymbol{\nu},\boldsymbol{\upsilon}) & =\tilde{\ell}_{p}(\boldsymbol{\beta},\boldsymbol{\rho},\boldsymbol{\delta})\\
 & +\sum_{t=1}^{T}\sum_{i<j}\langle\boldsymbol{\nu}_{ij,t},\boldsymbol{\beta}_{it}-\boldsymbol{\beta}_{jt}-\boldsymbol{\rho}_{ij,t}\rangle+\frac{\psi}{2}\sum_{t=1}^{T}\sum_{i<j}\Vert\boldsymbol{\beta}_{it}-\boldsymbol{\beta}_{jt}-\boldsymbol{\rho}_{ij,t}\Vert^{2}\\
 & {\displaystyle +\sum_{i=1}^{N}\sum_{t<t^{\prime}}\langle\boldsymbol{\upsilon}_{i,tt^{\prime}},\boldsymbol{\beta}_{it}-\boldsymbol{\beta}_{it^{\prime}}-\boldsymbol{\delta}_{i,tt^{\prime}}\rangle+\frac{\phi}{2}\sum_{i=1}^{N}\sum_{t<t^{\prime}}\Vert\ \boldsymbol{\beta}_{it}-\boldsymbol{\beta}_{it^{\prime}}-\boldsymbol{\delta}_{i,tt^{\prime}}\Vert^{2}},
\end{array}
\]
where the dual varibles $\boldsymbol{\nu}=\{\boldsymbol{\nu}_{ij,t}^{\top},i<j,t=1\cdots T\}^{\top}$
and $\boldsymbol{\upsilon}=\{\boldsymbol{\upsilon}_{i,tt^{\prime}}^{\top},t<t^{\prime},i=1\cdots N\}^{\top}$
are Lagrangian multipliers, $\psi$ and $\phi$ are fixed tuning parameters.

The ADMM method iteratively updates $\boldsymbol{\beta}$, $\boldsymbol{\rho}$,
$\boldsymbol{\delta}$, $\boldsymbol{\nu}$, and $\boldsymbol{\upsilon}$
based on the following three steps: (1) For given values of $\left(\boldsymbol{\beta},\boldsymbol{\nu},\boldsymbol{\upsilon}\right)$,
we update $\boldsymbol{\rho}$, and $\boldsymbol{\delta}$. (2) Then,
we update $\left(\boldsymbol{\nu},\boldsymbol{\upsilon}\right)$ given
other parameters. (3) Finally, the regression parameters $\boldsymbol{\beta}$
can be updated based on $(\boldsymbol{\rho},\boldsymbol{\delta},\boldsymbol{\nu},\boldsymbol{\upsilon})$.

More specifically, given $\boldsymbol{\beta}^{(s)}$, $\boldsymbol{\nu}^{(s)}$,
$\boldsymbol{\upsilon}^{(s)}$ at the $s$th step, we obtain $\boldsymbol{\beta}^{(s+1)}$,
$\boldsymbol{\nu}^{(s+1)}$, $\boldsymbol{\upsilon}^{(s+1)}$, $\boldsymbol{\rho}^{(s+1)}$,
$\boldsymbol{\delta}^{(s+1)}$ in the $(s+1)$th step, by using the
following ADMM iterative algorithm. First, we update $\boldsymbol{\rho}^{(s+1)}$
and $\boldsymbol{\delta}^{(s+1)}$, by solving \eqref{shrinkage1}
and \eqref{shrinkage2} below, i.e., 
\begin{equation}
\boldsymbol{\rho}^{(s+1)}=\mathrm{argmin}_{\boldsymbol{\rho}}L\left(\boldsymbol{\rho},\boldsymbol{\beta}^{(s)},\boldsymbol{\nu}^{(s)}\right),\label{shrinkage1}
\end{equation}
where 
\begin{equation}
L\left(\boldsymbol{\rho},\boldsymbol{\beta}^{(s)},\boldsymbol{\nu}^{(s)}\right)=\frac{\psi}{2}\sum_{t=1}^{T}\sum_{i<j}\left\Vert \boldsymbol{\beta}_{it}^{(s)}-\boldsymbol{\beta}_{jt}^{(s)}+\psi^{-1}\boldsymbol{\nu}_{ij,t}^{(s)}-\boldsymbol{\rho}_{ij,t}\right\Vert ^{2}+\sum_{t=1}^{T}\sum_{i<j}\mathcal{P}_{\lambda}(\Vert\boldsymbol{\rho}_{ij,t}\Vert),\label{L}
\end{equation}
\begin{equation}
\boldsymbol{\delta}^{(s+1)}=\mathrm{argmin}_{\boldsymbol{\delta}}H\left(\boldsymbol{\delta},\boldsymbol{\beta}^{(s)},\boldsymbol{\upsilon}^{(s)}\right),\label{shrinkage2}
\end{equation}
and 
\begin{equation}
H\left(\boldsymbol{\delta},\boldsymbol{\beta}^{(s)},\boldsymbol{\upsilon}^{(s)}\right)=\frac{\phi}{2}\sum_{i=1}^{N}\sum_{t=2}^{T}\left\Vert \boldsymbol{\beta}_{it}^{(s)}-\boldsymbol{\beta}_{it^{\prime}}^{(s)}+\phi^{-1}\boldsymbol{\upsilon}_{i,tt^{\prime}}^{(s)}-\boldsymbol{\delta}_{i,tt^{\prime}}\right\Vert ^{2}+\sum_{i=1}^{N}\sum_{t<t^{\prime}}\mathcal{P}_{\gamma}(\Vert\boldsymbol{\delta}_{i,tt^{\prime}}\Vert).\label{H}
\end{equation}
By arguments similar to \citet{Ma2016,MaHuang2017}, under \eqref{L}
and \eqref{H}, the elements $\boldsymbol{\rho}_{ij,t}^{(s+1)}$ of
$\boldsymbol{\rho}^{(s+1)}$ and the elements $\boldsymbol{\delta}_{i,tt^{\prime}}^{(s+1)}$
of $\boldsymbol{\delta}^{(s+1)}$ are the minimizers of $\frac{\varphi}{2}\Vert\boldsymbol{\xi}_{ij,t}^{(s)}-\boldsymbol{\rho}_{ij,t}\Vert^{2}+\mathcal{P}_{\lambda}(\Vert\boldsymbol{\rho}_{ij,t}||)$
, $\frac{\phi}{2}\Vert\boldsymbol{\vartheta}_{i,tt^{\prime}}^{(s)}-\boldsymbol{\delta}_{i,tt^{\prime}}\Vert^{2}+\mathcal{P}_{\gamma}\left(\Vert\boldsymbol{\delta}_{i,tt^{\prime}}\Vert\right)$,
respectively, where $\boldsymbol{\xi}_{ij,t}^{(s)}=\boldsymbol{\beta}_{it}^{(s)}-\boldsymbol{\beta}_{jt}^{(s)}+\varphi^{-1}\boldsymbol{\nu}_{ij,t}^{(s)}$
and $\boldsymbol{\vartheta}_{i,tt^{\prime}}^{(s)}=\boldsymbol{\beta}_{it}^{(s)}-\boldsymbol{\beta}_{it^{\prime}}^{(s)}+\phi^{-1}\boldsymbol{\upsilon}_{i,tt^{\prime}}^{(s)}$.
For different threshold operators $\mathcal{P}_{\lambda}(\cdot)$
and $\mathcal{P}_{\gamma}(\cdot)$, the estimates $\boldsymbol{\rho}_{ij,t}^{(s+1)}$
and $\boldsymbol{\delta}_{i,tt^{\prime}}^{(s+1)}$ are updated based
on different formula corresponding to that operator. In particular, 
\begin{itemize}
\item for the Lasso penalty, 
\[
\boldsymbol{\rho}_{ij,t}^{(s+1)}=S\left(\boldsymbol{\xi}_{ij,t}^{(s)},\lambda/\varphi\right);\boldsymbol{\delta}_{i,tt^{\prime}}^{(s+1)}=S\left(\boldsymbol{\vartheta}_{i,tt^{\prime}}^{(s)},\gamma/\phi\right);
\]
\item for the SCAD penalty with $a>\max(1/\varphi+1,1/\phi+1)$, 
\[
\boldsymbol{\rho}_{ij,t}^{(s+1)}=\left\{ \begin{array}{ll}
{S\left(\xi_{ij,t}^{(s)},\lambda/\varphi\right),} & {\text{if }\|\boldsymbol{\xi}_{ij,t}^{(s)}\|\leq\lambda+\lambda/\varphi}\\
{\displaystyle {\boldsymbol{\xi}_{ij,t}^{(s)},}} & {\text{if }\|\boldsymbol{\xi}_{ij,t}^{(s)}\|>a\lambda}\\
{\displaystyle {\frac{S\left(\boldsymbol{\xi}_{ij,t}^{(s)},a\lambda/((a-1)\varphi)\right)}{1-1/((a-1)\varphi)},}} & {\text{otherwise}},
\end{array}\right.
\]

\[
\boldsymbol{\delta}_{i,tt^{\prime}}^{(s+1)}=\left\{ \begin{array}{ll}
{S\left(\boldsymbol{\vartheta}_{k}^{(s)},\gamma/\phi\right),} & {\text{if }\|\boldsymbol{\vartheta}_{i,tt^{\prime}}^{(s)}\|\leq\gamma+\gamma/\phi}\\
{\displaystyle {\boldsymbol{\vartheta}_{i,tt^{\prime}}^{(s)},}} & {\text{if }\|\boldsymbol{\vartheta}_{i,tt^{\prime}}^{(s)}\|>a\gamma}\\
{\displaystyle {\frac{S\left(\boldsymbol{\vartheta}_{i,tt^{\prime}}^{(s)},a\gamma/((a-1)\phi)\right)}{1-1/((a-1)\phi)},}} & {\text{otherwise}},
\end{array}\right.
\]

\item for MCP with $a>\max(1/\varphi,1/\phi)$,

\[
\rho_{ij,t}^{(s+1)}=\left\{ \begin{array}{ll}
{\displaystyle {\frac{S\left(\boldsymbol{\xi}_{ij,t}^{(s)},\lambda/\varphi\right)}{1-1/(a\varphi)},}} & {\text{if }\Vert\boldsymbol{\xi}_{ij,t}^{(s)}\Vert\leq a\lambda}\\
{\displaystyle {\boldsymbol{\xi}_{ij,t}^{(s)},}} & \text{otherwise,}
\end{array}\right.
\]

\[
\boldsymbol{\delta}_{i,tt^{\prime}}^{(s+1)}=\left\{ \begin{array}{ll}
{\displaystyle {\frac{S\left(\boldsymbol{\vartheta}_{k}^{(s)},\gamma/\phi\right)}{1-1/(a\phi)},}} & {\text{if }\Vert\boldsymbol{\vartheta}_{i,tt^{\prime}}^{(s)}\Vert\leq a\gamma}\\
{\displaystyle {\boldsymbol{\vartheta}_{i,tt^{\prime}}^{(s)},}} & \text{ otherwise,}
\end{array}\right.
\]

\end{itemize}
where $\varphi$ is turning parameter and 
\[
S(w,t)=\left\{ \begin{array}{ll}
{(1-t/\Vert w\Vert)w,} & {\text{if }t/\Vert w\Vert<1}\\
{0,} & {\text{otherwise}}.
\end{array}\right.
\]

Next, we update $\boldsymbol{\nu}^{(s+1)}$ and $\boldsymbol{\upsilon}^{(s+1)}$
by 
\begin{equation}
\boldsymbol{\nu}_{ij,t}^{(s+1)}=\boldsymbol{\nu}_{ij,t}^{(s)}+\psi\left(\boldsymbol{\beta}_{it}^{(s)}-\boldsymbol{\beta}_{jt}^{(s)}-\boldsymbol{\rho}_{ij,t}^{(s+1)}\right)\label{fusion1}
\end{equation}
and 
\begin{equation}
\boldsymbol{\upsilon}_{i,tt^{\prime}}^{(s+1)}=\boldsymbol{\upsilon}_{i,tt^{\prime}}^{(s)}+\phi\left(\boldsymbol{\beta}_{it}^{(s)}-\boldsymbol{\beta}_{it^{\prime}}^{(s)}-\boldsymbol{\delta}_{i,tt^{\prime}}^{(s+1)}\right).\label{fusion2}
\end{equation}

At last, we update the coefficients $\boldsymbol{\beta}^{(s+1)}$
via 
\begin{equation}
\boldsymbol{\beta}^{(s+1)}=\mathrm{argmin}_{\boldsymbol{\beta}}Q\left(\boldsymbol{\beta},\boldsymbol{\rho}^{(s+1)},\boldsymbol{\delta}^{(s+1)},\boldsymbol{\nu}^{(s+1)},\boldsymbol{\upsilon}^{(s+1)}\right),\label{estimation}
\end{equation}
where 
\begin{eqnarray*}
 &  & Q\left(\boldsymbol{\beta},\boldsymbol{\rho}^{(s+1)},\boldsymbol{\delta}^{(s+1)},\boldsymbol{\nu}^{(s+1)},\boldsymbol{\upsilon}^{(s+1)}\right)\\
 & = & \tilde{\ell}_{p}\left(\boldsymbol{\beta},\boldsymbol{\rho}^{(s+1)},\boldsymbol{\delta}^{(s+1)}\right)+\sum_{i<j}\langle\nu_{ij}^{(s+1)},\boldsymbol{\beta}_{i}-\boldsymbol{\beta}_{j}-\rho_{ij}^{(s+1)}\rangle\\
 &  & +\frac{\psi}{2}\sum_{i<j}\left\Vert \boldsymbol{\beta}_{i}-\boldsymbol{\beta}_{j}-\boldsymbol{\rho}_{ij}^{(s+1)}\right\Vert ^{2}+\sum_{t<t^{\prime}}\langle\boldsymbol{\upsilon}_{tt^{\prime}}^{(s+1)},\boldsymbol{\beta}_{t}-\boldsymbol{\beta}_{t^{\prime}}-\boldsymbol{\delta}_{tt^{\prime}}^{(s+1)}\rangle+\frac{\phi}{2}\sum_{t<t^{\prime}}\left\Vert \boldsymbol{\beta}_{t}-\boldsymbol{\beta}_{t^{\prime}}-\boldsymbol{\delta}_{tt^{\prime}}^{(s+1)}\right\Vert ^{2}.
\end{eqnarray*}
Minimizing the objective function (\ref{estimation}) with respect
to $\boldsymbol{\beta}$ is equivalent to minimizing 
\begin{eqnarray}
h\left(\boldsymbol{\beta},{\boldsymbol{\rho}}^{(s+1)},{\boldsymbol{\delta}}^{(s+1)},{\boldsymbol{\nu}}^{(s+1)},{\boldsymbol{\upsilon}}^{(s+1)}\right) & = & \frac{1}{2}\sum_{i=1}^{N}\sum_{t=1}^{T}(y_{it}-\boldsymbol{x}_{it}^{\top}\boldsymbol{\beta}_{it})^{2}\label{EstObject}\\
 &  & +\frac{\psi}{2}{\left\Vert \Omega\boldsymbol{\beta}-{\boldsymbol{\rho}}^{(s+1)}+{\psi}^{-1}{\boldsymbol{\nu}}^{(s+1)}\right\Vert }^{2}\nonumber \\
 &  & +\frac{\phi}{2}{\left\Vert \Phi\boldsymbol{\beta}-\boldsymbol{\delta}^{(s+1)}+\phi^{-1}\boldsymbol{\upsilon}^{(s+1)}\right\Vert }^{2},
\end{eqnarray}
where $\Omega=(\mathcal{E}\otimes\boldsymbol{I}_{T})\otimes\boldsymbol{I}_{P}$,
$\Phi=(\boldsymbol{I}_{N}\otimes\mathcal{D})\otimes\boldsymbol{I}_{P}$,
$\mathcal{E}=\{(e_{i}-e_{j}),i<j\}_{\frac{N(N-1)}{2}\times N}^{\top}$
with $e_{i}$ being the $i$th unit vector whose $i$th element is
1 and the remaining elements are 0 and $\mathcal{D}=\{(e_{t}-e_{t^{\prime}}),t<t^{\prime}\}_{\frac{T(T-1)}{2}\times T}^{\top}$
with $e_{t}$ being the $t$th unit vector whose $t$th element is
1 and the remaining elements are 0. The integrative or fusion matrix
$\mathcal{E}$ aims to calculate the difference of coefficients between
each pairwise individuals, similarly to fusion matrix $\mathcal{D}$
for temporal dimension. Then, we get 
\begin{equation}
\boldsymbol{\beta}^{(s+1)}=\left(\boldsymbol{X}^{\top}\boldsymbol{X}+\psi\Omega^{\top}\Omega+\phi\Phi^{\top}\Phi\right)^{-1}\left\{ \boldsymbol{X}^{\top}\boldsymbol{Y}+\Omega^{\top}\left(\psi\boldsymbol{\rho}^{(s+1)}-\boldsymbol{\nu}^{(s+1)}\right)+\Phi^{\top}\left(\phi\boldsymbol{\delta}^{(s+1)}-\boldsymbol{\upsilon}^{(s+1)}\right)\right\} ,\label{Est}
\end{equation}
where $\boldsymbol{Y}=\left(y_{11},\cdots,y_{1T},\cdots,y_{N1},\cdots,y_{NT}\right)^{\top}$,
$\boldsymbol{X}=\mathrm{diag}(\boldsymbol{X}_{1},\cdots,\boldsymbol{X}_{N})$
and $\boldsymbol{X}_{i}=\mathrm{diag}(\boldsymbol{x}_{i1}^{\top},\cdots,\boldsymbol{x}_{iT}^{\top})$
with $\boldsymbol{x}_{it}=(1,\boldsymbol{z}_{it}^{\top})^{\top}$.

Explicit solution of $\boldsymbol{\beta}$ in (\ref{Est}) involves
computational burden caused by calculating the inverse of a $NTP\times NTP$
dimensional matrix, especially with large $N$ and $T$. It is also
noted that the design matrix $\boldsymbol{X}$, the fusion matrix
$\Omega$ and $\Phi$ contain amounts of sparsity part, motivating
us to accelerate the calculation process by by saving memory space
and employing some equivalent algebra. Let $\widetilde{\boldsymbol{X}}_{NT\times P}=\left(\boldsymbol{x}_{11},\cdots,\boldsymbol{x}_{1T},\cdots,\boldsymbol{x}_{N1},\cdots,\boldsymbol{x}_{NT}\right)^{\top}$
with $\boldsymbol{x}_{it}$ being $P\times1$ regressors under given
individual $i$ and period $t$, $\widetilde{\boldsymbol{\beta}}_{NT\times P}=\left(\boldsymbol{\beta}_{11},\cdots,\boldsymbol{\beta}_{1T},\cdots,\boldsymbol{\beta}_{N1},\cdot,\boldsymbol{\beta}_{NT}\right)^{\top}$
with $\boldsymbol{\beta}_{it}$ being $P\times1$ coefficients under
given individual $i$ and period $t$, as the dense regressors and
coefficients, rearranging the nonzero element of $\boldsymbol{X}$
and $\boldsymbol{\beta}$. Correspondingly, we set another form of
dual variables $\widetilde{\boldsymbol{\nu}}$ and $\widetilde{\boldsymbol{\rho}}$,
which are $\frac{N\times(N-1)}{2}\times TP$ dimensional matrices,
and $\widetilde{\boldsymbol{\upsilon}}$, $\widetilde{\boldsymbol{\delta}}$
are $\frac{T\times(T-1)}{2}\times NP$ dimensional matrix. Therefore,
(\ref{EstObject}) can be also rewritten as 
\begin{eqnarray}
 &  & h(\widetilde{\boldsymbol{\beta}},\widetilde{\boldsymbol{\rho}}^{(s+1)},\widetilde{\boldsymbol{\delta}}^{(s+1)},\widetilde{\boldsymbol{\nu}}^{(s+1)},\widetilde{\boldsymbol{\upsilon}}^{(s+1)})\nonumber \\
 & = & \frac{1}{2}\sum_{i=1}^{N}\sum_{t=1}^{T}(y_{it}-\boldsymbol{x}_{it}^{\top}\boldsymbol{\beta}_{it})^{2}\nonumber \\
 &  & +\frac{\psi}{2}{\left\Vert (\mathcal{E}\otimes\boldsymbol{I}_{T})\widetilde{\boldsymbol{\beta}}-\widetilde{\boldsymbol{\rho}}^{(s+1)}+{\psi}^{-1}\widetilde{\boldsymbol{\nu}}^{(s+1)}\right\Vert }_{F}^{2}+\frac{\phi}{2}{\left\Vert (\boldsymbol{I}_{N}\otimes\mathcal{D})\widetilde{\boldsymbol{\beta}}-\widetilde{\boldsymbol{\delta}}^{(s+1)}+{\phi}^{-1}\widetilde{\boldsymbol{\upsilon}}^{(s+1)}\right\Vert }_{F}^{2}.
\end{eqnarray}
Let $A=\psi\Omega^{\top}\Omega+\phi\Phi^{\top}\Phi=\left[\psi(\mathcal{E}^{\top}\mathcal{E}\otimes\boldsymbol{I}_{T})+\phi(\boldsymbol{I}_{N}\otimes\mathcal{D}^{\top}\mathcal{D})\right]\otimes\boldsymbol{I}_{P}$,
$\mathcal{E}^{\top}\mathcal{E}=N\boldsymbol{I}_{N}-1_{N}1_{N}^{\top}$,
$\mathcal{D}^{\top}\mathcal{D}=T\boldsymbol{I}_{T}-1_{T}1_{T}^{\top}$.
Applying the Sherman-Morrison-Woodbury formula, we can solve the above
matrix inverse by 
\[
(\boldsymbol{X}^{\top}\boldsymbol{X}+A)^{-1}=A^{-1}-A^{-1}\boldsymbol{X}^{\top}(\boldsymbol{I}_{NT}+\boldsymbol{X}A^{-1}\boldsymbol{X}^{\top})^{-1}\boldsymbol{X}A^{-1}.
\]
We have $A^{-1}=D\otimes\boldsymbol{I}_{P}$, where 
\begin{eqnarray}
D & = & \left[\psi(\mathcal{E}^{\top}\mathcal{E}\otimes\boldsymbol{I}_{T})+\phi(\boldsymbol{I}_{N}\otimes\mathcal{D}^{\top}\mathcal{D})\right]^{-1}\nonumber \\
 & = & \left\{ \psi\left[(N\boldsymbol{I}_{N}-1_{N}1_{N}^{\top})\otimes\boldsymbol{I}_{T}\right]+\phi\left[(\boldsymbol{I}_{N}\otimes(T\boldsymbol{I}_{T}-1_{T}1_{T}^{\top})\right]\right\} ^{-1}\nonumber \\
 & = & \left\{ (\psi N+\phi T)\boldsymbol{I}_{NT}-\left[\psi(1_{N}1_{N}^{\top})\otimes\boldsymbol{I}_{T}+\phi\boldsymbol{I}_{N}\otimes(1_{T}1_{T}^{\top})\right]\right\} ^{-1}.
\end{eqnarray}
Through setting 
\[
M=\left(I_{NT}+\left(\widetilde{\boldsymbol{X}}\widetilde{\boldsymbol{X}}^{\top}\right)\circ D\right)^{-1},
\]
\[
b^{(s+1)}=\widetilde{\boldsymbol{X}}\circ\boldsymbol{Y}+\mathcal{E}^{\top}(\psi\breve{\boldsymbol{\rho}}^{(s+1)}-\breve{\boldsymbol{\nu}}^{(s+1)})+\mathcal{D}^{\top}(\psi\breve{\boldsymbol{\delta}}^{(s+1)}-\breve{\boldsymbol{\upsilon}}^{(s+1)}),
\]
where $\breve{\boldsymbol{\rho}}^{(s+1)}$ and $\breve{\boldsymbol{\nu}}^{(s+1)}$
are matrices by rearranging $\widetilde{\boldsymbol{\rho}}^{(s+1)}$
and $\widetilde{\boldsymbol{\nu}}^{(s+1)}$ into a $\frac{N(N-1)}{2}\times TP$
matrix whose rows store the fused values between each individuals,
sequentially. Similarly, $\breve{\boldsymbol{\delta}}^{(s+1)}$ and
$\breve{\boldsymbol{\upsilon}}^{(s+1)}$ are matrices by rearranging
$\widetilde{\boldsymbol{\delta}}^{(s+1)}$ and $\widetilde{\boldsymbol{\upsilon}}^{(s+1)}$
into a $\frac{T(T-1)}{2}\times NP$ matrix whose rows store the fused
values between each periods, sequentially. Finally, let 
\[
B^{(s+1)}=\widetilde{\boldsymbol{X}}\circ\left\{ M\left[\widetilde{\boldsymbol{X}}\circ\left(Db^{(s+1)}\right)\right]^{+}\right\} ,
\]
we get 
\[
\boldsymbol{\beta}^{(s+1)}=\mathrm{vec}\left\{ \left[D\left(b^{(s+1)}-B^{(s+1)}\right)\right]^{+}\right\} .
\]

\section{Asymptotic Properties}

\subsection{Preliminary}

\begin{figure}[ht!]
\centering \subfloat[Group-Cohort Pattern]{\includegraphics[width=6cm,height=6cm]{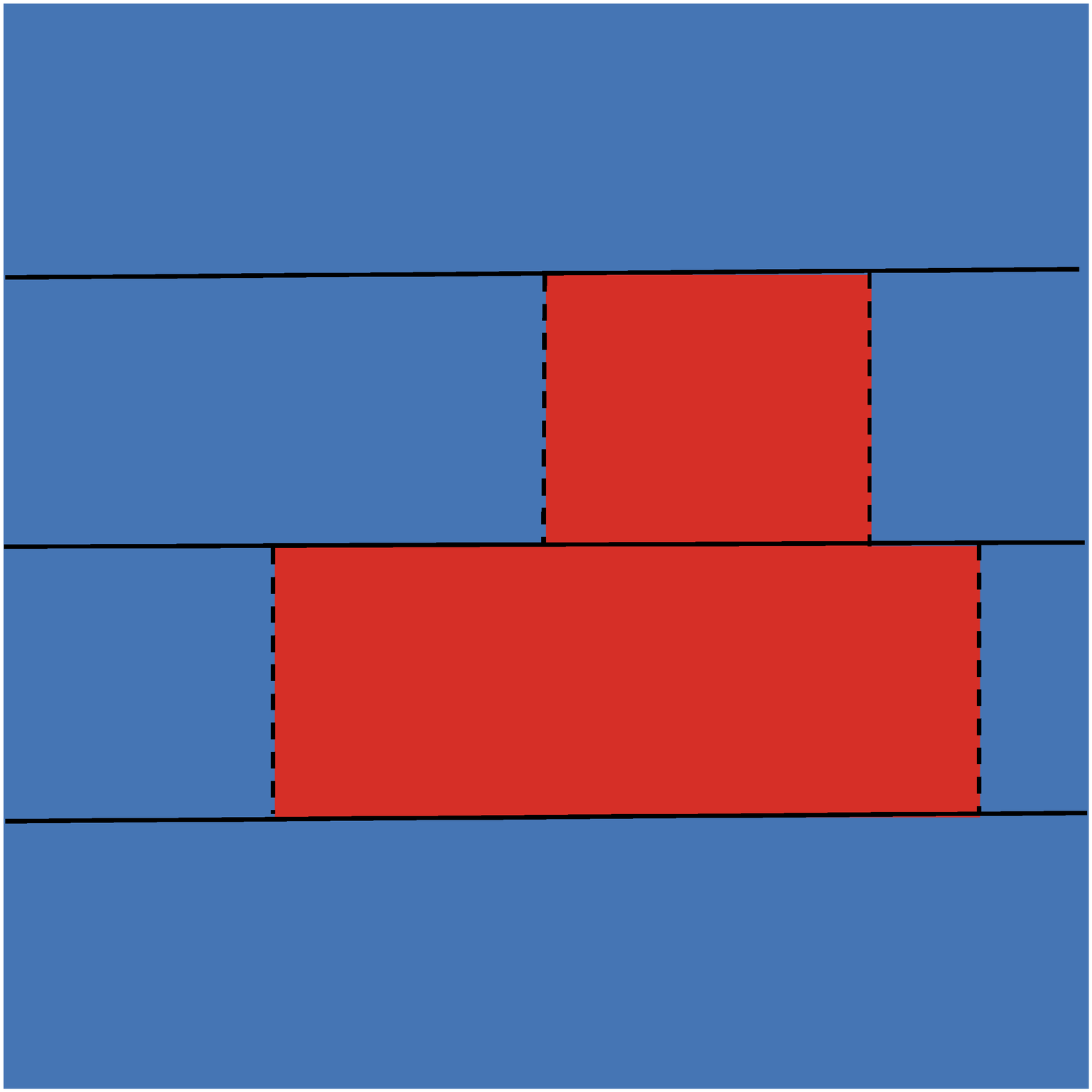}\label{Partitions1}

}\hspace{30pt} \subfloat[Cohort-Group Pattern]{\includegraphics[width=6cm,height=6cm]{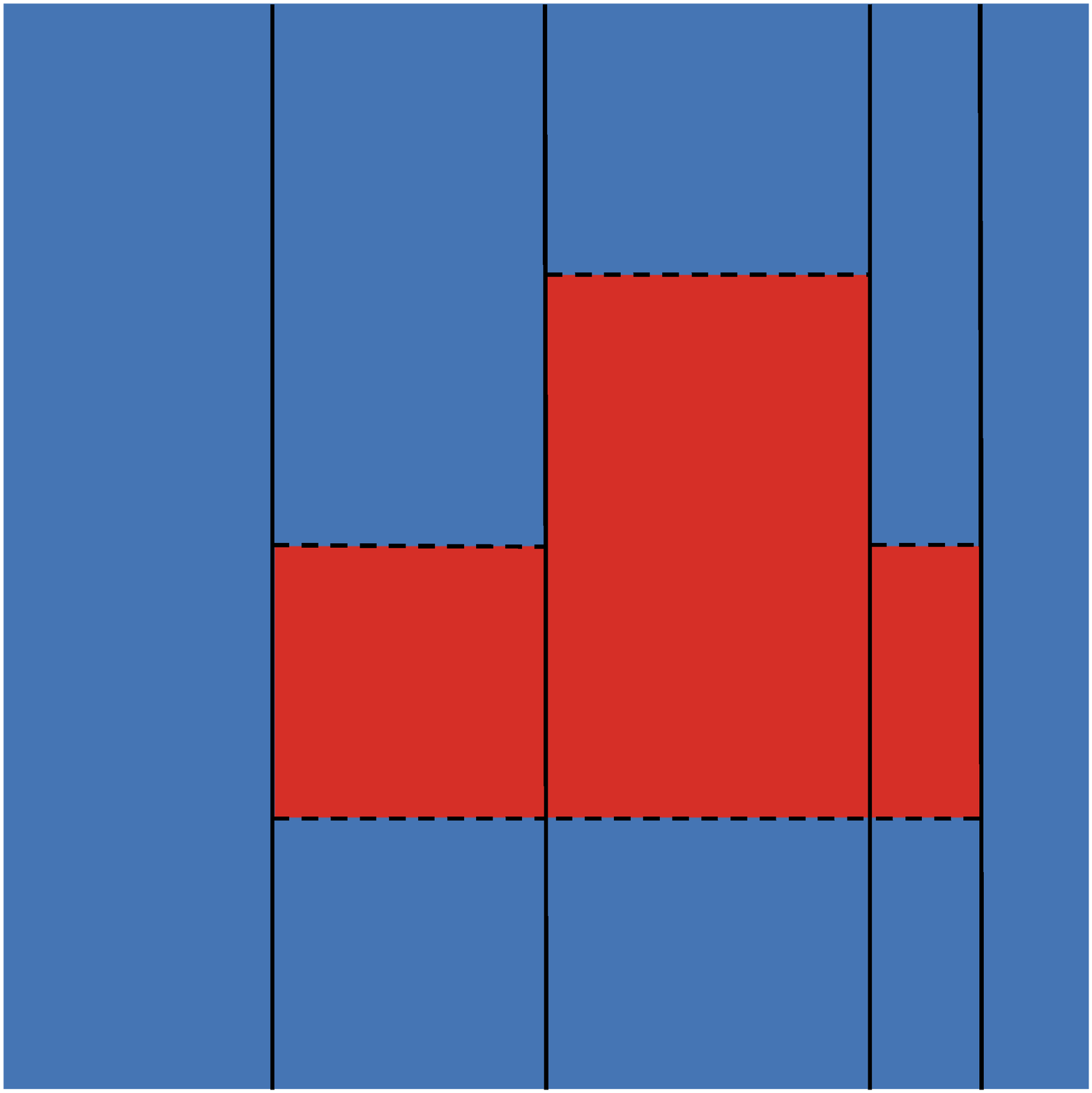}\label{Partitions2}

}\caption{Partitioned block structure on regression coefficients: firstly partition
the blocks along the dash line and then along the dotted line.}
\label{fig:Partitions} 
\end{figure}

In order to characterize the block structure on regression coefficients,
we may first partition the grouped structure among individuals, and
then determine the time structure of breaks in each group, as described
in Figure (\ref{fig:Partitions})(a), we call this the group-cohort
pattern. Alternatively, we may capture the block structure using a
cohort-group pattern (described in Figure (\ref{fig:Partitions})(b))
which firstly partitions the structural breaks along the time dimension
and then determines the group membership in each cohorts. Fortunately,
it does not matter which pattern we select, since they depict exactly
the same block structure by different structural matrices.

We firstly introduce the group-cohort pattern in detail below. Let
$K$ denote the split number of groups and $\mathcal{G}_{0k}$ denote
the individual memberships for the $kth$ group for $k=1,\cdots,K$.
Further, $R(k)$ denotes the number of blocks in the $kth$ group
and $\mathcal{H}_{0r}(k)$ denote the temporal memberships for the
$rth$ block in the $kth$ group $r=1,\cdots,R(k)$. Let $\boldsymbol{\Pi}$
denotes the grouped structure across individuals and $\widetilde{\Pi}=\{\pi_{ik},i=1,\cdots,N\}$
denotes an $N\times K$ matrix with $\pi_{ik}=1$ for $i\in\mathcal{G}_{0k}$
and $\pi_{ik}=0$ for $i\notin\mathcal{G}_{0k}$, indicating the group
structure. Then, $\boldsymbol{\Pi}_{NTP\times KTP}=(\widetilde{\Pi}\otimes I_{T})\otimes I_{P}$.
Furthermore, we let $\boldsymbol{W}$ denote the structural breaks
in each group and for $k=1,\cdots,K,$ set $\widetilde{W}(k)=\{w_{tr}\}$
denote an $T\times R(k)$ matrix with $w_{tr}=1$ for $t\in\mathcal{H}_{0r}(k)$
and $w_{tr}=0$ for $r\notin\mathcal{H}_{0r}(k)$, which depicts the
structural breaks under each group. Then, let $R=\sum_{k=1}^{K}R(k)$
and $\widetilde{\boldsymbol{W}}_{KT\times R}=\mathrm{diag}(\widetilde{W}(1),\cdots,\widetilde{W}(K))$
and $\boldsymbol{W}_{KTP\times RP}=\widetilde{\boldsymbol{W}}\otimes I_{P}$.
As we can see, the number of blocks partitioned by group-cohort is
not smaller than the number of real blocks, at least. Therefore, there
is also a structural matrix to depict the relationship between group-cohort
and real blocks. We use $\boldsymbol{Q}$ to depict the partitioned
structure and $\boldsymbol{\eta}$ is the vector of values for split
blocks with $L^{0}$ different values under the group-cohort pattern.
It is obvious that some of the split blocks belong to same true block.
Therefore, we set $\widetilde{\boldsymbol{Q}}_{R\times L^{0}}=\{q_{rl}\}$,
with $r=\{1,\cdots,R\}$ and $l=\{1,\cdots,L^{0}\}$, denote an $R\times L^{0}$
matrix with $q_{rl}=1$ for $\eta_{r}=\alpha_{l}$, which depicts
a structural matrix that integrate between spitted blocks and then
$\boldsymbol{Q}_{RP\times L^{0}P}=\widetilde{\boldsymbol{Q}}\otimes I_{P}$.
.As a result, $\boldsymbol{\beta}^{0}=\boldsymbol{\Pi}\boldsymbol{W}\boldsymbol{Q}\boldsymbol{\alpha}^{0}$
and the design matrix with known structural information $\mathbb{X}=\boldsymbol{X}\boldsymbol{\Pi}\boldsymbol{W}\boldsymbol{Q}$.

Similarly, we can also firstly set the cohorts structural matrix denoted
by $\bar{\boldsymbol{W}}$ and then use $\bar{\boldsymbol{\Pi}}$
to describe the grouped structure in each cohorts in cohort-group
pattern. Furthermore, $\bar{\boldsymbol{Q}}$ denotes the block integration
in cohort-group pattern. Obviously, $\boldsymbol{\beta}^{0}=\Pi\boldsymbol{W}\boldsymbol{Q}\alpha^{0}=\bar{\boldsymbol{W}}\bar{\boldsymbol{\Pi}}\bar{\boldsymbol{Q}}\boldsymbol{\alpha^{0}}$,
implying that we only need to select one of patterns. The two dimensional
heterogeneous structure is more general than that in \citet{OW2020},
due to the existence of structural matrices $\boldsymbol{Q}$ or $\bar{\boldsymbol{Q}}$,
which depict the integration between splitted blocks and can be viewed
as a mediator between different pattern.

To study the theoretical results of the proposed block regression
estimator, we first investigate the asymptotic properties of the estimator
with known block structure. Although, in practice, $L$ is generally
unknown and such an estimator is infeasible. This infeasible procedure
provides important information to which we should compare our feasible
estimator. Let $\boldsymbol{\beta}^{0}$, $\boldsymbol{\alpha}^{0}$,
$\mathcal{A}_{0}$ and $L_{0}$ denote the true values of $\boldsymbol{\beta}$,
$\boldsymbol{\alpha}$, $\mathcal{A}$ and $L$, respectively. We
also let $|\mathcal{A}_{l}|$ to signify the amount of elements in
$\mathcal{A}_{l}$. $\mathcal{A}_{\min}=\min_{1\leq l\leq L}|\mathcal{A}_{l}|$
and $\mathcal{A}_{\max}=\max_{1\leq l\leq L}|\mathcal{A}_{l}|$, respectively
represent the true minimum and maximum sample sizes among all blocks.

\subsection{Asymptotic Property of the Infeasible Estimator with Known Two-Dimensional
Heterogeneous Structure}

If the underlying block structure $\mathcal{A}=\{\mathcal{A}_{l}:l=1,\cdots,L^{0}\}$
is known, which is equivalent to know the prior information of matrices
$\boldsymbol{\Pi}$ and $\boldsymbol{W}$, $\boldsymbol{Q}$, and
notice that $\boldsymbol{\beta}=\boldsymbol{\Pi}\boldsymbol{W}\boldsymbol{Q}\boldsymbol{\alpha}$,
it is equivalent to consider $\widetilde{\boldsymbol{\alpha}}$ or
$\widetilde{\boldsymbol{\beta}}=\boldsymbol{\Pi}\boldsymbol{W}\boldsymbol{Q}\widetilde{\boldsymbol{\alpha}}$.
In this case, the post bi-integrative estimator is defined by: 
\begin{eqnarray}
\widetilde{\boldsymbol{\alpha}} & = & \mathop{{\rm argmin}}_{\boldsymbol{\alpha}\in\mathbb{R}^{PL}}\left\{ \sum_{l=1}^{L}\sum_{(i,t)\in\mathcal{A}_{l}}(y_{it}-x_{it}^{\top}\boldsymbol{\alpha}_{l})^{2}\right\} \nonumber \\
 & = & \mathop{{\rm argmin}}_{\boldsymbol{\alpha}\in\mathbb{R}^{PL}}\left\{ \Vert\boldsymbol{Y}-\mathbb{X}\boldsymbol{\alpha}\Vert^{2}\right\} \nonumber \\
 & = & \left(\mathbb{X}^{\top}\mathbb{X}\right)^{-1}\mathbb{X}^{\top}\boldsymbol{Y},\label{oracle}
\end{eqnarray}
where $\widetilde{\boldsymbol{\alpha}}=(\widetilde{\alpha}_{1}^{\top},\cdots,\widetilde{\alpha}_{L^{0}}^{\top})^{\top}$.

Due to the block structure information, i.e., $\mathcal{A}$ is generally
unknown in advance, the block-oracle estimators are infeasible in
practice. However, it can shed light on the theoretical properties
of the proposed estimators.

For investigating the statistical properties of the induced minimizer
$\widetilde{\boldsymbol{\alpha}}$, we impose the following conditions, 
\begin{itemize}
\item[(C1)] The noise vector $\boldsymbol{\epsilon}$ has sub-Gaussian tails
such that $P(|\tau^{\top}\boldsymbol{\epsilon}|<\|\tau\|x)\geq1-2\exp(-c_{1}x^{2})$
for any vector $\tau\in\mathbb{R}^{NT}$, $0<c_{1}<\infty$ and $x>0$,
and $\epsilon_{it}$ is a sequence of independent random variables
with $E(\epsilon_{it})=0$, $E(\epsilon_{it}^{2})=\sigma^{2}$ for
$i=1,\cdots,N;t=1,\cdots,T$. 
\item[(C2)] (i) $\gamma_{\mathrm{min}}(\mathbb{X}^{\top}\mathbb{X})\ge c_{2}\mathcal{A}_{\mathrm{min}}$,
$\gamma_{\mathrm{max}}(\mathbb{X}^{\top}\mathbb{X})\le c_{3}NT.$
(ii) $\sum_{(i,t)\in\mathcal{A}_{l}}x_{it,p}^{2}=\left|\mathcal{A}_{l}\right|$,
for $1\leq p\leq P$. (iii) $\sup_{it}\left\Vert \boldsymbol{x}_{it}\right\Vert \leq c_{4}\sqrt{P}$,
(iv) $\left\vert \mathcal{A}_{\min}\right\vert \gg(L^{0}P)^{1/2}(NT)^{3/4}$,
for some positive constants $c_{2}$, $c_{3}$ and $c_{4}$. 
\end{itemize}
\begin{remark} Condition (C1) about sub-Gaussian tails of the error
is widely used in the literature of high-dimensional regressions.
For Condition (C2), since 
\[
\mathbb{X}^{\top}\mathbb{X}=\mathrm{diag}(\sum_{(i,t)\in\mathcal{A}_{l}}x_{it}x_{it}^{\top},l=1,\cdots,L^{0}),
\]
$\gamma_{\mathrm{min}}(\mathbb{X}^{\top}\mathbb{X})\geq\gamma_{\mathrm{min}}(\sum_{(i,t)\in\mathcal{A}_{l}}x_{it}x_{it}^{\top})\geq c_{2}\mathcal{A}_{\min}$,
namely, the smallest eigenvalue of $\mathbb{X}^{\top}\mathbb{X}$
bounded by the smallest cardinal number of all blocks. Without loss
of generality, we standardize the covariates in every sub-population,
which is assumed in Condition (C2) (ii). Condition (C2) (iv) implies
there should be enough observations within each block. \end{remark}

\begin{remark} Usually, the proof of asymptotic normality on coefficients
needs a little stronger assumption than consistency. For example,
we only need finite second moment of $\epsilon_{it}$ to obtain consistency
and finite fourth moment to obtain asymptotic normality. For simplicity,
we are using the same assumptions for results (i) and (ii) in Theorem
\ref{th1} below. \end{remark}

\begin{theorem} \label{th1}(Asymptotic properties of the post bi-integrative
estimator $\widetilde{\alpha}$) 
\begin{itemize}
\item[(i)] (Consistency and Rate of convergence) Under Conditions (C1) and (C2),
we have 
\[
\left\Vert \widetilde{\boldsymbol{\alpha}}-\boldsymbol{\alpha}^{0}\right\Vert \leq\Delta_{n}\text{,}\;\left\Vert \widetilde{\boldsymbol{\beta}}-\boldsymbol{\beta}^{0}\right\Vert \leq\sqrt{\left\vert \mathcal{A}_{\max}\right\vert }\Delta_{n}\text{, }\mathrm{and}\;\sup_{i,t}\left\Vert \widetilde{\boldsymbol{\beta}}_{it}-\boldsymbol{\beta}_{it}^{0}\right\Vert \leq\Delta_{n},
\]
where $\Delta_{n}=c_{1}^{-\frac{1}{2}}c_{2}^{-1}\sqrt{PL^{0}}\sqrt{NT\log(NT)}\left\vert \mathcal{A}_{\min}\right\vert ^{-1}$. 
\item[(ii)] (Asymptotic normality) Under Conditions (C1) and (C2), we have 
\[
s_{n}(\boldsymbol{d}_{n})^{-1}\boldsymbol{d}_{n}^{\top}(\widetilde{\boldsymbol{\alpha}}-\boldsymbol{\alpha}^{0})\overset{D}{\rightarrow}N(0,1),
\]
where 
\[
s_{n}(\boldsymbol{d}_{n})=\sigma\{\boldsymbol{d}_{n}^{\top}(\mathbb{X}^{\top}\mathbb{X})^{-1}\boldsymbol{d}_{n}\}^{1/2},
\]
and $\sigma$ is the standard deviation for the error term, $\boldsymbol{d}_{n}$
is a $PL\times1$ vector such that $\Vert\boldsymbol{d}_{n}\Vert=1$. 
\end{itemize}
\end{theorem}

Theorem \ref{th1} states the post bi-integrative estimator $\widetilde{\boldsymbol{\alpha}}$
and the estimator $\widetilde{\boldsymbol{\beta}}$ with known block
structure are consistent as both $N$ and $T$ $\rightarrow\infty$.
Furthermore, the estimator $\widetilde{\boldsymbol{\beta}}$ is uniformly
consistent across the samples.

\subsection{Asymptotic Property of the Proposed Estimator with Unknown Block
Structure}

In practice, the block structure is unknown. In this section, we study
the asymptotic properties of our proposed estimator with unknown block
structure. We show that, under appropriate conditions, the induced
local minimizer of the objective function (\ref{Objec}) is asymptotically
equivalent to the post bi-integrative estimator under a prior knowledge
of block structure $\widetilde{\boldsymbol{\alpha}}$. %We also derive the lower bound of the minimum difference of coefficients between subpopulation in order to be able to estimate the  coefficients.

Let 
\[
b_{n}=\min_{\substack{(i,t)\in\mathcal{A}_{l}\\
(j,t^{\prime})\in\mathcal{A}_{l^{\prime}}
}
}\left\Vert \boldsymbol{\beta}_{it}^{0}-\boldsymbol{\beta}_{jt^{\prime}}^{0}\right\Vert =\min_{l\neq l^{\prime}}\left\Vert \alpha_{l}^{0}-\alpha_{l^{\prime}}^{0}\right\Vert 
\]
be the minimum difference of the coefficients between any two blocks.
In addition, we give assumption (C3): 
\begin{itemize}
\item[(C3)] The scaled penalty functions $\rho_{\lambda}(s)=\lambda^{-1}\mathcal{P}_{\lambda}(s)$
and $\rho_{\gamma}(s)=\gamma^{-1}\mathcal{P}_{\gamma}(s)$ are symmetric,
non-decreasing and concave on $[0,\infty)$. They are constant for
$s\geq a\lambda$ or $s\geq a^{\prime}\gamma$ with some small constant
$a>0$, $a^{\prime}>0$, and $\rho_{\lambda}(0)=\rho_{\gamma}(0)=0$.
In addition, the first derivatives $\rho_{\lambda}^{\prime}(s)$ and
$\rho_{\gamma}^{\prime}(s)$ exist and are continuous except for a
finite number values for $s$ and $\rho_{\lambda}^{\prime}(0+)=\rho_{\gamma}^{\prime}(0+)=1$. 
\end{itemize}
\begin{remark} Condition (C3) is commonly given in the literature
of concave penalties and penalized high-dimensional models such as
, SCAD\citep{Fan2001} and MCP \citep{Zhang2010}. In addition, Lasso
\citep{tibshirani2005} also satisfies (C1) and (C3) and just falls
at the boundary of the class of penalty functions. \end{remark}

\begin{theorem} \label{th2} Under Conditions (C1), (C2) and (C3)
and $b_{n}>\max(a\lambda,a^{\prime}\gamma)$ with $\lambda\gg\Delta_{n}$
and $\gamma\gg\Delta_{n}$, the block-oracle estimator is a local
minimizer of the objective function with probability tending to one,
i.e., as both $N$ and $T$ $\rightarrow\infty$, 
\[
\boldsymbol{P}\left(\widehat{\boldsymbol{\beta}}(\lambda,\gamma)=\widetilde{\boldsymbol{\beta}}\right)\rightarrow1,
\]
where $\widehat{\boldsymbol{\beta}}(\lambda,\gamma)$ is the estimator
by the integrative analysis.

\end{theorem}

The result in Theorem \ref{th2} implies that if the minimal difference
of the coefficients between any two blocks is restricted by a lower
bound, our proposed double penalized least square estimator can attain
the block-oracle estimator and actually recover the true block structure
with probability tending to one. Since the local minimizer $\widehat{\boldsymbol{\alpha}}$
of the objective function just attains the block-oracle estimator
$\widetilde{\boldsymbol{\alpha}}$ , we can conclude the following
corollary.

\begin{corollary} Let $\widehat{\boldsymbol{\alpha}}$ being the
estimated coefficient vector of blocks, corresponding to $\widehat{\boldsymbol{\beta}}(\lambda,\gamma)$.
Under conditions of Theorems \ref{th2}, we obtain 
\[
s_{n}(\boldsymbol{d}_{n})^{-1}\boldsymbol{d}_{n}^{\top}(\widehat{\boldsymbol{\alpha}}-\boldsymbol{\alpha}^{0})\overset{D}{\rightarrow}N(0,1),
\]
where 
\[
s_{n}(\boldsymbol{d}_{n})=\sigma\{\boldsymbol{d}_{n}^{\top}(\mathbb{X}^{\top}\mathbb{X})^{-1}\boldsymbol{d}_{n}\}^{1/2},
\]
and $\sigma$ is the standard deviation for the error term, $\boldsymbol{d}_{n}$
is a $PL\times1$ vector such that $\|\boldsymbol{d}_{n}\|=1$. In
practice, the $\sigma$ is unknown in prior. The $\hat{\sigma}$ is
estimated 
\[
\hat{\sigma}^{2}=\left(NT-\hat{L}P\right)^{-1}\sum_{i=1}^{N}\sum_{t=1}^{T}\left(y_{it}-\boldsymbol{x}_{it}^{\top}\hat{\boldsymbol{\beta}}_{it}\right)^{2},
\]
with $\hat{\sigma}^{2}\overset{p}{\rightarrow}\sigma^{2}$.

\end{corollary}

The asymptotic distribution of the estimator provides a theoretical
foundation for further statistical inference, such as the testing
of heterogeneity. Next, we present an asymptotic $\chi^{2}$ test
for hypothesis based on the estimators $\widehat{\boldsymbol{\alpha}}$.
Specifically, we consider the null $H_{0}:\mathcal{B}\boldsymbol{\alpha}=0$
versus the alternative hypothesis $H_{1}:\mathcal{B}\boldsymbol{\alpha}\neq0$,
where $\mathcal{B}$ is a $q\times LP$ matrix and $q=$ rank$(\mathcal{B})$.
Many important special cases belong to this hypothesis. For example,
$H_{0lj}$: $\alpha_{lj}=0$, $l\in\{1,\ldots,L\}$ and $j\in\{1,\ldots,P\}$,
which can be used to test the significance of the $j$th component
of coefficients in the $l$th block; The null hypothesis $H_{0}:\alpha_{l}-\alpha_{l^{\prime}}=0$,
$l,l^{\prime}\in\{1,\ldots,L\}$ can be used to test the existence
of coefficients heterogeneity among blocks.

A standard $\chi^{2}$-test statistic for testing $H_{0}$: $\mathcal{B}\boldsymbol{\alpha}=0$
can be constructed as follows: 
\begin{equation}
\mathcal{T}(\mathcal{B})=(\mathcal{B}\widehat{\boldsymbol{\alpha}})^{\top}(\mathcal{B}\widehat{\mathcal{V}}\mathcal{B}^{\top})^{-1}(\mathcal{B}\widehat{\boldsymbol{\alpha}}),\label{Testing}
\end{equation}
where $\widehat{\mathcal{V}}=\widehat{\sigma}^{2}(\mathbb{X}^{\top}\mathbb{X})^{-1}$.

\begin{theorem} \label{TestS} Under the null hypothesis and conditions
in Theorem \ref{th2}, $\mathcal{T}(\mathcal{B})\overset{D}{\rightarrow}\chi_{q}^{2}$,
as both $N$ and $T\rightarrow\infty$. \end{theorem}

Theorem \ref{TestS} provides the asymptotic distribution of the test
statistic $\mathcal{T}(\mathcal{B})$ under the null hypothesis $H_{0}$.
Therefore, the $100(1-\tau)\%$ confidence interval for $\mathcal{B}\boldsymbol{\alpha}$
is given by 
\[
\mathbb{R}_{\tau}=\left\{ \iota:(\mathcal{B}\widehat{\boldsymbol{\alpha}}-\iota)^{\top}(\mathcal{B}\widehat{\mathcal{V}}\mathcal{B}^{\top}-\iota)^{-1}(\mathcal{B}\widehat{\boldsymbol{\alpha}})\leq\chi_{q}^{2}(1-\tau)\right\} ,
\]
where $\chi_{q}^{2}(1-\tau)$ is the $(1-\tau)$-quantile of the $\chi^{2}$
distribution with $q$ degrees of freedom.

\section{Determination of the initial values and turning parameters}

\subsection{Initial values}

The proposed ADMM algorithm requires an initialization. Initial value
matters in accelerating the convergence of the iteration. In this
paper, we propose the ridge fusion criterion to select initial parameters,
since it has closed-form solution. Let 
\begin{equation}
\ell_{R}(\boldsymbol{\beta})=\frac{1}{2}\sum_{i=1}^{N}\sum_{t=1}^{T}\left(y_{it}-\boldsymbol{x}_{it}^{\top}\boldsymbol{\beta}_{it}\right)^{2}+\frac{\lambda^{\ast}}{2}\sum_{i<j}\left\Vert \boldsymbol{\beta}_{i}-\boldsymbol{\beta}_{j}\right\Vert ^{2}+\frac{\gamma^{\ast}}{2}\sum_{t<t^{\prime}}\left\Vert \boldsymbol{\beta}_{t}-\boldsymbol{\beta}_{t^{\prime}}\right\Vert ^{2},\label{initial}
\end{equation}
which can be written in matrix form 
\begin{equation}
\ell_{R}(\boldsymbol{\beta})=\frac{1}{2}\left\Vert \boldsymbol{Y}-\boldsymbol{X}\boldsymbol{\beta}\right\Vert ^{2}+\frac{\lambda^{\ast}}{2}\left\Vert \Omega\boldsymbol{\beta}\right\Vert ^{2}+\frac{\gamma^{\ast}}{2}\left\Vert \Phi\boldsymbol{\beta}\right\Vert ^{2},\label{initialmatrix}
\end{equation}
where $\lambda^{\ast}$, $\gamma^{\ast}$ are the tuning parameters
and chosen as $\lambda^{\ast}=\gamma^{\ast}=0.001$ in determination
of initial values. By minimizing objective function (\ref{initialmatrix}),
the initial value of $\boldsymbol{\beta}$ is given by 
\[
\boldsymbol{\beta}^{(1)}=\mathrm{vec}\left\{ \{D^{\ast}(\widetilde{\boldsymbol{X}}\circ\boldsymbol{Y}-\widetilde{\boldsymbol{X}}\circ(M^{\ast}[\widetilde{\boldsymbol{X}}\circ(D^{\ast}(\widetilde{\boldsymbol{X}}\circ\boldsymbol{Y}))]^{+}))\}^{\top}\right\} ,
\]
where 
\[
D^{\ast}=\left\{ (\lambda^{\ast}N+\gamma^{\ast}T)\boldsymbol{I}_{NT}-\left[\psi(1_{N}1_{N}^{\top})\otimes\boldsymbol{I}_{T}+\phi\boldsymbol{I}_{N}\otimes(1_{T}1_{T}^{\top})\right]\right\} ^{-1}
\]
\[
M^{\ast}=\left(I_{NT}+\left(\widetilde{\boldsymbol{X}}\widetilde{\boldsymbol{X}}^{\top}\right)\circ D^{\ast}\right)^{-1}.
\]

\subsection{Optimal tuning parameters}

The proposed estimation is based on a penalized procedure that entails
choices of tuning parameters $\lambda$ and $\gamma$. \ Unsuitable
choices of tuning parameters can produce poor estimates. Motivated
by \citet{Wang2009,Ma2016}, we select the optimal tuning parameters
$\widehat{\lambda}$ and $\widehat{\gamma}$ by minimizing the following
modified BIC: 
\begin{equation}
\mathrm{BIC}(\lambda,\gamma)=\log\left(\frac{1}{NT}\left\Vert \boldsymbol{Y}-\boldsymbol{X}\widehat{\boldsymbol{\beta}}(\lambda,\gamma)\right\Vert ^{2}\right)+\mathcal{C}_{NT}\frac{\log(NT)}{NT}\left(\widehat{L}(\lambda,\gamma)P\right),\label{tuningparas}
\end{equation}
where $\mathcal{C}_{NT}$ is a constant or depending on $N$ and $T$.
Following \citet{Ma2016,MaHuang2017}, we select $\mathcal{C}_{NT}=\log(NTP)$
in the Monte Carlo simulation and empirical analysis.

For convenience of analysis, we introduce some additional notations.
Let $\mathcal{L}=\{1,2,\cdots,L_{\max}\}$, its three subsets $\mathcal{L}_{0}=\{L\in\mathcal{L}:L=L_{0}\}$,
$\mathcal{L}_{\_{}}=\{L\in\mathcal{L}:L<L_{0}\}$, $\mathcal{L}_{+}=\{L\in\mathcal{L}:L>L_{0}\}$,
represent cases of the true, under and over-fitting bi-integration,
respectively. We establish asymptotic validity of the proposed BIC
criterion in the following theorem.

\begin{theorem} Supposing that all conditions of Theorem \ref{th2}
hold, Then 
\begin{equation}
p\left(\inf_{L\in\mathcal{L}_{\_{}}\cup\mathcal{L}_{+}}\mathrm{BIC}\left(L;\lambda,\gamma\right)>\mathrm{BIC}\left(L_{0};\lambda,\gamma\right)\right)\longrightarrow1,\qquad\mathrm{as}\;(N,T)\longrightarrow\infty.
\end{equation}
\end{theorem} 

\section{Monte Carlo simulation}

In this section, we perform Monte Carlo simulation with $\mathcal{R}$
replications to investigate the finite-sample performance of the proposed
bi-integration procedure with two data generating processes with various
heterogeneous block structures on regression coefficients, measured
by two aspects that one is the evaluation of the estimated regression
coefficients and the other is the accuracy of bi-integration or recovery
of block structures. We evaluate the performance of the estimated
regression coefficients by root mean square error (RMSE) and its bias,
measured by $\frac{1}{\mathcal{R}}\sum_{r=1}^{\mathcal{R}}\sqrt{\frac{1}{NTP}\|\widehat{\boldsymbol{\beta}}^{r}-\boldsymbol{\beta}^{0}\|^{2}}$
and $\frac{1}{\mathcal{R}}\sum_{r=1}^{\mathcal{R}}\left[\frac{1}{NTP}\sum_{i=1}^{N}\sum_{t=1}^{T}\sum_{j=1}^{P}(\widehat{\beta}_{itj}^{r}-\beta_{itj}^{0})\right]$
respectively, where $\widehat{\boldsymbol{\beta}}^{r}$ is the estimated
coefficients vector in the $r$th replicate.

We evaluate the estimated numbers of blocks $\widehat{L}$ by the
percentage (Per) of $\widehat{L}$ equal to the true number of blocks
by the proposed BIGCORE procedure, calculated by $\frac{1}{\mathcal{R}}\sum_{r=1}^{\mathcal{R}}I(\widehat{L}^{r}=L^{0})$,
where $\widehat{L}^{r}$ is the calculated number of blocks in the
$r$th replicate. We also use the extended rand index(ERI), which
measures percentage of correctly membership in each blocks. The Rand
Index (RI) is used to evaluate the accuracy of clustering, which lies
between 0 and 1, where higher values indicate better performance.
Motivated by the formation of RI, we can get individual or period-specified
RIs, denoted by $\text{RI}_{t}$ or $\text{RI}_{i}$ and define the
ERI(T) and ERI(N) as the average of the whole periods and individuals
respectively, i.e., ERI(T) $=\frac{1}{T}\sum_{t=1}^{T}\text{RI}_{t}$
and ERI(N)$=\frac{1}{N}\sum_{i=1}^{N}\text{RI}_{i}$. At last we adopt
ERI=$\frac{1}{2}${[}ERI(T)+ERI(N){]} to evaluate the accuracy of
BIGCORE procedure.

\subsection{Data Generating Process}

In this sections, we generate the simulated panel data observations
$\{y_{it},x_{it}\}$, $i=1,\cdots,N$ and $t=1,\cdots,T$ by two data
generating processes (DGP) with different block structures on regression
coefficients and set the sample size as $N=20,40,60$ with $T=20,40,60$.
In order to present the wide applicability of the proposed bi-integrating
procedure, we consider the complex block structure in the example
DGP1 and classical grouped structure in DGP2, respectively.

\textbf{DGP1}:(Block structure)

In this example, we generated observations from a two-dimensional
heterogeneous panel data model, 
\[
y_{it}=\mu_{it}+x_{it}\eta_{it}+\epsilon_{it},\qquad i=1,\cdots,N;\quad t=1,\cdots,T.
\]
Both the time-varying individual fixed effect $\mu_{it}$ and one-dimensional
slope coefficient $\eta_{it}$ have the same structure of time-varying
group memberships with common structural break depicted in Figure
(\ref{subfig:4}). Then two blocks are considered and coefficient
vector of the first block is set as $\boldsymbol{\alpha}_{1}=(-2,3)$
and let that of the second one be $\boldsymbol{\alpha}_{2}=(2,5)$
with the components corresponding to fixed effect $\mu_{it}$ and
slope coefficient $\eta_{it}$, respectively. The block-based two-dimensional
heterogeneous structure can be depicted by group-cohort pattern, such
as, under the case of $N=40$, $T=40$, we consider group structure
with $\mathcal{G}_{01}=\{1,\cdots,10,31,\cdots,40\}$, $\mathcal{G}_{02}=\{11,\cdots,20\}$,
$\mathcal{G}_{03}=\{21,\cdots,30\}$ and the corresponding cohort
structures under the assumed group formation are set by $\mathcal{H}_{01}(1)=\{1,\cdots,40\}$,
$\mathcal{H}_{01}(2)=\{1,\cdots,19,30,\cdots,40\}$, $\mathcal{H}_{02}(2)=\{20,\cdots,29\}$,
$\mathcal{H}_{01}(3)=\{1,\cdots,9,35,\cdots,40\}$, $\mathcal{H}_{02}(3)=\{10,\cdots,34\}$
and $\mathcal{H}_{01}(4)=\{1,\cdots,40\}$. Therefore, the ratio of
number of block-specified observations is about $|\mathcal{A}_{1}|:|\mathcal{A}_{2}|\approx3:1$.
Other cases setting the sample sizes in the block structure under
different combination of $N$ and $T$ has the similar way. The regressor
$x_{it}$ is generated by 
\[
x_{it}=1+0.5\mu_{it}+\epsilon_{it},
\]
where $\epsilon_{it}$ was taken from the standard normal distribution.

We consider the settings of homoscedasticity and heteroscedasticity
on the error term by respectively generating $\epsilon_{it}\sim\ N(0,\sigma^{2})$
with $\sigma^{2}=0.5$, $\sigma^{2}=1$ and 
\[
\epsilon_{it}=\sigma_{it}e_{it},\sigma_{it}=\tau(0.05+0.05x_{it}^{2})^{1/2},
\]
where $\tau=2$ or $\tau=1$ and $e_{it}\sim N(0,1)$.

\textbf{DGP2}(Grouped structure)

In this example, we consider the performance of proposed BIGCORE analysis
in panel data model with grouped individual fixed effect and grouped
slope coefficients. The Datasets are generated as: 
\[
y_{it}=\mu_{i}+x_{it}\eta_{i}+\epsilon_{it}.
\]
Here, both the fixed effects and slope coefficients have identical
grouped structure by randomly dividing the individuals into three
groups with the proportion that $|\mathcal{G}_{1}|:|\mathcal{G}_{2}|:|\mathcal{G}_{3}|=3:3:4$,
in which the true coefficients are $\boldsymbol{\alpha}_{1}=\{-2,3\}$,
$\boldsymbol{\alpha}_{1}=\{2,6\}$ and $\boldsymbol{\alpha}_{3}=\{6,-1\}$,
respectively. The regressor $x_{it}$ are generated as 
\[
x_{it}=1+0.5\mu_{i}+\epsilon_{it}.
\]
Lastly, the error term is set as in DGP 1.

\subsection{Simulation Results}

In the simulation, we select the number of replicates $\mathcal{R}=100$.
The grid of values of both tuning parameters $\lambda$ and $\gamma$
is set in the range of {[}0.1, 1.5{]} with step size 0.1. The fact
that increasing grid of tuning parameters apparently improves integrative
results is unconsidered here due to reducing computational cost. In
order to accelerate the convergence of the proposed ADMM algorithm,
we regard the converged value under the given combined tuning parameters
as the initial value of the next iteration, instead of adopting identical
initial values under different combined tuning parameters in the grid.

\begin{figure}[!ht]
\centering \subfloat[Estimated fixed effect in DGP1]{\includegraphics[width=5.3cm,height=4.5cm]{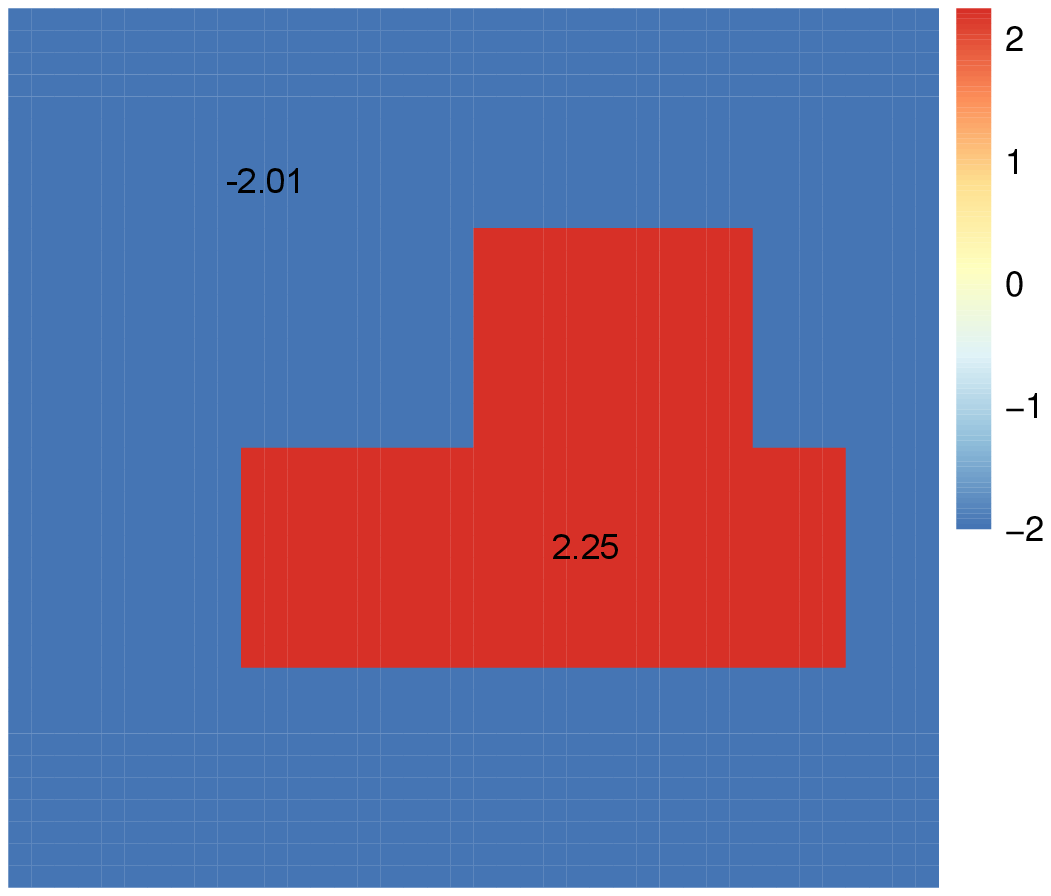}\label{EstDGP1:1}

}\hspace{30pt} \subfloat[Estimated slope efficient in DGP1]{\includegraphics[width=5.3cm,height=4.5cm]{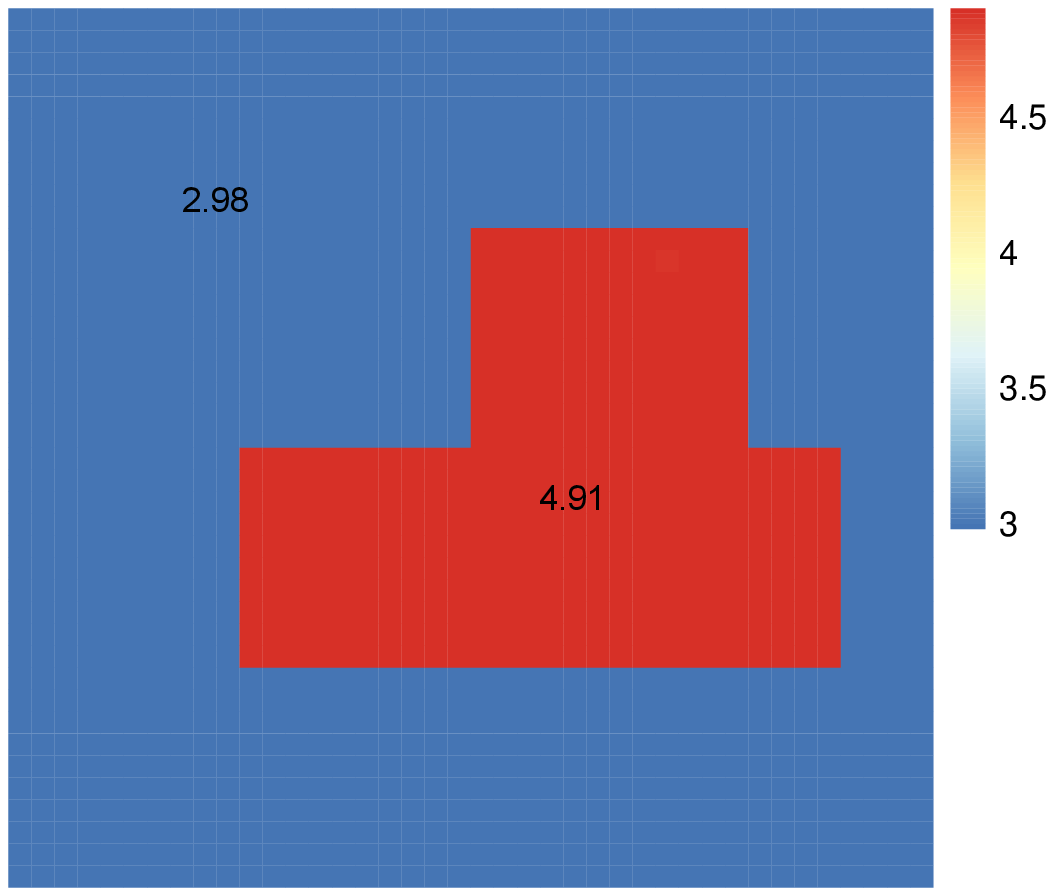}\label{EstDGP1:2}

}\\
 \subfloat[Estimated fixed effect in DGP2]{\includegraphics[width=5.3cm,height=4.5cm]{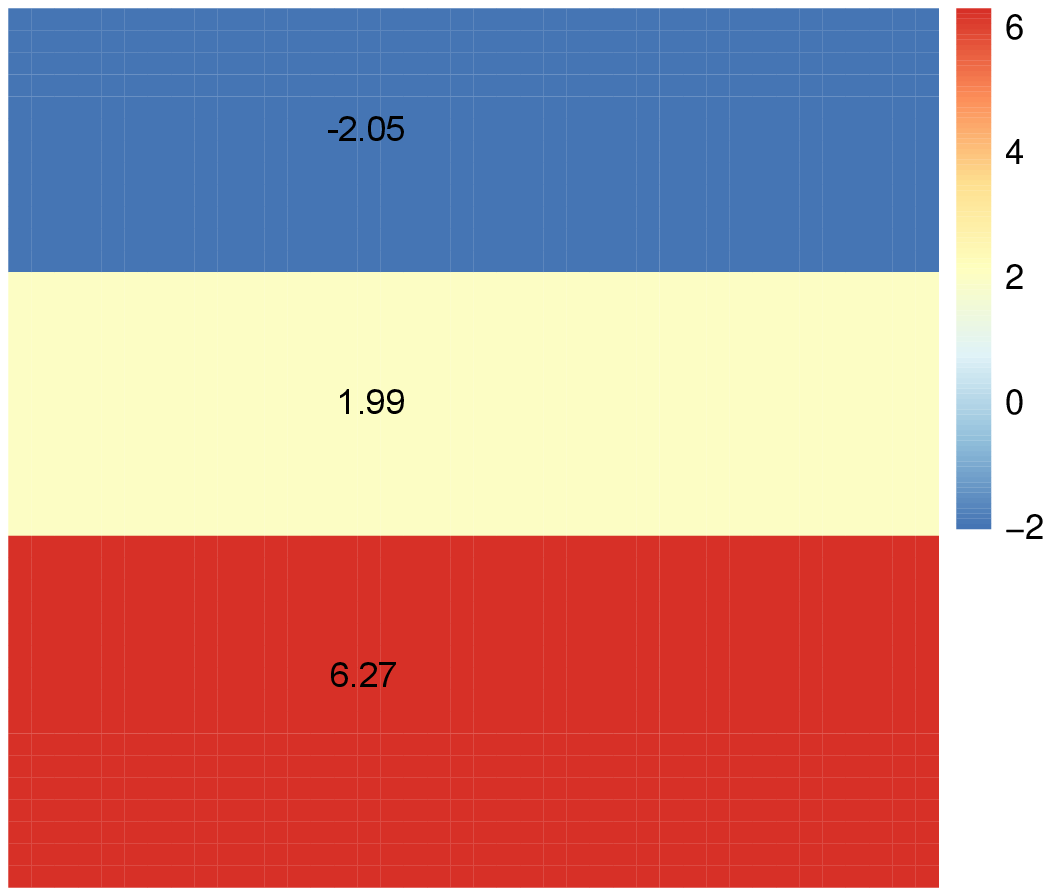}\label{EstDGP2:1}

}\hspace{30pt} \subfloat[Estimated slope efficient in DGP2]{\includegraphics[width=5.3cm,height=4.5cm]{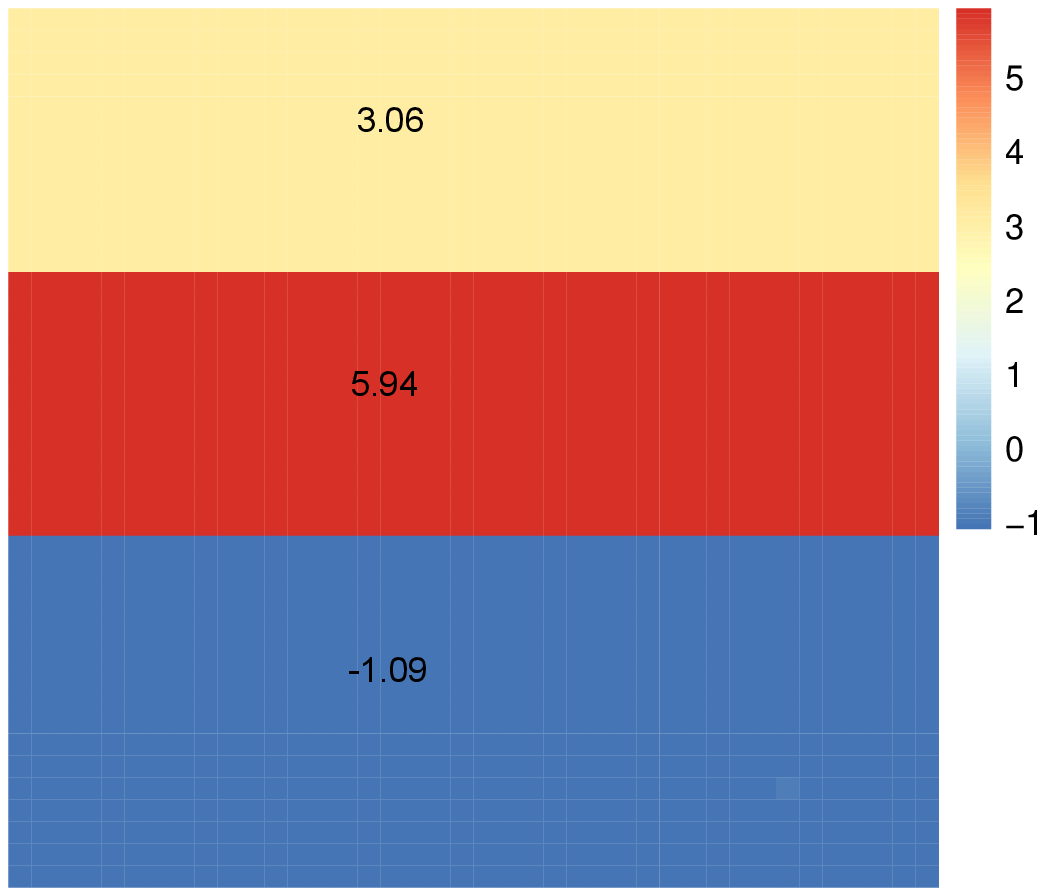}\label{EstDGP2:2}

}\caption{The estimates of result of fixed effect and slope efficient in DGP1
and DGP2 by one replicate.}
\label{EstOneTime} 
\end{figure}

After one replicate, Figures (\ref{EstOneTime}) vividly presents
the performance of our proposed BIGCORE analysis in estimating coefficients
and block structure in DGP1 and DGP2 settings under $N=40$, $T=40$
and heteroscedasticity with $\tau=2$ and the figure elucidates that
the developed method can achieve expected outcome because it can recover
the true block or grouped structure correctly with consistent value
of coefficient estimators. Another inevitable fact we should admit
is that although Figures (\ref{EstOneTime}) actually shows ideal
simulated results in one replicate, several worse BIGCORE results
still occur occasionally, which is reflected by the misintegration
that a little of observations may be wrongly partitioned, such as,
it may occur in the replicates with the cases of large standard deviation
of error term. Specifically, a little of observations originally belonging
to red block are mistakenly bi-integrated into the blue block. 

For checking the representation of coefficients estimators, we report
the RMSE and Bias of the slope coefficient for examples DGP1 and DGP2
in Tables (\ref{TabBias1}) and Table (\ref{TabBias2}), respectively,
which elucidates that (i) both SCAD and MCP penalties present similar
BIGCORE behaviors in terms of close values of RMSE and Bias, and both
are also close to the oracle results, which is the reason we reject
arguing the effective combination of the forms of double concave penalties;
(ii) with increasing number of individuals or periods, the values
of RMSE and the Bias decrease remarkably in all cases; (iii) the values
of RMSE and Bias in DGP1 are relatively larger than that in DGP2.
It may be owe to the more complex formation of block structure than
that of grouped pattern and the group-specified cohort structure should
be integrated in DGP1; (iv) the post estimators is recommended due
to that it attains much smaller RMSE and Bias, especially as $N$
or $T$ increases. It is also noted that the estimation performance
on slop coefficients under post-MCP is usually the same to that of
post-SCAD, which is attributed to the common block structure recovery
by both penalties.

For evaluating the accuracy of BIGCORE procedure in estimating the
number of blocks, Table (\ref{TabCount}) and Table (\ref{TabCount-1})
reports the percentage of the estimated numbers of blocks equal to
the true number of blocks by the SCAD and MCP shrinkage procedures
under different cases of DGP1 and DGP2. In all cases, the percentage
of correctly selecting the number of blocks increases as $N$ and
$T$ are enlarged. The two concave penalties SCAD and MCP procedures
have similar performance.

Another results under the evaluation criterion extended Rand index,
which is used to measure the bi-integration ability of recovering
the true underlying structures, are reported in Table (\ref{TabCount2}),
Table (\ref{TabCount2-1}) and the results show that the extended
Rand index are mostly close to one, which indicates the effectiveness
of the proposed BIGCORE method. The results also imply that the BIGCORE
performance was worsen with serious heteroscedasticity such as larger
$\sigma^{2}$.

\begin{table}
\caption{\label{TabBias1}The root mean square error (RMSE) and bias of the
estimator of $\eta_{it}$ in DGP1}

\centering

\begin{tabular}{cccccccccc}
\hline 
 &  & \multicolumn{4}{c}{Homoscedasticity} & \multicolumn{4}{c}{Heteroscedasticity}\tabularnewline
\hline 
 &  & \multicolumn{2}{c}{$\sigma^{2}=0.5$} & \multicolumn{2}{c}{$\sigma^{2}=1$} & \multicolumn{2}{c}{$\tau=1$} & \multicolumn{2}{c}{$\tau=2$}\tabularnewline
\hline 
(N,T) & Methods & RMSE & Bias & RMSE & Bias & RMSE & Bias & RMSE & Bias\tabularnewline
\hline 
(20, 20) & SCAD & 0.0718 & -0.0005 & 0.1442 & 0.0094 & 0.0645 & -0.0066 & 0.3288 & 0.0286\tabularnewline
 & Post-SCAD & 0.0517 & 0.0002 & 0.0842 & 0.0165 & 0.0529 & 0.0005 & 0.1616 & 0.0070\tabularnewline
 & MCP & 0.0713 & -0.001 & 0.1431 & 0.0159 & 0.0656 & -0.0052 & 0.3258 & 0.0245\tabularnewline
 & Post-MCP & 0.1122 & 0.0001 & 0.0894 & 0.0189 & 0.0529 & -0.0009 & 0.1644 & 0.0071\tabularnewline
 & Oracle & 0.0440 & 0.0002 & 0.0587 & 0.0122 & 0.0491 & -0.0009 & 0.1473 & 0.0085\tabularnewline
(40, 20) & SCAD & 0.0596 & -0.0033 & 0.1576 & 0.0126 & 0.0531 & -0.0033 & 0.3984 & 0.0312\tabularnewline
 & Post-SCAD & 0.0354 & 0.0015 & 0.0716 & 0.0082 & 0.0336 & 0.0026 & 0.2259 & 0.0034\tabularnewline
 & MCP & 0.0585 & -0.0037 & 0.1558 & 0.0149 & 0.0558 & -0.0033 & 0.3903 & 0.0309\tabularnewline
 & Post-MCP & 0.0354 & 0.00015 & 0.0725 & 0.0087 & 0.0379 & 0.0023 & 0.2216 & 0.0031\tabularnewline
 & Oracle & 0.0300 & 0.0007 & 0.0443 & 0.0037 & 0.0324 & 0.0021 & 0.2048 & 0.0041\tabularnewline
(20, 40) & SCAD & 0.0728 & -0.0039 & 0.1429 & 0.0057 & 0.0608 & -0.0055 & 0.2622 & 0.1058\tabularnewline
 & Post-SCAD & 0.0344 & 0.0002 & 0.0458 & 0.0029 & 0.0429 & -0.0044 & 0.1360 & 0.0195\tabularnewline
 & MCP & 0.0732 & -0.0016 & 0.1426 & 0.0064 & 0.0623 & -0.0048 & 0.2576 & 0.0999\tabularnewline
 & Post-MCP & 0.0345 & 0.0002 & 0.0463 & 0.0027 & 0.0411 & -0.0037 & 0.1301 & 0.0046\tabularnewline
 & Oracle & 0.0331 & 0.0002 & 0.0415 & 0.0029 & 0.0382 & -0.0038 & 0.0727 & 0.0016\tabularnewline
(40, 40) & SCAD & 0.0541 & -0.0058 & 0.1315 & 0.0030 & 0.0414 & -0.0057 & 0.1238 & 0.0319\tabularnewline
 & Post-SCAD & 0.0226 & -0.0026 & 0.0306 & 0.0002 & 0.0226 & 0.0013 & 0.0491 & 0.0016\tabularnewline
 & MCP & 0.1179 & -0.0075 & 0.1306 & 0.0035 & 0.0489 & -0.0029 & 0.1217 & 0.0341\tabularnewline
 & Post-MCP & 0.0226 & -0.0027 & 0.0301 & 0.0002 & 0.0263 & 0.0018 & 0.0473 & 0.0022\tabularnewline
 & Oracle & 0.0226 & -0.0026 & 0.0298 & 0.0001 & 0.0226 & 0.0013 & 0.0470 & 0.0014\tabularnewline
(60, 40) & SCAD & 0.0569 & -0.0037 & 0.1336 & 0.0034 & 0.0447 & -0.0037 & 0.1748 & 0.0844\tabularnewline
 & Post-SCAD & 0.0184 & 0.0012 & 0.0278 & -0.0015 & 0.0203 & 0.0014 & 0.0496 & 0.0016\tabularnewline
 & MCP & 0.0561 & -0.0030 & 0.1332 & 0.0041 & 0.0448 & -0.0037 & 0.0415 & 0.0013\tabularnewline
 & Post-MCP & 0.0186 & 0.0012 & 0.0253 & -0.0018 & 0.0202 & 0.00014 & 0.0481 & 0.0015\tabularnewline
 & Oracle & 0.0184 & 0.0012 & 0.0260 & -0.0017 & 0.0203 & -0.0009 & 0.0382 & -0.0013\tabularnewline
(40, 60) & SCAD & 0.0549 & -0.0066 & 0.1294 & 0.0030 & 0.0406 & -0.0025 & 0.1701 & 0.0805\tabularnewline
 & Post-SCAD & 0.0197 & 0.0008 & 0.0295 & 0.0004 & 0.0190 & -0.0009 & 0.0699 & 0.0027\tabularnewline
 & MCP & 0.0544 & -0.0062 & 0.2676 & -0.0015 & 0.0407 & -0.0027 & 0.1702 & 0.0110\tabularnewline
 & Post-MCP & 0.0197 & 0.0008 & 0.0298 & 0.0004 & 0.0190 & -0.0009 & 0.0624 & 0.0013\tabularnewline
 & Oracle & 0.0197 & 0.0008 & 0.0257 & 0.0004 & 0.0190 & -0.0009 & 0.0482 & -0.0013\tabularnewline
(60, 60) & SCAD & 0.0455 & -0.0013 & 0.1158 & -0.0024 & 0.0385 & -0.0019 & 0.1178 & 0.0156\tabularnewline
 & Post-SCAD & 0.0197 & 0.0063 & 0.0284 & 0.0019 & 0.0188 & -0.0007 & 0.0485 & -0.0012\tabularnewline
 & MCP & 0.0827 & -0.0011 & 0.1212 & -0.0031 & 0.0382 & -0.0019 & 0.1193 & 0.0163\tabularnewline
 & Post-MCP & 0.0197 & 0.0063 & 0.0285 & 0.0019 & 0.0188 & -0.0007 & 0.0488 & -0.0014\tabularnewline
 & Oracle & 0.0197 & 0.0063 & 0.0284 & 0.0019 & 0.0188 & -0.0007 & 0.0281 & 0.0041\tabularnewline
\hline 
\end{tabular}
\end{table}

\begin{table}
\caption{\label{TabBias2}The root mean square error (RMSE) and bias of the
estimator of $\eta_{i}$ in DGP2.}

\centering

\begin{tabular}{cccccccccc}
\hline 
 &  & \multicolumn{4}{c}{Homoscedasticity} & \multicolumn{4}{c}{Heteroscedasticity}\tabularnewline
\hline 
 &  & \multicolumn{2}{c}{$\sigma^{2}=0.5$} & \multicolumn{2}{c}{$\sigma^{2}=1$} & \multicolumn{2}{c}{$\tau=1$} & \multicolumn{2}{c}{$\tau=2$}\tabularnewline
\hline 
(N,T) & Methods & RMSE & Bias & RMSE & Bias & RMSE & Bias & RMSE & Bias\tabularnewline
\hline 
(20,20) & SCAD & 0.0924 & 0.0031 & 0.2424 & 0.0183 & 0.1030 & 0.0051 & 0.3288 & 0.0286\tabularnewline
 & Post-SCAD & 0.0480 & -0.0036 & 0.0881 & -0.0142 & 0.0666 & -0.0041 & 0.1615 & 0.0069\tabularnewline
 & MCP & 0.0924 & 0.0036 & 0.2428 & 0.0187 & 0.1046 & 0.0051 & 0.2129 & -0.0189\tabularnewline
 & Post-MCP & 0.0474 & -0.0036 & 0.0867 & -0.0132 & 0.0666 & -0.0042 & 0.0812 & -0.0018\tabularnewline
 & Oracle & 0.0457 & -0.0037 & 0.0655 & -0.0109 & 0.0666 & -0.0041 & 0.1473 & 0.0085\tabularnewline
(40, 20) & SCAD & 0.0978 & 0.0039 & 0.2443 & -0.0153 & 0.0998 & -0.0054 & 0.3102 & -0.0231\tabularnewline
 & Post-SCAD & 0.0759 & 0.0011 & 0.0754 & 0.0172 & 0.0674 & 0.0042 & 0.1561 & 0.0054\tabularnewline
 & MCP & 0.0962 & 0.0034 & 0.2453 & -0.0153 & 0.1009 & -0.0059 & 0.3154 & -0.0231\tabularnewline
 & Post-MCP & 0.0714 & 0.0037 & 0.0763 & 0.0172 & 0.0647 & 0.0045 & 0.1567 & 0.0058\tabularnewline
 & Oracle & 0.0314 & 0.0038 & 0.0639 & 0.0111 & 0.0638 & -0.0019 & 0.1248 & 0.0041\tabularnewline
(20, 40) & SCAD & 0.0344 & 0.0025 & 0.0773 & 0.0055 & 0.0509 & 0.0025 & 0.0966 & 0.0042\tabularnewline
 & Post-SCAD & 0.0315 & -0.0014 & 0.0483 & -0.0121 & 0.0507 & 0.0008 & 0.0954 & -0.0079\tabularnewline
 & MCP & 0.0345 & 0.0023 & 0.1811 & 0.0051 & 0.0513 & 0.0028 & 0.0972 & 0.0042\tabularnewline
 & Post-MCP & 0.0315 & -0.0012 & 0.1392 & -0.0121 & 0.0492 & 0.0008 & 0.0955 & -0.0079\tabularnewline
 & Oracle & 0.0315 & -0.0014 & 0.0483 & -0.0121 & 0.0507 & 0.0008 & 0.0954 & -0.0079\tabularnewline
(40,40) & SCAD & 0.0646 & 0.0034 & 0.1635 & 0.0064 & 0.0688 & 0.0046 & 0.2869 & 0.0595\tabularnewline
 & Post-SCAD & 0.0209 & -0.0002 & 0.0317 & -0.0002 & 0.0346 & 0.0025 & 0.0932 & 0.0019\tabularnewline
 & MCP & 0.0646 & 0.0036 & 0.1628 & 0.0064 & 0.0686 & 0.0040 & 0.2828 & 0.0560\tabularnewline
 & Post-MCP & 0.0208 & -0.0003 & 0.0317 & -0.0002 & 0.0344 & 0.0027 & 0.0936 & 0.0020\tabularnewline
 & Oracle & 0.0209 & -0.0002 & 0.0317 & -0.0002 & 0.0346 & 0.0025 & 0.0707 & 0.0016\tabularnewline
(60, 40) & SCAD & 0.0998 & 0.0084 & 0.2182 & 0.0210 & 0.0924 & 0.0054 & 0.2414 & -0.0247\tabularnewline
 & Post-SCAD & 0.0189 & 0.0014 & 0.0279 & -0.0015 & 0.0189 & -0.0009 & 0.0248 & 0.0041\tabularnewline
 & MCP & 0.0977 & 0.0081 & 0.2179 & 0.0100 & 0.0934 & 0.0058 & 0.2399 & -0.0247\tabularnewline
 & Post-MCP & 0.0169 & 0.0014 & 0.0279 & -0.0015 & 0.0189 & -0.0009 & 0.0249 & 0.0041\tabularnewline
 & Oracle & 0.0189 & 0.0014 & 0.0278 & -0.0015 & 0.0188 & -0.0009 & 0.0167 & 0.0041\tabularnewline
(40, 60) & SCAD & 0.0352 & 0.0021 & 0.0545 & 0.0025 & 0.0354 & 0.0015 & 0.0828 & 0.0047\tabularnewline
 & Post-SCAD & 0.0189 & -0.0001 & 0.0334 & -0.0001 & 0.0132 & 0.0017 & 0.0827 & 0.0010\tabularnewline
 & SCAD & 0.0349 & 0.0020 & 0.0542 & 0.0023 & 0.0348 & 0.0019 & 0.0833 & 0.0049\tabularnewline
 & Post-SCAD & 0.0188 & -0.0001 & 0.0329 & -0.0001 & 0.0132 & 0.0017 & 0.0845 & 0.0011\tabularnewline
 & Oracle & 0.0188 & -0.0001 & 0.0276 & -0.0001 & 0.0132 & 0.0017 & 0.0845 & 0.0009\tabularnewline
(60, 60) & SCAD & 0.0443 & 0.0029 & 0.1153 & 0.0055 & 0.0414 & 0.0037 & 0.2469 & 0.0315\tabularnewline
 & Post-SCAD & 0.0189 & -0.0001 & 0.0299 & -0.0002 & 0.0162 & 0.0021 & 0.0724 & 0.0011\tabularnewline
 & MCP & 0.0438 & 0.0030 & 0.1152 & 0.0054 & 0.0414 & 0.0037 & 0.2421 & 0.0315\tabularnewline
 & Post-MCP & 0.0189 & -0.0001 & 0.0294 & -0.0002 & 0.0162 & 0.0021 & 0.0724 & 0.0017\tabularnewline
 & Oracle & 0.0189 & -0.0001 & 0.0217 & -0.0002 & 0.0162 & 0.0027 & 0.0619 & 0.0009\tabularnewline
\hline 
\end{tabular}
\end{table}

\begin{table}
\caption{\label{TabCount}The percentage(Per) of $\widehat{L}$ equal to the
true number of blocks in DGP1.}

\centering

\begin{tabular}{ccccccc}
\hline 
 &  & \multicolumn{2}{c}{Homoscedasticity} &  & \multicolumn{2}{c}{Heteroscedasticity}\tabularnewline
\hline 
 &  & \multicolumn{1}{c}{$\sigma^{2}=0.5$} & \multicolumn{1}{c}{$\sigma^{2}=1$} &  & \multicolumn{1}{c}{$\tau=1$} & \multicolumn{1}{c}{$\tau=2$}\tabularnewline
\hline 
(N,T) & Methods & Per & Per &  & Per & Per\tabularnewline
\hline 
(20, 20) & SCAD & 0.99 & 0.89 &  & 1.00 & 0.80\tabularnewline
 & MCP & 0.99 & 0.89 &  & 1.00 & 0.81\tabularnewline
(40, 20) & SCAD & 1.00 & 0.90 &  & 0.97 & 0.84\tabularnewline
 & MCP & 1.00 & 0.90 &  & 0.97 & 0.84\tabularnewline
(20, 40) & SCAD & 0.99 & 0.85 &  & 0.97 & 0.89\tabularnewline
 & MCP & 0.99 & 0.87 &  & 0.97 & 0.88\tabularnewline
(40, 40) & SCAD & 0.98 & 0.89 &  & 1.00 & 0.92\tabularnewline
 & MCP & 0.98 & 0.89 &  & 1.00 & 0.92\tabularnewline
(60, 40) & SCAD & 0.98 & 0.93 &  & 0.99 & 0.94\tabularnewline
 & MCP & 0.98 & 0.93 &  & 0.98 & 0.93\tabularnewline
(40, 60) & SCAD & 1.00 & 0.91 &  & 0.99 & 0.91\tabularnewline
 & MCP & 1.00 & 0.92 &  & 0.99 & 0.92\tabularnewline
(60, 60) & SCAD & 1.00 & 0.95 &  & 0.99 & 0.94\tabularnewline
 & MCP & 1.00 & 0.95 &  & 0.99 & 0.94\tabularnewline
\hline 
\end{tabular}
\end{table}

\begin{table}
\caption{\label{TabCount-1}The percentage(Per) of $\widehat{L}$ equal to
the true number of blocks in DGP2.}

\centering

\begin{tabular}{ccccccc}
\hline 
 &  & \multicolumn{2}{c}{Homoscedasticity} &  & \multicolumn{2}{c}{Heteroscedasticity}\tabularnewline
\hline 
 &  & \multicolumn{1}{c}{$\sigma^{2}=0.5$} & \multicolumn{1}{c}{$\sigma^{2}=1$} &  & \multicolumn{1}{c}{$\tau=1$} & \multicolumn{1}{c}{$\tau=2$}\tabularnewline
\hline 
(N,T) & Methods & Per & Per &  & Per & Per\tabularnewline
\hline 
(20, 20) & SCAD & 1.00 & 1.00 &  & 1.00 & 0.99\tabularnewline
 & MCP & 1.00 & 1.00 &  & 1.00 & 1.00\tabularnewline
(40, 20) & SCAD & 1.00 & 1.00 &  & 1.00 & 0.98\tabularnewline
 & MCP & 1.00 & 1.00 &  & 1.00 & 1.00\tabularnewline
(20, 40) & SCAD & 1.00 & 1.00 &  & 1.00 & 1.00\tabularnewline
 & MCP & 1.00 & 1.00 &  & 1.00 & 1.00\tabularnewline
(40, 40) & SCAD & 1.00 & 1.00 &  & 1.00 & 0.97\tabularnewline
 & MCP & 1.00 & 1.00 &  & 1.00 & 0.98\tabularnewline
(60, 40) & SCAD & 1.00 & 1.00 &  & 1.00 & 1.00\tabularnewline
 & MCP & 1.00 & 1.00 &  & 1.00 & 1.00\tabularnewline
(40, 60) & SCAD & 1.00 & 1.00 &  & 1.00 & 1.00\tabularnewline
 & MCP & 1.00 & 1.00 &  & 1.00 & 1.00\tabularnewline
(60, 60) & SCAD & 1.00 & 1.00 &  & 1.00 & 1.00\tabularnewline
 & MCP & 1.00 & 1.00 &  & 1.00 & 1.00\tabularnewline
\hline 
\end{tabular}
\end{table}

\begin{table}
\caption{\label{TabCount2}The values of Extended Rand Index (ERI) in DGP1}

\centering

\begin{tabular}{cccccc}
\hline 
 &  & \multicolumn{2}{c}{Homoscedasticity} & \multicolumn{2}{c}{Heteroscedasticity}\tabularnewline
\hline 
 &  & \multicolumn{1}{c}{$\sigma^{2}=0.5$} & \multicolumn{1}{c}{$\sigma^{2}=1$} & \multicolumn{1}{c}{$\tau=1$} & \multicolumn{1}{c}{$\tau=2$}\tabularnewline
\hline 
(N,T) & Methods & ERI & ERI & ERI & ERI\tabularnewline
\hline 
(20, 20) & SCAD & 0.9974 & 0.9892 & 0.9991 & 0.9880\tabularnewline
 & MCP & 0.9973 & 0.9897 & 0.9994 & 0.9884\tabularnewline
(40, 20) & SCAD & 0.9978 & 0.9863 & 0.9983 & 0.9742\tabularnewline
 & MCP & 0.9978 & 0.9854 & 0.9982 & 0.9731\tabularnewline
(20, 40) & SCAD & 0.9968 & 0.9879 & 0.9981 & 0.9801\tabularnewline
 & MCP & 0.9969 & 0.9878 & 0.9979 & 0.9995\tabularnewline
(40, 40) & SCAD & 0.9982 & 0.9902 & 0.9990 & 0.9926\tabularnewline
 & MCP & 0.9982 & 0.9898 & 0.9990 & 0.9926\tabularnewline
(60, 40) & SCAD & 0.9980 & 0.9899 & 0.9989 & 0.9954\tabularnewline
 & MCP & 0.9979 & 0.9893 & 0.9989 & 0.9956\tabularnewline
(40, 60) & SCAD & 0.9981 & 0.9903 & 0.9990 & 0.9949\tabularnewline
 & MCP & 0.9980 & 0.9900 & 0.9990 & 0.9953\tabularnewline
(60, 60) & SCAD & 0.9981 & 0.9903 & 0.9990 & 0.9949\tabularnewline
 & MCP & 0.9980 & 0.9900 & 0.9990 & 0.9953\tabularnewline
\hline 
\end{tabular}

\end{table}

\begin{table}
\caption{\label{TabCount2-1}The values of Extended Rand Index(ERI) in DGP2.}

\centering

\begin{tabular}{cccccc}
\hline 
 &  & \multicolumn{2}{c}{Homoscedasticity} & \multicolumn{2}{c}{Heteroscedasticity}\tabularnewline
\hline 
 &  & \multicolumn{1}{c}{$\sigma^{2}=0.5$} & \multicolumn{1}{c}{$\sigma^{2}=1$} & \multicolumn{1}{c}{$\tau=1$} & \multicolumn{1}{c}{$\tau=2$}\tabularnewline
\hline 
(N,T) & Methods & ERI & ERI & ERI & ERI\tabularnewline
\hline 
(20, 20) & SCAD & 0.9986 & 0.9932 & 0.9993 & 0.9894\tabularnewline
 & MCP & 0.9986 & 0.8910 & 0.9993 & 0.9142\tabularnewline
(40, 20) & SCAD & 0.9647 & 0.9274 & 0.9814 & 0.9415\tabularnewline
 & MCP & 0.9659 & 0.9243 & 0.9820 & 0.9409\tabularnewline
(20, 40) & SCAD & 0.9999 & 0.9989 & 1.0000 & 0.9995\tabularnewline
 & MCP & 0.9999 & 0.9981 & 0.9998 & 0.9986\tabularnewline
(40, 40) & SCAD & 0.9992 & 0.9963 & 0.9996 & 0.9894\tabularnewline
 & MCP & 0.9993 & 0.9965 & 0.9996 & 0.9142\tabularnewline
(60, 40) & SCAD & 0.9816 & 0.9428 & 0.9883 & 0.9512\tabularnewline
 & MCP & 0.9799 & 0.9449 & 0.9901 & 0.9564\tabularnewline
(40, 60) & SCAD & 0.9999 & 0.9992 & 1.0000 & 0.9894\tabularnewline
 & MCP & 0.9999 & 0.9990 & 1.0000 & 0.9887\tabularnewline
(60, 60) & SCAD & 0.9996 & 0.9982 & 0.9998 & 0.9901\tabularnewline
 & MCP & 0.9996 & 0.9979 & 0.9998 & 0.9912\tabularnewline
\hline 
\end{tabular}
\end{table}

\section{Empirical application}

In this section, we apply the proposed BIGCORE procedure to measure
the heterogeneous impact of inputs on the economic output. Based on
the classical Solow model about economic growth equation, the economic
output is mainly determined by technological progress, Capital and
Labor, and we establish the following regression given by 
\begin{equation}
\log(\mathrm{GDP}_{it})=\mu_{it}+\beta_{1,it}\log(\mathrm{Hc}_{it})+\beta_{2,it}\log(\mathrm{Ck}_{it})+\beta_{3,it}\log(\mathrm{Ngd}_{it})+\epsilon_{it},
\end{equation}
where $\mathrm{GDP}_{it}$ denotes the real gross domestic product,
technological progress are usually represented by human capital(Hc)
in the empirical literature, Ck is physical capital stock and Ngd
denotes population growth plus break even investments of $5\%$. The
coefficients represent different economic meanings, such as, the slope
coefficient $\beta_{2,it}$ is the elasticity of investment on output.
In the whole world, countries with different resource endowments and
technical power are at different stages of development. For the developing
countries, the elasticity of investment on output is larger than that
of developed countries according to the law of economic development.
From the perspective of period, the improvement of technological progress
on economic development is diminishing marginally or remain stable,
and the marginal effect shifts to higher level as the emergence of
new technologies. Therefore, the slope coefficients are heterogeneous
across countries and may exist structural breaks in the long span.

\begin{figure}
\center{\includegraphics[width=14cm,height=9cm]{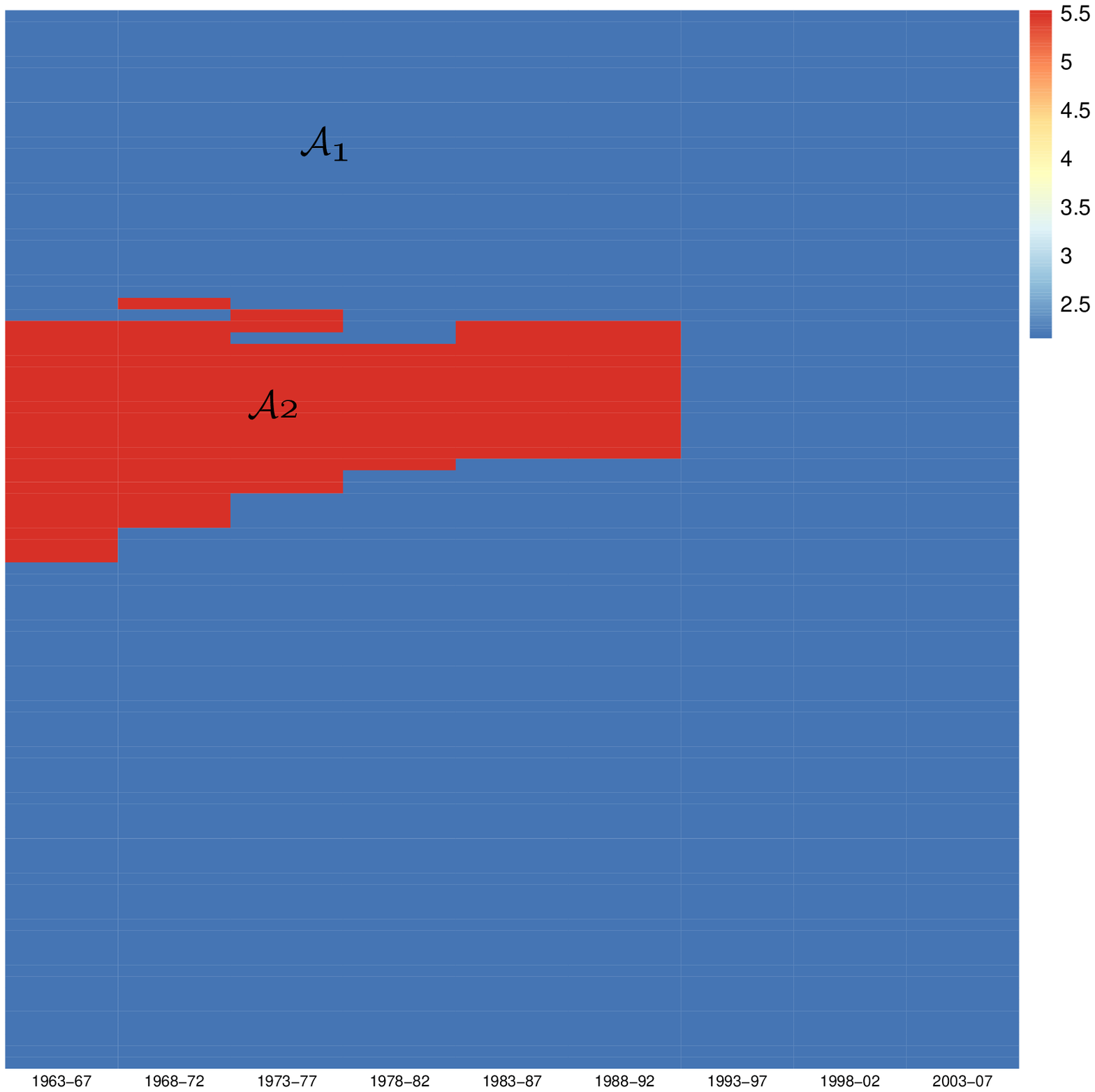}
\caption{The estimated block structure and the corresponded estimation of intercept.}
\label{EmpiricalPlot}} 
\end{figure}

The original data are available form the Penn World Tables 8 and can
be directly obtained from the package xtdcce2 of Stata. The dataset
contains panel data with 92 countries and its yearly observations
from 1963 until 2007. Same as \citet{QianSu2016}, the observations
are averaged by each 5 years. Thus, we finally get panel data with
$N=92$, $T=9$ and $P=4$. In addition, we set a grid of turning
parameters $\lambda$ and $\gamma$ from 0.2 to 3 with interval of
0.2. Finally, we get two blocks $\hat{L}=2$ shown in Figure (\ref{EmpiricalPlot})
that the blue block is denoted by $\mathcal{A}_{1}$ and the red block
is denoted by $\mathcal{A}_{2}$. In the estimation process, optimal
$\lambda$ and $\gamma$ are selected by 2.6 and 1.6 respectively,
according to the modified BIC. The penalized estimators and post estimators
based on the estimated block structure are reported in the Table (\ref{EmpiricalTable})
and the results imply that all the penalized estimators are statistically
significant and most post estimators are statistically significant,
except for $\beta_{2,it}$ and $\beta_{3,it}$ in block $\mathcal{A}_{2}$.
What's more, the standard errors of post estimators are smaller than
that of penalized estimators, which is consistent to that of the Monte
Carlo simulation. In conclusion, the block heterogeneity is remarkable
in Solow model based on the BIGCORE method that the elasticity or
marginal coefficient are significant across blocks, owing to different
endowments across countries, continuous development and progress.

\begin{table}
\caption{\label{EmpiricalTable}Empirical Results.}

\centering

\begin{tabular}{@{\extracolsep{5pt}}lD{.}{.}{-3}D{.}{.}{-3}D{.}{.}{-3}D{.}{.}{-3}}
\hline 
 & \multicolumn{4}{c}{Dependent Variable: $\log({\rm GDP})$}\tabularnewline
\hline 
 & \multicolumn{2}{c}{\textit{Penalized Estimators}} & \multicolumn{2}{c}{\textit{Post Estimators}}\tabularnewline
\hline 
 & \multicolumn{1}{c}{$\hat{\mathcal{A}}_{1}$} & \multicolumn{1}{c}{$\hat{\mathcal{A}}_{2}$} & \multicolumn{1}{c}{$\hat{\mathcal{A}}_{1}$} & \multicolumn{1}{c}{$\hat{\mathcal{A}}_{2}$}\tabularnewline
\hline 
Intercept & 2.144{{*}{*}{*}} & 5.521{{*}{*}{*}} & 1.719{{*}{*}{*}} & 5.540{{*}{*}{*}}\tabularnewline
 & (0.708) & (1.978) & (0.514) & (1.436)\tabularnewline
$\log$(Hc) & 2.525{{*}{*}{*}} & 1.336{{*}{*}{*}} & 2.221{{*}{*}{*}} & 2.735{{*}{*}{*}}\tabularnewline
 & (0.165) & (0.429) & (0.120) & (0.312)\tabularnewline
$\log$(Ck) & 0.184{{*}{*}{*}} & 3.005{{*}{*}{*}} & 0.127{{*}{*}{*}} & -0.010\tabularnewline
 & (0.021) & (0.163) & (0.016) & (0.119)\tabularnewline
$\log$(Ngd) & -0.898{{*}{*}{*}} & -9.645{{*}{*}{*}} & -1.389{{*}{*}{*}} & -0.321\tabularnewline
 & (0.306) & (0.836) & (0.222) & (0.607)\tabularnewline
\hline 
Obs(df = 820) & \multicolumn{1}{c}{828} & \multicolumn{1}{c}{828} & \multicolumn{1}{c}{828} & \multicolumn{1}{c}{828}\tabularnewline
Std. Error & \multicolumn{1}{c}{0.924} & \multicolumn{1}{c}{0.924} & \multicolumn{1}{c}{0.671} & \multicolumn{1}{c}{0.671}\tabularnewline
\hline 
\textit{Note:} & \multicolumn{2}{r}{$^{*}$p$<$0.1; $^{**}$p$<$0.05; $^{***}$p$<$0.01} &  & \tabularnewline
 & \multicolumn{2}{c}{\textit{Standard error are in parentheses}} &  & \tabularnewline
\end{tabular}
\end{table}

\section{Conclusion}

In this work, we consider a general panel data model with two-dimensional
heterogeneous coefficients and propose a novel BIGCORE procedure to
discover the assumed block structure and obtain the estimators simultaneously.
An ADMM algorithm is developed to iteratively solve the objective
function with double concave fused penalties. Simulated studies suggest
that our method is effective and show fine performance in conducting
BIGCORE analysis by correctly estimating the structure and coefficients.
A modified Bayesian information criteria is proposed to get rid of
the knotty issue that the double tuning parameters are sensitive to
the estimators. However, computational complexity is growing with
the increasing number of samples and time periods, causing the burden
of the ADMM algorithm. What's more, this work assumes that all coefficients
have identical block structure under two-dimensional heterogeneity,
which motivates us to plan to extend the BIGCORE analysis to the case
of covariate-specified block structure. All or other issues, including
the consideration of two-dimensional heterogeneous panel model with
interactive effects, are worthy to be studied in the further research.

 \bibliographystyle{econ-aea}
\phantomsection\addcontentsline{toc}{section}{\refname}\bibliography{mybibfile}

@article{RobinOkuiWang2020,
	title={Estimation of Panel Group Structure Models with Structural Breaks in Group Memberships and Coefficients},
	author={Lumsdaine, Robin L and Okui, Ryo and Wang, Wendun},
	journal={Available at SSRN 3617416},
	year={2020}
}

@article{BM2015,
   author = {Bonhomme, Stéphane and Manresa, Elena},
   title = {Grouped Patterns of Heterogeneity in Panel Data},
   journal = {Econometrica},
   volume = {83},
   number = {3},
   pages = {1147-1184},
   ISSN = {00129682},
   year = {2015},
   type = {Journal Article}
}

@article{BH2002,
  title={Unobserved ability and the return to schooling},
  author={Belzil, Christian and Hansen, J{\"o}rgen},
  journal={Econometrica},
  volume={70},
  number={5},
  pages={2075--2091},
  year={2002},
  publisher={JSTOR}
}

@article{cornett2011,
  title={Liquidity risk management and credit supply in the financial crisis},
  author={Cornett, Marcia Millon and McNutt, Jamie John and Strahan, Philip E and Tehranian, Hassan},
  journal={Journal of Financial Economics},
  volume={101},
  number={2},
  pages={297--312},
  year={2011},
  publisher={Elsevier}
}

@article{Li2011,
   author = {Li, Degui and Chen, Jia and Gao, Jiti},
   title = {Non-parametric time-varying coefficient panel data models with fixed effects},
   journal = {The Econometrics Journal},
   volume = {14},
   number = {3},
   pages = {387-408},
   year = {2011},
   type = {Journal Article}
}

@article{Pei2018,
   author = {Pei, Youquan and Huang, Tao and You, Jinhong},
   title = {Nonparametric fixed effects model for panel data with locally stationary regressors},
   journal = {Journal of Econometrics},
   volume = {202},
   number = {2},
   pages = {286-305},
   ISSN = {0304-4076},
   year = {2018},
   type = {Journal Article}
}

@article{Bai2010,
   author = {Bai, Jushan},
   title = {Common breaks in means and variances for panel data},
   journal = {Journal of Econometrics},
   volume = {157},
   number = {1},
   pages = {78-92},
   ISSN = {0304-4076},
   year = {2010},
   type = {Journal Article}
}

@article{Kim2011,
   author = {Kim, Dukpa},
   title = {Estimating a common deterministic time trend break in large panels with cross sectional dependence},
   journal = {Journal of Econometrics},
   volume = {164},
   number = {2},
   pages = {310-330},
   ISSN = {0304-4076},
   year = {2011},
   type = {Journal Article}
}

@article{QianSu2016,
   author = {Qian, Junhui and Su, Liangjun},
   title = {Shrinkage estimation of common breaks in panel data models via adaptive group fused Lasso},
   journal = {Journal of Econometrics},
   volume = {191},
   number = {1},
   pages = {86-109},
   ISSN = {0304-4076},
   year = {2016},
   type = {Journal Article}
}

@article{Li2017,
   author = {Li, Degui and Qian, Junhui and Su, Liangjun},
   title = {Panel Data Models With Interactive Fixed Effects and Multiple Structural Breaks},
   journal = {Journal of the American Statistical Association},
   volume = {111},
   number = {516},
   pages = {1804-1819},
   ISSN = {0162-1459
1537-274X},
   year = {2017},
   type = {Journal Article}
}

@article{Wooldridge2005,
   author = {Wooldridge, Jeffrey M},
   title = {Fixed-effects and related estimators for correlated random-coefficient and treatment-effect panel data models},
   journal = {Review of Economics and Statistics},
   volume = {87},
   number = {2},
   pages = {385-390},
   ISSN = {0034-6535},
   year = {2005},
   type = {Journal Article}
}

@article{MW2008,
   author = {Murtazashvili, Irina and Wooldridge, Jeffrey M},
   title = {Fixed effects instrumental variables estimation in correlated random coefficient panel data models},
   journal = {Journal of Econometrics},
   volume = {142},
   number = {1},
   pages = {539-552},
   ISSN = {0304-4076},
   year = {2008},
   type = {Journal Article}
}

@article{Boneva2015,
   author = {Boneva, Lena and Linton, Oliver and Vogt, Michael},
   title = {A semiparametric model for heterogeneous panel data with fixed effects},
   journal = {Journal of Econometrics},
   volume = {188},
   number = {2},
   pages = {327-345},
   year = {2015},
   type = {Journal Article}
}

@article{Vogt2017,
   author = {Vogt, Michael and Linton, Oliver},
   title = {Classification of non-parametric regression functions in longitudinal data models},
   journal = {Journal of the Royal Statistical Society: Series B (Statistical Methodology)},
   volume = {79},
   number = {1},
   pages = {5-27},
   year = {2017},
   type = {Journal Article}
}

@article{Pesaran2006,
   author = {Pesaran, M. Hashem},
   title = {Estimation and Inference in Large Heterogeneous Panels with a Multifactor Error Structure},
   journal = {Econometrica},
   volume = {74},
   number = {4},
   pages = {967-1012},
   year = {2006},
   type = {Journal Article}
}

@article{CPT2011,
   author = {Chudik, Alexander and Pesaran, M. Hashem and Tosetti, Elisa},
   title = {Weak and strong cross-section dependence and estimation of large panels},
   journal = {The Econometrics Journal},
   volume = {14},
   number = {1},
   pages = {C45-C90},
   year = {2011},
   type = {Journal Article}
}

@article{KHJW2017,
   author = {Karabiyik, Hande and Urbain, Jean-Pierre and Westerlund, Joakim},
   title = {CCE estimation of factor-augmented regression models with more factors than observables},
   journal = {Journal of Applied Econometrics},
   volume = {0},
   number = {0},
   year = {2017},
   type = {Journal Article}
}

@article{Bester2016,
   author = {Bester, C. Alan and Hansen, Christian B.},
   title = {Grouped effects estimators in fixed effects models},
   journal = {Journal of Econometrics},
   volume = {190},
   number = {1},
   pages = {197-208},
   ISSN = {03044076},
   year = {2016},
   type = {Journal Article}
}

@article{FGPZ2017,
   author = {Feng, Guohua and Gao, Jiti and Peng, Bin and Zhang, Xiaohui},
   title = {A varying-coefficient panel data model with fixed effects: Theory and an application to US commercial banks},
   journal = {Journal of Econometrics},
   volume = {196},
   number = {1},
   pages = {68-82},
   ISSN = {0304-4076},
   year = {2017},
   type = {Journal Article}
}

@article{KS2009,
   author = {Kasahara, Hiroyuki and Shimotsu, Katsumi},
   title = {Nonparametric identification of finite mixture models of dynamic discrete choices},
   journal = {Econometrica},
   volume = {77},
   number = {1},
   pages = {135-175},
   ISSN = {0012-9682},
   year = {2009},
   type = {Journal Article}
}

@article{BC2010,
   author = {Browning, Martin and Carro, Jesus M},
   title = {Heterogeneity in dynamic discrete choice models},
   journal = {The Econometrics Journal},
   volume = {13},
   number = {1},
   pages = {1-39},
   ISSN = {1368-4221},
   year = {2010},
   type = {Journal Article}
}

@article{Su2016,
   author = {Su, Liangjun and Shi, Zhentao and Phillips, Peter C. B.},
   title = {Identifying Latent Structures in Panel Data},
   journal = {Econometrica},
   volume = {84},
   number = {6},
   pages = {2215-2264},
   ISSN = {0012-9682},
   year = {2016},
   type = {Journal Article}
}

@article{SuJu2018,
   author = {Su, Liangjun and Ju, Gaosheng},
   title = {Identifying latent grouped patterns in panel data models with interactive fixed effects},
   journal = {Journal of Econometrics},
   volume = {206},
   number = {2},
   pages = {554-573},
   ISSN = {03044076},
   year = {2018},
   type = {Journal Article}
}

@article{Huang2018,
   author = {Huang, Wenxin and Phillips, Peter C. B. and Su, Liangjun},
   title = {Nonstationary Panel Model with Latent Group Structures and Cross-sectional Dependence},
   journal = {Working Paper},
   year = {2018},
   type = {Journal Article}
}

@article{WangSu2018,
   author = {Wang, Wuyi and Phillips, Peter C. B. and Su, Liangjun},
   title = {Homogeneity pursuit in panel data models: Theory and application},
   journal = {Journal of Applied Econometrics},
   volume = {33},
   number = {6},
   pages = {797-815},
   year = {2018},
   type = {Journal Article}
}

@article{Baltagi2016,
   author = {Baltagi, Badi H. and Feng, Qu and Kao, Chihwa},
   title = {Estimation of heterogeneous panels with structural breaks},
   journal = {Journal of Econometrics},
   volume = {191},
   number = {1},
   pages = {176-195},
   ISSN = {0304-4076},
   year = {2016},
   type = {Journal Article}
}

@article{OW2020,
	title={Heterogeneous structural breaks in panel data models},
	author={Okui, Ryo and Wang, Wendun},
	journal={Journal of Econometrics},
	year={2020},
	publisher={Elsevier}
}

@article{Smith2018,
  title={Forecasting Panel Data with Structural Breaks and Regime-Specific Grouped Heterogeneity},
  author={Smith, Simon},
  journal={USC-INET Research Paper},
  number={18-20},
  year={2018}
}

@article{Su2019,
  title={Sieve estimation of time-varying panel data models with latent structures},
  author={Su, Liangjun and Wang, Xia and Jin, Sainan},
  journal={Journal of Business \& Economic Statistics},
  volume={37},
  number={2},
  pages={334--349},
  year={2019},
  publisher={Taylor \& Francis}
}

@article{Ma2016,
  title={Estimating subgroup-specific treatment effects via concave fusion},
  author={Ma, Shujie and Huang, Jian},
  journal={arXiv preprint arXiv:1607.03717},
  year={2016}
}

@article{MaHuang2017,
   author = {Ma, Shujie and Huang, Jian},
   title = {A Concave Pairwise Fusion Approach to Subgroup Analysis},
   journal = {Journal of the American Statistical Association},
   volume = {112},
   number = {517},
   pages = {410-423},
   ISSN = {0162-1459
1537-274X},
   year = {2017},
   type = {Journal Article}
}

@article{Fan2001,
   author = {Fan, Jianqing and Li, Runze},
   title = {Variable Selection via Nonconcave Penalized Likelihood and its Oracle Properties},
   journal = {Journal of the American Statistical Association},
   volume = {96},
   number = {456},
   pages = {1348-1360},
   ISSN = {0162-1459
1537-274X},
   year = {2001},
   type = {Journal Article}
}

@article{Zhang2010,
   author = {Zhang, Cun-Hui},
   title = {NEARLY UNBIASED VARIABLE SELECTION UNDER MINIMAX CONCAVE PENALTY},
   journal = {The Annals of Statistics},
   volume = {38},
   number = {2},
   pages = {894-942},
   ISSN = {0090-5364},
   year = {2010},
   type = {Journal Article}
}

@article{tibshirani2005,
  title={Sparsity and smoothness via the fused lasso},
  author={Tibshirani, Robert and Saunders, Michael and Rosset, Saharon and Zhu, Ji and Knight, Keith},
  journal={Journal of the Royal Statistical Society: Series B (Statistical Methodology)},
  volume={67},
  number={1},
  pages={91--108},
  year={2005},
  publisher={Wiley Online Library}
}

@article{Wang2009,
   author = {Wang, Hansheng and Li, Bo and Leng, Chenlei},
   title = {Shrinkage tuning parameter selection with a diverging number of parameters},
   journal = {Journal of the Royal Statistical Society: Series B (Statistical Methodology)},
   volume = {71},
   number = {3},
   pages = {671-683},
   ISSN = {1369-7412},
   year = {2009},
   type = {Journal Article}
}

@article{chernozhukov2018,
	title={Inference for Heterogeneous Effects using Low-Rank Estimation of Factor Slopes},
	author={Chernozhukov, Victor and Hansen, Christian and Liao, Yuan and Zhu, Yinchu},
	journal={arXiv preprint arXiv:1812.08089},
	year={2018}
}

@article{LuSu2019,
	title={Uniform Inference in Linear Panel Data Models with Two-Dimensional Heterogeneity},
	author={Lu, Xun and Su, Liangjun},
	journal={Working Paper},
	year={2019}
}

@article{Neal2018,
	title={Multidimensional Slope Heterogeneity in Panel Data Models},
	author={Neal, Timothy},
	journal={UNSW Business School Research Paper},
	number={2016-15A},
	year={2018}
}

\pagebreak{}

\appendix
%dummy comment inserted by tex2lyx to ensure that this paragraph is not emptyAppendix

\global\long\def\theequation{A.\arabic{equation}}%
\setcounter{equation}{0}

\subsection*{Proof of Theorem \protect\ref{th1}}

The post bi-integrative estimator defined in \eqref{oracle} has expression
given by 
\[
\widetilde{\boldsymbol{\alpha}}=(\mathbb{X}^{\top}\mathbb{X})^{-1}\mathbb{X}^{\top}\boldsymbol{Y},
\]
and 
\[
\widetilde{\boldsymbol{\alpha}}-\boldsymbol{\alpha}^{0}=(\mathbb{X}^{\top}\mathbb{X})^{-1}\mathbb{X}^{\top}\boldsymbol{\epsilon},
\]
Hence 
\[
\left\Vert \widetilde{\boldsymbol{\alpha}}-\boldsymbol{\alpha}^{0}\right\Vert \leq\|(\mathbb{X}^{\top}\mathbb{X})^{-1}\|\|\mathbb{X}^{\top}\boldsymbol{\epsilon}\|.
\]
Condition (C2) (i) implies that 
\[
\|(\mathbb{X}^{\top}\mathbb{X})^{-1}\|\leq c_{2}^{-1}(\mathcal{A}_{\min})^{-1}.
\]
Moreover, $\boldsymbol{X}\Pi\boldsymbol{W}\boldsymbol{Q}=\left[x_{it}^{\top}\boldsymbol{1}_{\left\{ (i,t)\in\mathcal{A}_{l}\right\} }\right]_{i=1,t=1,l=1}^{N,T,L}$,
$\sum_{i=1}^{N}\sum_{t=1}^{T}x_{it}^{2}\boldsymbol{1}_{\left\{ (i,t)\in\mathcal{A}_{l}\right\} }=\left|\mathcal{A}_{l}\right|,$
\[
\left\Vert (\boldsymbol{X}\Pi\boldsymbol{W}\boldsymbol{Q})^{\top}\boldsymbol{\epsilon}\right\Vert _{\infty}=\sup_{p,l}\Big|\sum_{i=1}^{N}\sum_{t=1}^{T}x_{itp}\epsilon_{it}\boldsymbol{1}_{\left\{ (i,t)\in\mathcal{A}_{l}\right\} }\Big|,
\]
\[
P\left(\|\mathbb{X}^{\top}\boldsymbol{\epsilon}\|_{\infty}>C\sqrt{NT\log(NT)}\right)=P\left(\|(\boldsymbol{X}\Pi\boldsymbol{W}\boldsymbol{Q})^{\top}\boldsymbol{\epsilon}\|_{\infty}>C\sqrt{NT\log(NT)}\right).
\]
For some constant $0<C<\infty$, by union bound, we have 
\begin{eqnarray*}
 &  & P\left(\|(\boldsymbol{X}\Pi\boldsymbol{W}\boldsymbol{Q})^{\top}\boldsymbol{\epsilon}\|_{\infty}>C\sqrt{NT\log(NT)}\right)\\
 & \leq & \sum_{p=1,l=1}^{P,L}P\left(\Big|\sum_{n=1}^{N}\sum_{t=1}^{T}x_{itp}\boldsymbol{1}_{\{(i,t)\in\mathcal{A}_{l}\}}\epsilon_{it}\Big|>C\sqrt{NT\log(NT)}\right)\\
 & \leq & \sum_{p=1,l=1}^{P,L}P\left(\Big|\sum_{i=1}^{N}\sum_{t=1}^{T}x_{itp}\boldsymbol{1}_{\{(i,t)\in\mathcal{A}_{l}\}}\epsilon_{it}\Big|>C\sqrt{|\mathcal{A}_{l}|\log(NT)}\right)\\
 & \leq & 2\sum_{p=1,l=1}^{P,L}\exp\left(-c_{1}C^{2}\log(NT)\right)\\
 & = & 2PL(NT)^{-c_{1}C^{2}}.
\end{eqnarray*}

Since $\|\mathbb{X}^{\top}\boldsymbol{\epsilon}\|\leq\sqrt{LP}\|\mathbb{X}^{\top}\boldsymbol{\epsilon}\|_{\infty},$
then 
\[
P\left(\|\mathbb{X}^{\top}\boldsymbol{\epsilon}\|>C\sqrt{PL}\sqrt{NT\log(NT)}\right)\leq2PL(NT)^{-c_{1}C^{2}}.
\]
Therefore, we have with probability at least $1-2PL(NT)^{-c_{1}C^{2}}$,
\[
\left\Vert \boldsymbol{\widetilde{\alpha}}-\boldsymbol{\alpha}^{0}\right\Vert \leq Cc_{2}^{-1}\sqrt{PL}\sqrt{NT\log(NT)}\left|\mathcal{A}_{\min}\right|^{-1}.
\]
Therefore, we need $\left|\mathcal{A}_{\min}\right|\gg(LP)^{1/2}(NT)^{3/4}$,
and the result of consistency in Theorem \ref{th1} (i) is proved
by letting $C=c_{1}^{-1/2}$, 
\[
\begin{aligned}\left\Vert \tilde{\beta}-\boldsymbol{\beta}^{0}\right\Vert ^{2} & =\sum_{l=1}^{L}\sum_{(i,t)\in\mathcal{A}_{l}}\left\Vert \widetilde{\boldsymbol{\alpha}}-\boldsymbol{\alpha}^{0}\right\Vert ^{2}\leq\left|\mathcal{A}_{\max}\right|\sum_{l=1}^{L}\left\Vert \widetilde{\boldsymbol{\alpha}}-\boldsymbol{\alpha}^{0}\right\Vert ^{2}\\
 & =\left|\mathcal{A}_{\max}\right|\left\Vert \widetilde{\boldsymbol{\alpha}}-\boldsymbol{\alpha}^{0}\right\Vert ^{2}\leq\left|\mathcal{A}_{\max}\right|\Delta_{n}^{2},
\end{aligned}
\]
\[
\sup_{i,t}\left\Vert \tilde{\beta}_{it}-\boldsymbol{\beta}_{it}^{0}\right\Vert =\sup_{l}\left\Vert \widetilde{\boldsymbol{\alpha}}-\boldsymbol{\alpha}_{l}^{0}\right\Vert \leq\left\Vert \widetilde{\boldsymbol{\alpha}}-\boldsymbol{\alpha}^{0}\right\Vert \leq\Delta_{n}.
\]
For any $\boldsymbol{d}_{n}\in\mathbb{R}^{LP}$ with $\|\boldsymbol{d}_{n}\|=1$
and $\mathbb{X}=(\mathbb{X}_{1\cdot},\cdots,\mathbb{X}_{(NT)\cdot})^{\top}$
\[
\boldsymbol{d}_{n}^{\top}(\widetilde{\boldsymbol{\alpha}}-\boldsymbol{\alpha}^{0})=\sum_{i=1}^{N}\sum_{t=1}^{T}\boldsymbol{d}_{n}^{\top}(\mathbb{X}^{\top}\mathbb{X})^{-1}\mathbb{X}_{(it)\cdot}\epsilon_{it}.
\]
Hence, $E\{\boldsymbol{d}_{n}^{\top}(\widetilde{\boldsymbol{\alpha}}-\boldsymbol{\alpha}^{0})\}=0$
\[
var\{\boldsymbol{d}_{n}^{\top}(\widetilde{\boldsymbol{\alpha}}-\boldsymbol{\alpha}^{0})\}=s_{n}(\boldsymbol{d}_{n})^{2}=\sigma^{2}\boldsymbol{d}_{n}^{\top}(\mathbb{X}^{\top}\mathbb{X})^{-1}\boldsymbol{d}_{n}\ge\sigma^{2}c_{3}^{-1}(NT)^{-1}.
\]
Moreover, for any $\epsilon>0$, 
\begin{eqnarray*}
 &  & \sum_{i=1}^{N}\sum_{t=1}^{T}E\Big\{\left(\boldsymbol{d}_{n}^{\top}(\mathbb{X}^{\top}\mathbb{X})^{-1}\mathbb{X}_{(it)\cdot}\epsilon_{it}\right)^{2}\boldsymbol{1}_{\{|\boldsymbol{d}_{n}^{\top}(\mathbb{X}^{\top}\mathbb{X})^{-1}\mathbb{X}_{(it)\cdot}\epsilon_{it}|>\epsilon s_{n}(\boldsymbol{d}_{n})\}}\Big\}\\
 & \leq & NT\Big\{ E\left(\boldsymbol{d}_{n}^{\top}(\mathbb{X}^{\top}\mathbb{X})^{-1}\mathbb{X}_{(it)\cdot}\epsilon_{it}\right)^{4}\Big\}^{1/2}\Big\{ P\{|\boldsymbol{d}_{n}^{\top}(\mathbb{X}^{\top}\mathbb{X})^{-1}\mathbb{X}_{(it)\cdot}\epsilon_{it}|>\epsilon s_{n}(\boldsymbol{d}_{n})\}\Big\}^{1/2}.
\end{eqnarray*}
Since $E(\epsilon_{it}^{4})\leq c$ by Condition (C1) and $\sup_{it}\|x_{it}\|\leq c_{4}\sqrt{P}$
by Condition (C3), then 
\[
\{E(\boldsymbol{d}_{n}^{\top}(\mathbb{X}^{\top}\mathbb{X})^{-1}\mathbb{X}_{(it)\cdot}\epsilon_{it})^{4}\}^{1/2}\leq\|\boldsymbol{d}_{n}^{\top}(\mathbb{X}^{\top}\mathbb{X})^{-1}\|^{2}\|\mathbb{X}_{it}\|^{2}[E(\epsilon_{it}^{4})]^{1/2}\leq c^{\prime}c_{2}^{-2}c_{4}^{2}|\mathcal{A}_{min}|^{-2}LP
\]
, for some constant $0<c^{\prime}<\infty$.

Similarly, $E\left(\boldsymbol{d}_{n}^{\top}(\mathbb{X}^{\top}\mathbb{X})^{-1}\mathbb{X}_{(it)\cdot}\epsilon_{it}\right)^{2}\leq c^{\prime\prime}c_{2}^{-2}c_{4}^{2}|\mathcal{A}_{min}|^{-2}LP$,
for some constant $0<c^{\prime\prime}<\infty$. Thus 
\begin{eqnarray*}
P\Big\{|\boldsymbol{d}_{n}^{\top}(\mathbb{X}^{\top}\mathbb{X})^{-1}\mathbb{X}_{i}^{\top}\epsilon_{i}|>\epsilon s_{n}(\boldsymbol{d}_{n})\Big\} & \leq & E\left(\boldsymbol{d}_{n}^{\top}(\mathbb{X}^{\top}\mathbb{X})^{-1}\mathbb{X}_{it}\epsilon_{it}\right)^{2}/\{\epsilon^{2}s_{n}(\boldsymbol{d}_{n})^{2}\}\\
 & \leq & c^{\prime\prime}c_{2}^{-2}c_{4}^{2}c_{3}^{-1}\sigma^{-2}\epsilon^{-2}\left|\mathcal{A}_{\min}\right|^{-2}LPNT,
\end{eqnarray*}
for some constant $c^{\prime}$. Therefore, by the above results,
we have 
\begin{eqnarray*}
 &  & s_{n}(\boldsymbol{d}_{n})^{-2}\sum_{i=1}^{N}\sum_{t=1}^{T}E\Big\{\left(\boldsymbol{d}^{\top}(\mathbb{X}^{\top}\mathbb{X})^{-1}\mathbb{X}_{(it)\cdot}\epsilon_{it}\right)^{2}\boldsymbol{1}_{\{|\boldsymbol{d}^{\top}(\mathbb{X}^{\top}\mathbb{X})^{-1}\mathbb{X}_{(it)\cdot}\epsilon_{it}|>\epsilon s_{n}(\boldsymbol{d})\}}\Big\}\\
 & \leq & O\left\{ (NT)^{3}\left|\mathcal{A}_{\min}\right|^{-4}(LP)^{2}\right\} =o(1).
\end{eqnarray*}
The last equality follows from the assumption that $\left|\mathcal{A}_{\min}\right|\gg(LP)^{1/2}(NT)^{3/4}$.
Then, the result in this Theorem follows from Lindeberg-Feller Central
Limit Theorem.

\subsection*{Proof of Theorem 2}

\begin{figure}[!ht]
\centering \includegraphics[width=6cm,height=6cm]{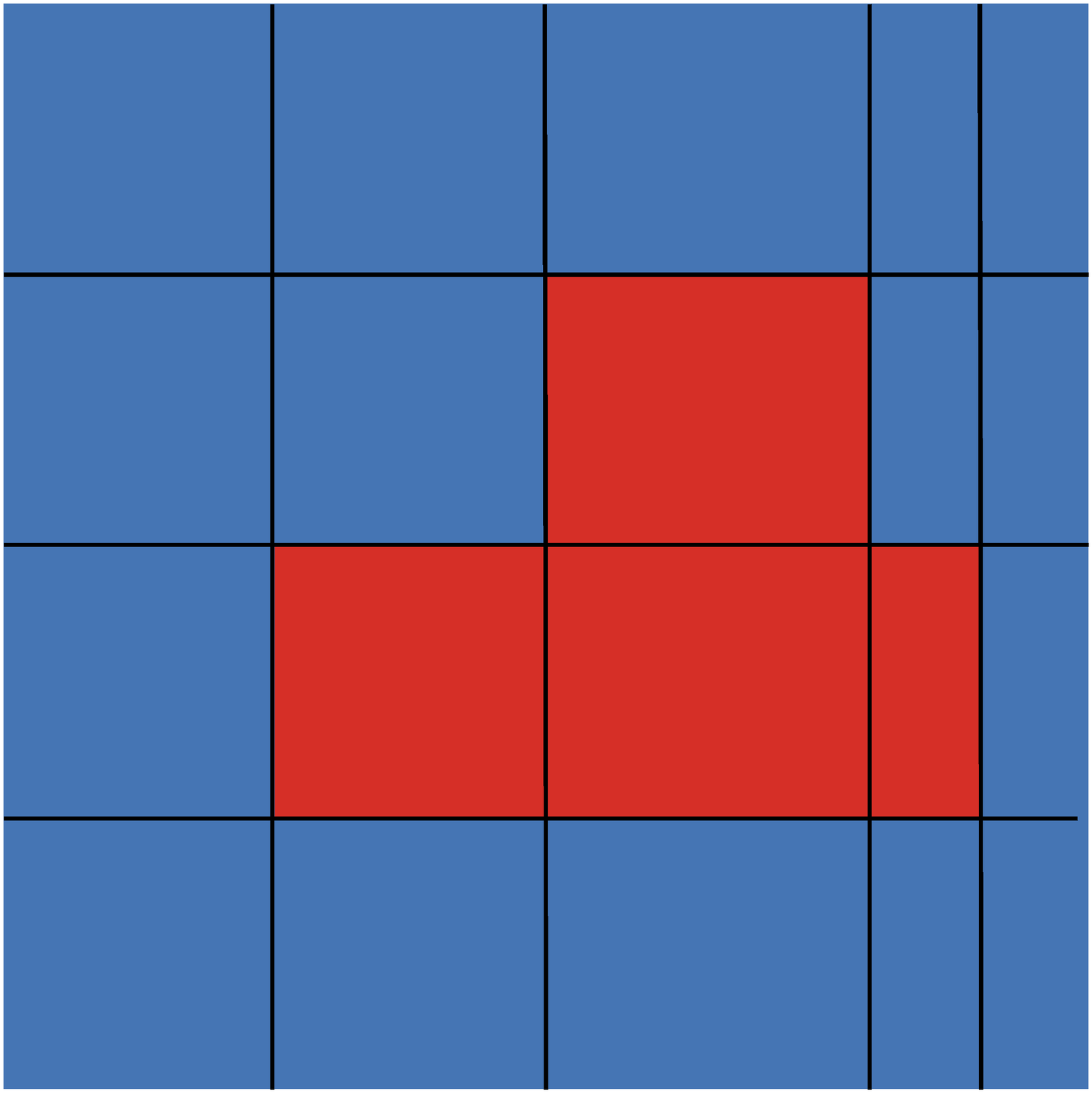} \caption{Partitioned block structure on regression coefficients.}
\label{Newsplit} 
\end{figure}

In order to prove of Theorem 2, we firstly split the true block structure
of coefficients as shown in Figure \ref{Newsplit}, which is different
from the pattern of group-cohort or cohort-group in Figure \ref{fig:Partitions}.
In this pattern, we firstly split the sample according to individuals
as the true group structure and then split the sample across the temporal
dimension. The difference lies how the cohorts are split. If there
exist structural breaks for any individuals at given time period,
we split the cohorts for all individuals, instead of splitting the
cohorts under given groups. As we can see, the number of split blocks
is larger. Let, 
\begin{equation}
L(\boldsymbol{\beta})=\frac{1}{2}\|\boldsymbol{Y}-\boldsymbol{X}\boldsymbol{\beta}\|^{2},\qquad\mathcal{P}(\boldsymbol{\beta})=\sum_{t=1}^{T}\sum\limits _{i<j}\mathcal{P}_{\lambda}(\Vert\boldsymbol{\beta}_{it}-\boldsymbol{\beta}_{jt}\Vert)+\sum_{i=1}^{N}\sum\limits _{t<t^{\prime}}\mathcal{P}_{\gamma}\left(\left\Vert \boldsymbol{\beta}_{it}-\boldsymbol{\beta}_{it^{\prime}}\right\Vert \right)
\end{equation}
\begin{equation}
L^{\mathcal{A}}(\alpha)=\frac{1}{2}\|\boldsymbol{Y}-\mathbb{X}\alpha\|^{2},\mathcal{P}^{\mathcal{A}}(\alpha)=\sum_{l<l^{\prime}}\left\{ \lambda\sum_{c=1}^{C}(|\mathcal{A}_{lc}||\mathcal{A}_{l^{\prime}c}|)\rho_{\lambda}(\|\alpha_{l}-\alpha_{l^{\prime}}\|)+\gamma\sum_{k=1}^{K}(|\mathcal{A}_{lk}||\mathcal{A}_{l^{\prime}k}|)\rho_{\gamma}(\|\alpha_{l}-\alpha_{l^{\prime}}\|)\right\} ,
\end{equation}
where $C$ denotes the number of groups split as Figure \ref{Newsplit},
$K$ denotes the number of cohorts split as Figure \ref{Newsplit},
$|\mathcal{A}_{lc}|$ denotes the cardinality of observations that
belong to both the $c$th cohorts and true $l$ blocks , $|\mathcal{A}_{lk}|$
denotes the cardinality of observations that belong to both the $k$th
cohorts and true $l$ blocks. 

Then, 
\[
Q(\boldsymbol{\beta})=L(\boldsymbol{\beta})+\mathcal{P}(\boldsymbol{\beta}),\qquad Q^{\mathcal{A}}(\alpha)=L^{\mathcal{A}}(\alpha)+\mathcal{P}^{\mathcal{A}}(\alpha).
\]
Let $\mathcal{M}_{\mathcal{A}}$ denote the set of $\mathbb{R}^{NTP}$
coefficients that have block structure and $T:\mathcal{M}_{\mathcal{A}}\longmapsto\mathbb{R}^{LP}$
be the mapping that $T(\boldsymbol{\beta})$ is the $LP\times1$ vector
consisting of $L$ vectors with dimension $P$ and its $l$th vector
is the values for the $l$th block. Let $T^{*}:\mathbb{R}^{NTP}\longmapsto\mathbb{R}^{LP}$
be the mapping that $T^{*}(\boldsymbol{\beta})=\{|\mathcal{A}_{l}|^{-1}\sum_{(i,t)\in\mathcal{A}_{l}}\beta_{it}^{\top},l=1,\cdots,L\}$.
In the following, we consider the neighborhood of $\boldsymbol{\beta^{0}}$,
$\Theta_{1}=\{\boldsymbol{\beta}\in\mathbb{R}^{NTP}:\sup_{it}\left\Vert \beta_{it}-\beta_{it}^{0}\right\Vert \leq\Delta_{n}\}$.

It is obvious that when $\boldsymbol{\beta}\in\mathcal{M}_{\mathcal{A}}$,
$T(\boldsymbol{\beta})=T^{*}(\boldsymbol{\beta})$. By calculation,
$\mathcal{P}(\boldsymbol{\beta})=\mathcal{P}^{\mathcal{A}}(T(\boldsymbol{\beta}))$
for every $\boldsymbol{\beta}\in\mathcal{M}_{\mathcal{A}}$ and $\mathcal{P}(T^{-1}(\alpha))=\mathcal{P}^{\mathcal{A}}(\alpha)$
according to split structure as shown in Figure \ref{Newsplit}. Therefore,
we conclude that

\begin{equation}
Q(\boldsymbol{\beta})=Q^{\mathcal{A}}(T(\boldsymbol{\beta})),\qquad Q^{\mathcal{A}}(\alpha)=Q(T^{-1}(\alpha)).\label{A5}
\end{equation}

We will verify the Theorem 2 by two steps. 

\subsubsection*{The first step:}

For any $\boldsymbol{\beta}\in\mathbb{R}^{NTP}$, Let $T^{*}(\boldsymbol{\beta})=\alpha$
and $\boldsymbol{\beta}^{*}=T^{-1}(T^{*}(\boldsymbol{\beta}))=\left(\boldsymbol{\beta}_{11}^{*\top},\cdots,\boldsymbol{\beta}_{1T}^{*\top},\cdots,\boldsymbol{\beta}_{N1}^{*\top},\cdots,\boldsymbol{\beta}_{NT}^{*\top}\right)^{*\top}$.
Since 
\[
\left\Vert \alpha_{l}-\alpha_{l^{\prime}}\right\Vert \geq\left\Vert \alpha_{l}^{0}-\alpha_{l^{\prime}}^{0}\right\Vert -2\sup_{l}\left\Vert \alpha_{l}-\alpha_{l}^{0}\right\Vert 
\]
and 
\[
\begin{aligned}\sup_{l}\left\Vert \alpha_{l}-\alpha_{l}^{0}\right\Vert ^{2} & =\sup_{l}\left\Vert \left|\mathcal{A}_{l}\right|^{-1}\sum_{(i,t)\in\mathcal{A}_{l}}\beta_{it}-\alpha_{l}^{0}\right\Vert ^{2}=\sup_{l}\left\Vert \left|\mathcal{A}_{l}\right|^{-1}\sum_{(i,t)\in\mathcal{A}_{l}}\left(\beta_{it}-\beta_{it}^{0}\right)\right\Vert ^{2}\\
 & =\sup_{l}\left|\mathcal{A}_{l}\right|^{-2}\left\Vert \sum_{(i,t)\in\mathcal{A}_{l}}\left(\beta_{it}-\beta_{it}^{0}\right)\right\Vert ^{2}\leq\sup_{l}\left|\mathcal{A}_{l}\right|^{-1}\sum_{(i,t)\in\mathcal{A}_{l}}\left\Vert \left(\beta_{it}-\beta_{it}^{0}\right)\right\Vert ^{2}\\
 & \leq\sup_{(i,t)}\left\Vert \beta_{it}-\beta_{it}^{0}\right\Vert ^{2}\leq\Delta_{n}^{2},
\end{aligned}
\]
then, for all $l$ and $l^{\prime}$ 
\[
\left\Vert \alpha_{l}-\alpha_{l^{\prime}}\right\Vert \geq\left\Vert \alpha_{l}^{0}-\alpha_{l^{\prime}}^{0}\right\Vert -2\sup_{l}\left\Vert \alpha_{l}-\alpha_{l}^{0}\right\Vert \geq b-2\Delta_{n}.
\]
By the assumption that $b>a\lambda\gg\Delta_{n}$ and $b>a^{\prime}\gamma\gg\Delta_{n}$,
we can get that $b-2\Delta_{n}>a\lambda$, $b-2\Delta_{n}>a^{\prime}\gamma$.
As a result, we have $\mathcal{P}^{\mathcal{A}}(T^{*}(\boldsymbol{\beta}))=\mathcal{C}$,
where $\mathcal{C}$ is a constant,under the assumption. Furthermore,
$Q^{\mathcal{A}}(T^{*}(\boldsymbol{\beta}))=L^{\mathcal{A}}(T^{*}(\boldsymbol{\beta}))+\mathcal{C}$.

Because that $\widetilde{\alpha}$ is the unique global minimizer
of $L^{\mathcal{A}}(\alpha)$, $L^{\mathcal{A}}(T^{*}(\boldsymbol{\beta}))>L^{\mathcal{A}}(\widetilde{\alpha})$.
Furthermore, $Q^{\mathcal{A}}(T^{*}(\boldsymbol{\beta}))>Q^{\mathcal{A}}(\widetilde{\alpha})$.
Implied by (\ref{A5}) that $Q^{\mathcal{A}}(\widetilde{\alpha})=Q(\widetilde{\boldsymbol{\beta}})$
and $Q^{\mathcal{A}}(T^{*}(\boldsymbol{\beta}))=Q(T^{-1}(T^{*}(\boldsymbol{\beta})))=Q(\boldsymbol{\beta}^{*})$.
Finally, $Q(\boldsymbol{\beta}^{*})>Q(\widetilde{\boldsymbol{\beta}})$
for all $\boldsymbol{\beta}^{*}\neq\widetilde{\boldsymbol{\beta}}$.

\subsubsection*{The second step:}

For a positive sequence $t$, let $\Theta_{2}=\{\beta_{it}:\sup_{it}\left\Vert \beta_{it}-\tilde{\beta}_{it}\right\Vert \leq t\}$.
For $\boldsymbol{\beta}\in\Theta_{1}\cap\Theta_{2}$, By the Taylor
expansion around $\boldsymbol{\beta}_{it}$, we can divide the difference
of objective function into three parts: 
\begin{equation}
Q(\boldsymbol{\beta})-Q(\boldsymbol{\beta}^{*})=\Gamma_{1}+\Gamma_{2}+\Gamma_{3}
\end{equation}
where 
\[
\Gamma_{1}=-(\boldsymbol{Y}-\boldsymbol{X}\boldsymbol{\beta}^{m})^{\top}\boldsymbol{X}(\boldsymbol{\beta}-\boldsymbol{\beta}^{*}),
\]
\[
\Gamma_{2}=\sum_{t=1}^{T}\sum_{i=1}^{N}\frac{\partial\mathcal{P}(\boldsymbol{\beta}^{m})}{\partial\boldsymbol{\beta}_{it}}(\boldsymbol{\beta}_{it}-\boldsymbol{\beta}_{it}^{*}),
\]
\[
\Gamma_{3}=\sum_{i=1}^{N}\sum_{t=1}^{T}\frac{\partial\mathcal{P}(\boldsymbol{\beta}^{m})}{\partial\boldsymbol{\beta}_{t}}(\boldsymbol{\beta}_{it}-\boldsymbol{\beta}_{it}^{*}),
\]
and $\boldsymbol{\beta}_{i}=(\beta_{i1}^{\top},\beta_{i2}^{\top},\cdots,\beta_{iT}^{\top})^{\top}$,
$\boldsymbol{\beta}_{t}=(\beta_{1t}^{\top},\beta_{2t}^{\top},\cdots,\beta_{Nt}^{\top})^{\top}$,\quad{}$\boldsymbol{\beta}^{m}=\theta\boldsymbol{\beta}+(1-\theta)\boldsymbol{\beta}^{*}$.
\begin{eqnarray*}
\Gamma_{2} & = & \sum_{t=1}^{T}\sum_{i<j}\mathcal{P}_{\lambda}\left(\left\Vert \boldsymbol{\beta}_{it}^{m}-\boldsymbol{\beta}_{jt}^{m}\right\Vert \right)-\sum_{t=1}^{T}\sum_{i<j}\mathcal{P}_{\lambda}\left(\left\Vert \boldsymbol{\beta}_{it}^{*}-\boldsymbol{\beta}_{jt}^{*}\right\Vert \right)\\
 & = & \lambda\sum_{t=1}^{T}\sum_{i<j}\left[\rho\left(\left\Vert \boldsymbol{\beta}_{it}^{m}-\boldsymbol{\beta}_{jt}^{m}\right\Vert \right)-\rho\left(\left\Vert \boldsymbol{\beta}_{it}^{*}-\boldsymbol{\beta}_{jt}^{*}\right\Vert \right)\right]\\
 & = & \lambda\sum_{t=1}^{T}\sum_{i<j}\left\{ \left[\frac{\partial\rho\left(\left\Vert \boldsymbol{\beta}_{it}^{m}-\boldsymbol{\beta}_{jt}^{m}\right\Vert \right)}{\partial\boldsymbol{\beta}_{it}}\right]^{\prime}\left(\boldsymbol{\beta}_{it}-\boldsymbol{\beta}_{it}^{*}\right)+\left[\frac{\partial\rho\left(\left\Vert \boldsymbol{\beta}_{it}^{m}-\boldsymbol{\beta}_{jt}^{m}\right\Vert \right)}{\partial\boldsymbol{\beta}_{jt}}\right]^{\prime}\left(\boldsymbol{\beta}_{jt}-\boldsymbol{\beta}_{jt}^{*}\right)\right\} .
\end{eqnarray*}
Since 
\begin{eqnarray*}
\frac{\partial\rho\left(\left\Vert \boldsymbol{\beta}_{it}^{m}-\boldsymbol{\beta}_{jt}^{m}\right\Vert \right)}{\partial\boldsymbol{\beta}_{it}} & = & \rho^{\prime}\left(\left\Vert \boldsymbol{\beta}_{it}^{m}-\boldsymbol{\beta}_{jt}^{m}\right\Vert \right)\frac{\partial\left\Vert \boldsymbol{\beta}_{it}^{m}-\boldsymbol{\beta}_{jt}^{m}\right\Vert }{\partial\boldsymbol{\beta}_{it}}\\
 & = & \rho^{\prime}\left(\left\Vert \boldsymbol{\beta}_{it}^{m}-\boldsymbol{\beta}_{jt}^{m}\right\Vert \right)\left\Vert \boldsymbol{\beta}_{it}^{m}-\boldsymbol{\beta}_{jt}^{m}\right\Vert ^{-1}\frac{\partial\left[\left\Vert \boldsymbol{\beta}_{it}^{m}-\boldsymbol{\beta}_{jt}^{m}\right\Vert ^{2}\right]}{\partial\boldsymbol{\beta}_{it}}\\
 & = & \rho^{\prime}\left(\left\Vert \boldsymbol{\beta}_{it}^{m}-\boldsymbol{\beta}_{jt}^{m}\right\Vert \right)\left\Vert \boldsymbol{\beta}_{it}^{m}-\boldsymbol{\beta}_{jt}^{m}\right\Vert ^{-1}\left(\boldsymbol{\beta}_{it}^{m}-\boldsymbol{\beta}_{jt}^{m}\right)^{\top}
\end{eqnarray*}
So we have 
\[
\Gamma_{2}=\sum_{t=1}^{T}\sum_{i<j}\left\{ \rho^{\prime}\left(\left\Vert \boldsymbol{\beta}_{it}^{m}-\boldsymbol{\beta}_{jt}^{m}\right\Vert \right)\left\Vert \boldsymbol{\beta}_{it}^{m}-\boldsymbol{\beta}_{jt}^{m}\right\Vert ^{-1}\left(\boldsymbol{\beta}_{it}^{m}-\boldsymbol{\beta}_{jt}^{m}\right)^{\top}\left[\left(\boldsymbol{\beta}_{it}-\boldsymbol{\beta}_{it}^{*}\right)+\left(\boldsymbol{\beta}_{jt}-\boldsymbol{\beta}_{jt}^{*}\right)\right]\right\} 
\]
When $(i,t),(j,t)\in\mathcal{\mathcal{A}}_{l},\boldsymbol{\beta}_{it}^{*}=\boldsymbol{\beta}_{jt}^{*}$
and $\boldsymbol{\beta}_{it}^{m}-\boldsymbol{\beta}_{jt}^{m}=\theta\left(\boldsymbol{\beta}_{it}-\boldsymbol{\beta}_{jt}\right)$
, 
\[
\begin{aligned}\Gamma_{2}= & \lambda\sum_{t=1}^{T}\sum_{l=1}^{L}\sum_{\substack{(i,t),(j,t)\in\mathcal{A}_{l}\\
i<j
}
}\rho^{\prime}\left(\left\Vert \beta_{it}^{m}-\beta_{jt}^{m}\right\Vert \right)\left\Vert \beta_{it}^{m}-\beta_{jt}^{m}\right\Vert ^{-1}\left(\beta_{it}^{m}-\beta_{jt}^{m}\right)^{\mathrm{T}}\left(\beta_{it}-\beta_{jt}\right)\\
+ & \lambda\sum_{t=1}^{T}\sum_{l<l^{\prime}}\sum_{\substack{(i,t)\in\mathcal{A}_{l}\\
(j,t)\in\mathcal{A}_{l^{\prime}}
}
}\rho^{\prime}\left(\left\Vert \beta_{it}^{m}-\beta_{jt}^{m}\right\Vert \right)\left\Vert \beta_{it}^{m}-\beta_{jt}^{m}\right\Vert ^{-1}\left(\beta_{it}^{m}-\beta_{jt}^{m}\right)^{\top}\left\{ \left(\beta_{it}-\beta_{it}^{*}\right)-\left(\beta_{jt^{\prime}}-\beta_{jt}^{*}\right)\right\} .
\end{aligned}
\]
Moreover, 
\[
\sup_{(i,t)}\left\Vert \beta_{it}^{*}-\beta_{it}^{0}\right\Vert ^{2}=\sup_{l}\left\Vert \alpha_{l}-\alpha_{l}^{0}\right\Vert ^{2}\leq\Delta_{n}^{2},
\]
and 
\[
\sup_{(i,t)}\left\Vert \beta_{it}^{m}-\beta_{it}^{0}\right\Vert \leq\theta\sup_{(i,t)}\|\beta_{it}-\beta_{it}^{0}\|+(1-\theta)\sup_{(i,t)}\|\beta_{it}^{*}-\beta_{it}^{0}\|\leq\theta\Delta_{n}+(1-\theta)\Delta_{n}=\Delta_{n}.
\]
Hence, for $l\neq l^{\prime},(i,t)\in\mathcal{A}_{l},(j,t)\in\mathcal{A}_{l^{\prime}}$,

\[
\left\Vert \beta_{it}^{m}-\beta_{jt}^{m}\right\Vert \geq\min_{(i,t)\in\mathcal{A}_{l},(j,t)\in\mathcal{A}_{l^{\prime}}}\left\Vert \beta_{it}^{0}-\beta_{jt}^{0}\right\Vert -2\max_{(i,t)}\left\Vert \beta_{it}^{m}-\beta_{it}^{0}\right\Vert \geq b-2\Delta_{n}>a\lambda,
\]
and thus $\rho^{\prime}\left(\left\Vert \beta_{it}^{m}-\beta_{jt}^{m}\right\Vert \right)=0$.
Therefore, 
\[
\begin{aligned}\Gamma_{2}= & \lambda\sum_{t=1}^{T}\sum_{l=1}^{L}\sum_{\substack{(i,t),(j,t)\in\mathcal{A}_{l}\\
i<j
}
}\rho^{\prime}\left(\left\Vert \beta_{it}^{m}-\beta_{jt}^{m}\right\Vert \right)\left\Vert \beta_{it}^{m}-\beta_{jt}^{m}\right\Vert ^{-1}\left(\beta_{it}^{m}-\beta_{jt}^{m}\right)^{\mathrm{T}}\left(\beta_{it}-\beta_{jt}\right)\\
= & \lambda\sum_{t=1}^{T}\sum_{l=1}^{L}\sum_{\substack{(i,t),(j,t)\in\mathcal{A}_{l}\\
i<j
}
}\rho^{\prime}\left(\left\Vert \beta_{it}^{m}-\beta_{jt}^{m}\right\Vert \right)\left\Vert \beta_{it}-\beta_{jt}\right\Vert .
\end{aligned}
\]
With that $\sup_{(i,t)}\|\beta_{it}^{*}-\widetilde{\beta}_{it}\|\leq\sup_{(i,t)}\|\beta_{it}-\widetilde{\beta}_{it}\|$,
we can obtain 
\[
\begin{aligned}\sup_{(i,t)}\left\Vert \beta_{it}^{m}-\beta_{jt^{\prime}}^{m}\right\Vert  & \leq2\sup_{(i,t)}\left\Vert \beta_{it}^{m}-\beta_{it}^{*}\right\Vert \leq2\sup_{i}\left\Vert \beta_{it}-\beta_{it}^{*}\right\Vert \\
 & \leq2\left\{ \sup_{(i,t)}\|\beta_{it}-\widetilde{\beta}_{it}\|+\sup_{(i,t)}\|\beta_{it}^{*}-\widetilde{\beta}_{it}\|\right\} \leq4\sup_{(i,t)}\left\Vert \beta_{it}-\widetilde{\beta}_{it}\right\Vert \leq4t.
\end{aligned}
\]
Therefore, $\rho^{\prime}\left(\left\Vert \beta_{it}^{m}-\beta_{jt}^{m}\right\Vert \right)\geq\rho^{\prime}\left(4t\right)$,
and by concavity of $\rho(\cdot)$, 
\begin{equation}
\Gamma_{2}\geq\lambda\sum_{t=1}^{T}\sum_{l=1}^{L}\sum_{\substack{(i,t),(j,t)\in\mathcal{A}_{l}\\
i<j
}
}\rho^{\prime}\left(4t\right)\left\Vert \beta_{it}-\beta_{jt}\right\Vert 
\end{equation}
Similarly, 
\begin{equation}
\Gamma_{3}\geq\gamma\sum_{i=1}^{N}\sum_{l=1}^{L}\sum_{\substack{(i,t),(i,t^{\prime})\in\mathcal{A}_{l}\\
t<t^{\prime}
}
}\rho^{\prime}\left(4t\right)\left\Vert \beta_{it}-\beta_{it}\right\Vert .
\end{equation}
Let, 
\[
\boldsymbol{Q}=\left(\boldsymbol{Q}_{1}^{\top},\ldots,\boldsymbol{Q}_{NT}^{\top}\right)^{\top}=\left[\left(\boldsymbol{Y}-\mathbf{X}\boldsymbol{\beta}^{m}\right)^{\top}\boldsymbol{X}\right]^{\top},
\]
then 
\[
\begin{aligned}\Gamma_{1} & =-\boldsymbol{Q}^{\top}\left(\boldsymbol{\beta}-\boldsymbol{\beta}^{*}\right)=-\sum_{l=1}^{L}\sum_{\left\{ (i,t),(j,t^{\prime})\in\mathcal{A}_{l}\right\} }\frac{\boldsymbol{Q}_{it}^{\top}\left(\beta_{it}-\beta_{jt^{\prime}}\right)}{\left|\mathcal{A}_{l}\right|}\\
 & =-\sum_{l=1}^{L}\sum_{\left\{ (i,t),(j,t^{\prime})\in\mathcal{A}_{l}\right\} }\frac{\boldsymbol{Q}_{it}^{\top}\left(\beta_{it}-\beta_{jt^{\prime}}\right)}{2\left|\mathcal{A}_{l}\right|}-\sum_{l=1}^{L}\sum_{\left\{ (i,t),(j,t^{\prime})\in\mathcal{A}_{l}\right\} }\frac{\boldsymbol{Q}_{it}^{\top}\left(\beta_{it}-\beta_{jt^{\prime}}\right)}{2\left|\mathcal{A}_{l}\right|}\\
 & =-\sum_{l=1}^{L}\sum_{\left\{ (i,t),(j,t^{\prime})\in\mathcal{A}_{l}\right\} }\frac{\left(\boldsymbol{Q}_{jt^{\prime}}-\boldsymbol{Q}_{it}\right)^{\top}\left(\beta_{jt^{\prime}}-\beta_{it}\right)}{2\left|\mathcal{A}_{l}\right|}\\
 & =-\sum_{l=1}^{L}\sum_{\substack{(i,t),(j,t^{\prime})\in\mathcal{A}_{l}\\
i<j,t<t^{\prime}
}
}\frac{\left(\mathbf{Q}_{jt^{\prime}}-\boldsymbol{Q}_{it}\right)^{\top}\left(\beta_{jt^{\prime}}-\beta_{it}\right)}{\left|\mathcal{A}_{l}\right|}.
\end{aligned}
\]
Moreover, 
\[
\boldsymbol{Q}_{it}=\left(y_{it}-x_{it}^{\top}\beta_{it}^{m}\right)x_{it}=\left(\epsilon_{it}+x_{it}^{\top}\left(\beta_{it}^{0}-\beta_{it}^{m}\right)\right)x_{it}
\]
and 
\[
\sup_{(i,t)}\left\Vert \boldsymbol{Q}_{it}\right\Vert \leq\sup_{(i,t)}\left\{ \left\Vert x_{it}\right\Vert \left(\|\epsilon\|_{\infty}+\left\Vert x_{it}\right\Vert \left\Vert \beta_{it}^{0}-\beta_{it}^{m}\right\Vert \right)\right\} ,
\]
then 
\[
\sup_{(i,t)}\left\Vert \boldsymbol{Q}_{it}\right\Vert \leq c_{4}\sqrt{P}\left(\|\epsilon\|_{\infty}+c_{4}\sqrt{P}\Delta_{n}\right).
\]
By condition (C2), 
\[
P\left(\|\epsilon\|_{\infty}>\sqrt{2c_{1}^{-1}}\sqrt{\log(NT)}\right)\leq\sum_{i=1}^{N}\sum_{t=1}^{T}P\left(\left|\epsilon_{it}\right|>\sqrt{2c_{1}^{-1}}\sqrt{\log(NT)}\right)\leq2(NT)^{-1}.
\]
Then 
\[
\sup_{(i,t)}\left\Vert \boldsymbol{Q}_{it}\right\Vert \leq c_{4}\sqrt{P}\left(\sqrt{2c_{1}^{-1}}\sqrt{\log(NT)}+c_{4}\sqrt{P}\Delta_{n}\right)
\]
and 
\begin{align}
 & \frac{\left(\mathbf{Q}_{jt^{\prime}}-\boldsymbol{Q}_{it}\right)^{\top}\left(\boldsymbol{\beta}_{jt^{\prime}}-\boldsymbol{\beta}_{it}\right)}{\left|\mathcal{A}_{l}\right|}\nonumber \\
 & \leq\left|\mathcal{A}_{\min}\right|^{-1}\left\Vert \boldsymbol{Q}_{jt^{\prime}}-\boldsymbol{Q}_{it}\right\Vert \left\Vert \beta_{it}-\beta_{jt^{\prime}}\right\Vert \leq\left|\mathcal{A}_{\min}\right|^{-1}2\sup_{it}\left\Vert \boldsymbol{Q}_{it}\right\Vert \left\Vert \beta_{it}-\beta_{jt^{\prime}}\right\Vert \nonumber \\
 & \leq2\left|\mathcal{A}_{\min}\right|^{-1}c_{4}\sqrt{P}\left(\sqrt{2c_{1}^{-1}}\sqrt{\log(NT)}+c_{4}\sqrt{P}\Delta_{n}\right)\left\Vert \beta_{it}-\beta_{jt^{\prime}}\right\Vert .
\end{align}
At last, we have 
\begin{eqnarray*}
\boldsymbol{Q}(\boldsymbol{\beta})-\boldsymbol{Q}(\boldsymbol{\beta}^{*}) & \geq & \sum_{l=1}^{L}\left[\sum_{t=1}^{T}\sum_{\substack{(i,t),(j,t)\in\mathcal{A}_{l}\\
i<j
}
}\lambda\rho^{\prime}(4t)\left\Vert \beta_{it}-\beta_{jt^{\prime}}\right\Vert +\sum_{i=1}^{N}\sum_{\substack{(i,t),(i,t^{\prime})\in\mathcal{A}_{l}\\
t<t^{\prime}
}
}\gamma\rho^{\prime}(4t)\right]\left\Vert \beta_{it}-\beta_{jt^{\prime}}\right\Vert -\\
 &  & \sum_{l=1}^{L}\sum_{\substack{(i,t),(j,t^{\prime})\in\mathcal{A}_{l}\\
i<j,t<t^{\prime}
}
}2\left|\mathcal{A}_{\min}\right|^{-1}c_{4}\sqrt{P}\left(\sqrt{2c_{1}^{-1}}\sqrt{\log(NT)}+c_{4}\sqrt{P}\Delta_{n}\right)\left\Vert \beta_{it}-\beta_{jt^{\prime}}\right\Vert \\
 & \geq & \sum_{l=1}^{L}\sum_{\substack{(i,t),(j,t^{\prime})\in\mathcal{A}_{l}\\
i<j,t<t^{\prime}
}
}min(\lambda,\gamma)\rho^{\prime}(4t)\left\Vert \beta_{it}-\beta_{jt^{\prime}}\right\Vert -\\
 &  & \sum_{l=1}^{L}\sum_{\substack{(i,t),(j,t^{\prime})\in\mathcal{A}_{l}\\
i<j,t<t^{\prime}
}
}2\left|\mathcal{A}_{\min}\right|^{-1}c_{4}\sqrt{P}\left(\sqrt{2c_{1}^{-1}}\sqrt{\log(NT)}+c_{4}\sqrt{P}\Delta_{n}\right)\left\Vert \beta_{it}-\beta_{jt^{\prime}}\right\Vert .
\end{eqnarray*}

If $t=o(1)$, then $\rho^{\prime}(4t)\rightarrow1$. Since $\lambda\gg\Delta_{n}$,
$\gamma\gg\Delta_{n}$, $P=o(NT)$, and $|\mathcal{A}_{min}|^{-1}P=o(1)$,
then $min(\lambda,\gamma)\gg\left|\mathcal{A}_{\min}\right|^{-1}\sqrt{P\log(NT)}$
and $min(\lambda,\gamma)\gg\left|\mathcal{A}_{\min}\right|^{-1}P\Delta_{n}$.
Therefore, $\boldsymbol{Q}(\boldsymbol{\beta})-\boldsymbol{Q}(\boldsymbol{\beta}^{*})\geq0$,
if $N$ or $T$ tend towards infinity.

\subsection*{Proof of COROLLARY 1}

According to asymptotic equivalence, we can conclude COROLLARY 1 based
on Theorem 1 and Theorem 2. Besides, we can also get $\hat{L}\overset{p}{\rightarrow}L^{0}$.

According to assumption (C2) (iv) $\left\vert \mathcal{A}_{\min}\right\vert \gg(LP)^{1/2}(NT)^{3/4}$,
we can get $L^{0}P=o_{p}\left(\frac{1}{(NT)^{1/6}}\right)$. Thus,

\[
{\displaystyle \lim_{(N,T)\to\infty}\hat{\sigma}^{2}={\displaystyle \lim_{(N,T)\to\infty}\frac{\boldsymbol{\epsilon}^{\top}\boldsymbol{\epsilon}}{NT}-{\displaystyle \lim_{(N,T)\to\infty}\frac{\boldsymbol{\epsilon}^{\top}\boldsymbol{X}}{NT}\left(\frac{\boldsymbol{X}^{\top}\boldsymbol{X}}{NT}\right)^{-1}\frac{\boldsymbol{X}^{\top}\boldsymbol{\epsilon}}{NT}}}}.
\]
According to assumption (C1) and assumption (C2) that $\frac{\boldsymbol{X}^{\top}\boldsymbol{\epsilon}}{NT}\overset{p}{\rightarrow}0$
and $\frac{\boldsymbol{X}^{\top}\boldsymbol{X}}{NT}=O_{p}(1)$, as
$(N,T)\to\infty$ , $\hat{\sigma}^{2}\overset{p}{\rightarrow}\sigma^{2}$
is obtained. 
\end{document}